\documentclass[final,authoryear,5p,twocolumn]{elsarticle}
\usepackage{a4,url} % remove on submission
\usepackage[cm]{fullpage}
\usepackage{doi}
\usepackage[all]{hypcap}
\usepackage[english]{babel}
\usepackage{xspace}
\usepackage{natbib}
\usepackage{amsmath,amssymb,graphicx,wrapfig}
\usepackage[hang]{subfigure}
\usepackage{color,xcolor}
\usepackage{lineno,multirow}

\definecolor{darkgreen}{RGB}{0,100,0}
\definecolor{bgcolour}{RGB}{150,0,250}

\xspaceaddexceptions{]}

\newcommand{\degree}{\ensuremath{^\circ}}
\newcommand{\dSi}{\text{Si}\ensuremath{_\text{diss}}\xspace}
\newcommand{\dAl}{\text{Al}\ensuremath{_\text{diss}}\xspace}
\newcommand{\ads}{\text{Al}\ensuremath{_\text{ads}}\xspace}

\newcommand{\bsi}{\text{bSiO}\ensuremath{_2}\xspace}

\begin{document}
\begin{frontmatter}
\journal{Journal of Marine Systems}
\title{Aluminium in an ocean general circulation model compared with the
West Atlantic Geotraces cruises}%\footnote{%
%If there are any updates to this article, these can be found at \url{http://arxiv.org/abs/1202.4679}.
%}}
%Published in \emph{Journal of Marine Systems}.  
%\href{http://dx.doi.org/10.1016/j.jmarsys.2012.05.005}{doi:10.1016/j.jmarsys.2012.05.005}.
%This work is licensed under \emph{Creative Commons Attribution-ShareAlike 3.0 Unported License} (CC BY-SA). This note overrides other licenses, giving you true freedom over this preprint.
%}}
\author[knmi]{M.M.P.~van~Hulten\corref{cor1}}
\ead{hulten@knmi.nl}
\author[knmi]{A.~Sterl}
\author[lsce,uct]{A.~Tagliabue}
\author[lsce]{J.-C.~Dutay}
\author[lsce]{M.~Gehlen}
\author[nioz,rug]{H.J.W.~de~Baar}
\author[nioz,ucsc]{R.~Middag}
%\author{M.M.P.\ van Hulten$^1$, {\small A.\ Sterl$^1$, A.\ Tagliabue$^2$, J.-C.\ Dutay$^3$, M.\ Gehlen$^3$, H.J.W.\ de Baar$^{4,5}$, R.\ Middag$^{5,6}$ }}
%\affil{$^1$Royal Netherlands Meteorological Institute (KNMI),
%$^2$University of Cape Town (UCT),
%$^3$Laboratoire des Sciences du Climat et de l'Environnement (LSCE),
%$^4$University of Groningen (RUG),
%$^5$Royal Netherlands Institute for Sea Research (NIOZ),
%$^6$University of California Santa Cruz (UCSC)
%}
\address[knmi]{Royal Netherlands Meteorological Institute (KNMI)}
\address[uct]{University of Cape Town (UCT)}
\address[lsce]{Laboratoire des Sciences du Climat et de l'Environnement (LSCE)}
\address[rug]{University of Groningen (RUG)}
\address[nioz]{Royal Netherlands Institute for Sea Research (NIOZ)}
\address[ucsc]{University of California Santa Cruz (UCSC)}
% Abstract, version 2.0
% 
\begin{abstract}
\let\thefootnote\relax\footnotetext{%
%\textbf{List of abbreviations:}\\
\begin{tabular}{rl}
    NEMO    &   {Nucleus for European Modelling of the Ocean}   \\
    OPA     &   {Oc\'ean PArall\'elis\'e}   \\
    BEC     &   {Biogeochemical Elemental Cycling}\\
    INCA    &   {Integrated Nitrogen model for multiple source assessment in CAtchments}\\
    LMDzT   &   {Laboratoire de M\'et\'eorologie Dynamique GCM}\\
    HAMOCC  &   {HAmburg Model for the Ocean Carbon Cycle}\\
    LSG     &   {(Hamburg) Large Scale Geostrophic model}\\
    OGCM    &   {Ocean General Circulation Model}\\
    IPY     &   {International Polar Year}\\
    BATS    &   {Bermuda Atlantic Time-series Study}\\
    SAFe    &   {Sampling and Analysis of Fe}\\
    POM     &   {Particulate Organic Matter}\\
    NADW    &   {North Atlantic Deep Water}\\
    AABW    &   {AntArctic Bottom Water}\\
    AAIW    &   {AntArctic Intermediate Water}\\
    SAAMW   &   {SubAntArctic Mode Water}\\
    MOC     &   {Meridional Overturning Circulation}\\
    OSF     &   {Overturning Stream Function}
\end{tabular}
}

% Model
A model of aluminium has been
developed and implemented in an Ocean General Circulation Model (NEMO-PISCES).
%
% Dust deposition and solubility
In the model, aluminium enters the ocean by means of dust deposition.
%
% Internal processes
The internal oceanic processes are described by advection, mixing and
reversible scavenging.
%
% Comparison with observations
The model has been evaluated against a number of selected high-quality datasets
covering much of the world ocean, especially those from the West Atlantic
Geotraces cruises of 2010 and 2011.  Generally, the model results are in fair
agreement with the observations.  However, the model does not describe well the
vertical distribution of dissolved Al in the North Atlantic Ocean.  The model
may require changes in the physical forcing and the vertical dependence of the
sinking velocity of biogenic silica to account for other discrepancies.
To explore the model behaviour,
sensitivity experiments have been performed, in which we changed the key
parameters of the scavenging process as well as the input of aluminium into the
ocean.
%
% Conclusion
This resulted in a better understanding of aluminium in the ocean, and it is
now clear which parameter has what effect on the dissolved aluminium
distribution and which processes might be missing in the model, among which
boundary scavenging and biological incorporation of aluminium into diatoms.

\vfill

\paragraph{Keywords}
aluminium, dust deposition, modelling, PISCES, biogenic silica, scavenging, GEOTRACES

%\paragraph{Notes}
%This work is licensed under \emph{Creative Commons Attribution-ShareAlike 3.0 Unported License} (CC BY-SA). This note overrides other licenses, giving you true freedom over this preprint.\\
%\noindent The paper has been accepted by \emph{Journal of Marine Systems} for the special issue \emph{Traces and Tracers}.\\
%\noindent If there are any updates to this article, these can be found at \url{http://arxiv.org/abs/1202.4679}.
\vskip .3cm
{\small
This paper is published in \emph{Journal of Marine Systems} for the special issue \emph{Traces and Tracers}.
The official publication can be found at \href{http://dx.doi.org/10.1016/j.jmarsys.2012.05.005}{doi:10.1016/j.jmarsys.2012.05.005}.
This work is licensed under \emph{Creative Commons Attribution-ShareAlike 3.0 Unported License} (CC BY-SA). This note overrides other licenses, giving you true freedom over this preprint.
}
\end{abstract}

\end{frontmatter}

\newpage
%\begin{linenumbers}
% Introduction, version 2.0
% 
\section{Introduction}   \label{sec:introduction}
The distribution and cycling of aluminium (Al) in the ocean has received
attention for a variety of reasons.
Firstly, if the Al cycle is understood well, aluminium surface concentrations
can be used to constrain atmospheric dust deposition fields
\citep{gehlen2003,measures2005,han2008,measures2010,han2010}, which
 are used to predict aeolian iron addition to the euphotic zone.  This is
important, since iron is an essential trace-nutrient for phytoplankton, thus
its availability has a direct consequence on primary production and air-sea
CO$_2$ exchange \citep{martin1990,debaar2005,boyd2007}.
Secondly, there is evidence that Al inhibits the solubility of sedimentary
biogenic silica
\citep{lewin1961,vanbennekom1991,dixit2001,sarmiento2006,emerson2006}.  If
less biogenic silica gets dissolved from sediments, eventually there will be
less silicic acid available in the euphotic zone, which will reduce diatom
production as silicon is an essential major nutrient for diatoms.  Modified
diatom productivity will impact ocean food webs and the export of organic
carbon to the ocean's interior.  For advancement in both of these fields of
interest a good understanding of the Al cycle is pertinent.

Currently it is assumed that the major source of Al to the ocean is via dust
deposition \citep[e.g.][]{orians1986,maring1987,kramer2004,measures2005}.  When
dust enters the ocean, a part of its aluminium content (1--15\%) dissolves in
the uppermost layer and is quickly distributed over the mixed layer by turbulent
mixing.  Most dust remains in the particulate phase and sinks to the bottom of
the ocean, while a small fraction might dissolve in the water column.
The dust that does not dissolve at all is buried in the sediment
\citep{gehlen2003,sarmiento2006}.

Arguments that dissolution occurs primarily in the upper layer of the ocean come
from shipboard experiments and atmospheric moisture considerations.
% 
% shipboard experiments:
\citet{maring1987} and \citet{measures2010} showed that within a day after
deposition, most of dissolvable Al will be dissolved.  Assuming a sinking speed
of dust of 30 m/day, most Al would then dissolve in the upper 30 m of the ocean.
This depth is shallower than the mixed layer depth, which means that there is
little dissolution below the mixed layer.
% 
% rain and atmospheric processes:
Even though some earlier studies showed that most deposition is dry
\citep{jickells1994,jickells1995}, more recent work shows that dust deposition
is mostly wet \citep{guerzoni1997,vink2001}.  It has been argued that most dust
passes through low $pH$ environments in the atmosphere, which means that for wet
deposition Al is already dissolved when it enters the ocean surface.  Since the
wetly deposited Al is most important, dissolution in the surface ocean is most
relevant (as a lower bound, since dry deposition results in both surface and
water column dissolution).  Furthermore, based on the same low $pH$ argument,
dissolvable Al from dry deposition is likely to instantaneously dissolve in the
surface ocean \citep[][and references herein]{measures2010}, making the relative
amount of surface dissolution even higher compared to water column dissolution.

Fluvial input can be thought to be important as well, since rivers carry large
concentrations of Al, but in estuaries and coastal regions this Al is removed
by scavenging of Al onto particles
\citep{mackin1986control,mackin1986,orians1986,brown2010aluminium}.  There are
also indications for Al input as a consequence of sediment remobilisation, as
in the Arctic Ocean \citep{middag2009} and North Atlantic Ocean
\citep{moran1991}.  However, the importance of sedimentary sources can vary by
basin (e.g.\ in the Southern Ocean these are small as shown by \citet{moran1992}
and \citet{middag2011}).  Finally, hydrothermal vents are thought to only play a
minor role (e.g.\ \citet{hydes1986,middag2010}).  In summary, the dominant
external source of aluminium in the ocean is atmospheric dust deposition.

% Scavenging
Dissolved aluminium (\dAl) is removed mainly by particle scavenging
\citep{stoffyn1982,orians1986,moran1989,bruland2006}.  This is the combination
of adsorption onto a solid surface, followed by sinking due to insufficient
buoyancy of the particulates in the seawater
\citep{goldberg1954,bacon1982,bruland2006}.  Typically, scavenging is deemed
to be reversible, which means that during sinking release of the adsorbed, or
particulate, aluminium (\ads) may occur.  This happens both directly (by
desorption) and indirectly (by dissolution of the biogenic carrier particles).
As a consequence, \dAl concentrations increase with depth
\citep{bacon1982,anderson2006}.  In this way \dAl is distributed over depth
more efficiently than due to mixing and water mass transport.
Aluminium is scavenged relatively efficiently and therefore has a relatively
short residence time in the ocean (100--200 yr) \citep{orians1985}.

Except for scavenging, there are strong suggestions from observations in
certain regions that Al is biologically incorporated into the siliceous cell
walls of diatoms \citep{mackenzie1978,stoffyn1979,moran1988,gehlen2002}.  It
seems that Al does not play an essential role for the diatoms, but it can be
incorporated functioning as a replacement for silicon (Si), since it is similar
in size.  Therefore it is likely that the incorporation ratio Al\,:\,Si is
close to that of the surrounding waters.  These regions include the Arctic
Ocean \citep{middag2009} and the Mediterranean Sea
\citep{mackenzie1978,hydes1988,chou1997}.  Given the ratio Al\,:\,Si of
incorporation into the diatom in the photic zone, after remineralisation
anywhere in the water column, the same dissolved Al\,:\,Si will be present, as
long as this is the only source of Al and Si.  When the dissolved Al and Si is
then advected into the Atlantic Ocean by the North Atlantic Deep Water (NADW),
this signal slowly disappears because of other sources of Al among which dust
deposition and possibly sediment resuspension \citep{middag2011} and a source
of Si from Antarctic Bottom Water.

% Previous work
Recent years have seen the development of models of the marine biogeochemical
cycle of Al. \citet{gehlen2003} implemented a basic scavenging model, while
\citet{han2008} also included a biological aluminium incorporation module.
%
% description of Gehlen
\citet{gehlen2003} had the objective of testing the sensitivity of modelled Al
fields to dust input and thus to evaluate the possibility for constraining dust
deposition via \dAl.  To this purpose they embedded an Al cycle in the HAMOCC2
biogeochemical model.  The model consists of an equilibrium relation between,
on the one hand, \ads and, on the other hand, \dAl.  In chemical equilibrium
\ads is proportional to the biogenic silica (\bsi) concentration.  In their
work, as well as this paper, the term \emph{biogenic silica} or \bsi refers to
the detrital fractions which is fuelled by diatoms and other silicifying
phytoplankton, which have no stable organic matter coating and sink.
When \bsi sinks to the seafloor (together with adsorbed Al), it is buried.  The
resulting concentration of modelled \dAl was of the same order as the then
published observations, but it suggested a significant overestimation of
Saharan dust input \citep{gehlen2003} when the dust deposition field of
\citet{mahowald1999} was used.
% description of Han
The main goal of \citet{han2008} was to better constrain the dust deposition
field.  For this purpose they used the Biogeochemical Elemental Cycling (BEC)
model improved by \citet{moore2008} as a starting point.  They used all
dissolved Al datasets used by \citet{gehlen2003} and added more datasets.
Except for scavenging \citet{han2008} added a biological Al uptake module where
the Al\,:\,Si uptake ratio is a function of the ambient Al and Si
concentrations \citep{han2008}.  The surface residence time of Al for both modelling
studies varies strongly between different locations (from less than one year to
almost 80 years).

% Open questions
Overall, there are a number of questions regarding the oceanic Al cycle that
remain to be fully addressed.  These touch on issues of ocean circulation, the
specific sources and sinks of Al in different parts of the world ocean and what
processes are needed to accurately simulate the oceanic distribution of Al.
Firstly, there is the question of the meridional (north to south) distribution
of Al through the Atlantic Ocean.  In the North Atlantic Ocean and northern
seas, water sinks and forms NADW, which is then transported southward (e.g.,
\citet{lozier2010:deconstructing,gary2011}).  In the deep Atlantic Ocean
dissolved silicon (Si) concentration increases from north to south
\citep{ragueneau2000,sarmiento2006}, while the concentration of \dAl stays
relatively constant until about 20\degree S and then decreases (\citet{middag_prep};
see also Section \ref{sec:results}).  Thus it has a generally
opposite behaviour compared to Si.  Since there are strong suggestions that the
processes controlling the distribution of Si and Al are linked, the question is
raised how this negative correlation is possible.
Secondly, there is the question about the observed profiles of \dAl at
different locations in the ocean.  Generally, profiles of \dAl have a
reversible-scavenging profile (increasing with depth) and often with a minimum
near 1\;km depth and a maximum at the surface because of dust deposition.
However, observations in the Mediterranean Sea \citep{hydes1988,chou1997} and
IPY-Geotraces-NL observations in the eastern Arctic \citep{middag2009} show
that there is a strong positive relation between aluminium and silicon.  This
supports the hypothesis of biological incorporation of aluminium into the cell
wall of diatoms.

%\paragraph{Modelling aluminium.}
These issues can be analysed further by the use of numerical models.  Since
there is a strong spatial variation in aluminium concentration (and its relation
to silicon), an ocean general circulation model should be used to simulate the
distribution of \dAl.  Potentially crucial parameters and sources can be
modified in the model to test its sensitivity to these changes.  In this way a
better understanding of the aluminium cycle can be reached.

% Goal
In this paper the observed distribution of Al is modelled and the
processes driving it are examined.
Based on new observations and previous work on aluminium modelling
\citep{gehlen2003,han2008} a model of aluminium based on dust deposition and
scavenging by biogenic silica is formulated.  This model and the
configuration of the simulations will be set out in the following section.  Then
the observations which are used to check and improve the model will be discussed.
The results of the several experiments follow in Section \ref{sec:results}, as
well as a comparison with the observations.  The discussion in Section
\ref{sec:discussion} comprises of a comparison between our model results and
\citet{gehlen2003}, a timescale analysis and based on that a discussion of our
simulations.  Our results are not compared with \citet{han2008}, since we
have not performed simulations with biological aluminium incorporation.
The last section (\ref{sec:conclusion}) sums up the conclusions, gives an
outline for further development of the model and suggests what further study is
needed.

% Methods, version 2.0
% 
\section{Methods}    \label{sec:methods}
\subsection{Model description}  \label{sec:model}
To model the three-dimensional distribution of \dAl, the Ocean General
Circulation Model (OGCM) called Nucleus for European Modelling of the Ocean
(NEMO) is used \citep{madec2008}.
For this study we use PISCES, one of the biogeochemical components available in
NEMO \citep{aumont2006,ethe2006}, which has been employed for many other
studies concerning trace metals, as well as large scale ocean biogeochemistry
\citep{aumont2006,dutay2009,arsouze2009,dutay2009,tagliabue2010,tagliabue2011}.  PISCES is
run off-line forced by a climatological year of physical fields including
subgrid turbulence.  The forcing frequency was set to five days.  These
physical fields were calculated by the circulation component of NEMO named OPA
\citep{madec1998}, which was forced by satellite derived wind stress data.
Details on the forcings and model configuration can be found in
\citet{arsouze2009} and \citet{dutay2009}.

\begin{figure}
    \centering
    \includegraphics[width=\columnwidth]{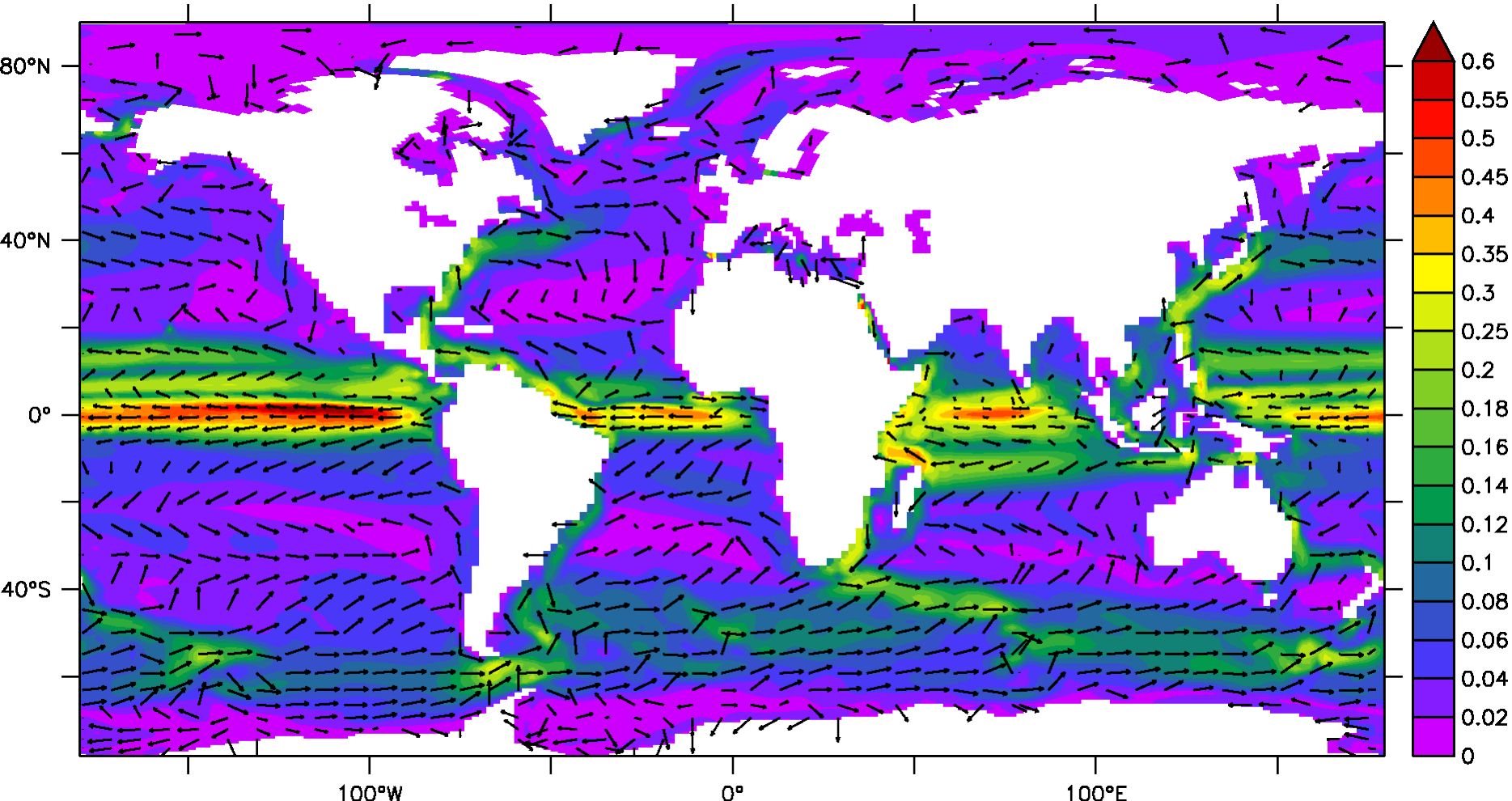} \caption{Velocity at
    the surface, yearly average.
    The colour indicates the speed in m/s, while the vectors represent
    the direction of the flow.  Vectors are plotted at every fourth gridpoint.}
    \label{fig:opa}
\end{figure}

All input and output fields are defined on the ORCA2 grid, an irregular grid
covering the whole world ocean with a nominal resolution of $2\degree\times 2\degree$, with the meridional
resolution increased near the equator and two `north poles' in Canada and
Russia to eliminate the coordinate singularity in the Arctic Ocean.  Its
vertical resolution is 10\,m in the upper 100\,m, increasing downward such that
there are 30 layers in total.  In Fig.\
\ref{fig:opa} the annually averaged surface velocity field of the OPA output is
plotted on a lon-lat grid.
It shows the main features of the ocean circulation such as the equatorial
current systems, the western boundary currents (Gulf Stream and Kuroshio) as
well as the Antarctic Circumpolar Current.

\subsubsection{Biogeochemical model}   \label{sec:model_pisces}
The aluminium cycle (see Section \ref{sec:model_aluminium}) is implemented in
PISCES, a biogeochemical model simulating the cycle of carbon and the major
nutrients (nitrate, phosphate, ammonium, silicic acid, and iron), along with two
phytoplankton types (diatoms and nanophytoplankton), two zooplankton grazers
(micro- and meso-zooplankton), two classes of particulate organic carbon (small
and large) of differential labilities and sinking speeds, as well as calcite and
biogenic silica.  The standard version of PISCES accounts for 24 tracers.  For a
more detailed description of PISCES see the auxiliary material of
\citet{aumont2006}.

The aluminium model interacts with the silicon cycle.  The model PISCES
distinguishes three silicon pools: the silicon content of living diatoms (diatom
Si); the silicon content of dead, sinking diatoms (biogenic Si) and dissolved
silicic acid.  In the model silicic acid and other nutrients are supplied to the
ocean by means of atmospheric dust deposition, rivers and sediment mobilisation
\citep{aumont2006,gehlen2007}.

\subsubsection{Aluminium model}    \label{sec:model_aluminium}
For the modelling of Al, two new tracers are introduced to PISCES: dissolved
aluminium (\dAl) and adsorbed aluminium (\ads).  Our Al model follows the
approach of \citet{gehlen2003}, i.e., the internal processes (described below) of
adsorption and desorption are identical to the model of \citet{gehlen2003},
although the associated biogeochemical model is more complex.

For the standard model configuration, the sole source of Al to the ocean is by
dissolution of dust particles in the ocean surface.  In one of the sensitivity
experiments, a small fraction is also dissolved in the water column (see Section
\ref{sec:model_sensitivity}).  The used Al\,:\,Fe dust fraction of
8.1\,:\,3.5 is based on the mass percentages of Al and Fe known to be
present in the Earth's crust \citep{wedepohl1995}.  The solubility of Al from
dust is not well constrained, with reported values ranging from 0.5--86\%, but
is probably in the range of 1--15\%
\citep{orians1986,measures2000,jickells2005}.  Furthermore, it is likely that
the source of dust has an impact on the solubility of the Al fraction
\citep[][and references herein]{baker2006,measures2010}.  The dissolution
of dust is taken as a constant with respect to the origin and location of deposition.
The dissolution occurs only in the upper model layer, and is described by the
following equation:
\begin{equation}
    \frac{\partial}{\partial t} [\dAl]\bigg|_{z=0} = \alpha \mathcal{D}_\text{Al}/H_1 \;,
\end{equation}
where $\mathcal{D}_\text{Al}$ is the Al flux into the ocean, $\alpha$ is the
fraction of Al that is dissolved, and $H_1$ is the thickness of first model
layer.  Square brackets denote the concentration of the tracer.  Since in the
mixed layer vertical mixing is rapid, adding \dAl in the first layer
is effectively the same as adding it spread out over the mixed layer.  The Al
that does not dissolve from dust, either at the surface or in the water column,
is assumed not to play any role in the biogeochemical cycle of Al on our
timescales of interest, and can be thought of as being buried in marine
sediments.

Dissolved Al is assumed to adsorb onto
biogenic silica particles and, aside from external inputs, the \dAl
concentration is essentially governed by adsorption and desorption.  The \dAl
and \ads concentrations follow the following reversible first-order adsorption
equation \citep{gehlen2003}:
\begin{equation}
    \frac{\partial}{\partial t} [\ads] \bigg|_\text{adsorption} = \kappa ([\text{Al}^\text{eq}_\text{ads}] - [\ads]) \;,
    \label{eqn:scav-gehlen}
\end{equation}
where
\begin{equation}   \label{eqn:eq-gehlen}
    [\textrm{Al}^\textrm{eq}_\textrm{ads}] = k_d \cdot [\dAl] \cdot [\bsi] \;,
\end{equation}
in which \bsi stands for biogenic silica and
[$\textrm{Al}^\textrm{eq}_\textrm{ads}$] is the chemical equilibrium
concentration of \ads.  The parameter $k_d$ is the partition coefficient and
$\kappa$ is the first order rate constant for equilibration of \ads to
$\textrm{Al}^\textrm{eq}_\textrm{ads}$.
Since total Al is conserved when only internal processes are concerned, the time
derivative of \dAl is equal to the negative of the time derivative of \ads.

Eq.\ \ref{eqn:eq-gehlen} describes the chemical equilibrium between \dAl
and \ads, and Eq.\ \ref{eqn:scav-gehlen} illustrates that some time is
needed before equilibrium is reached (it is not modelled as an instantaneous
process).
As noted by \citet{gehlen2003}, this is a purely empirical model and a more
mechanistic description of the adsorption/desorption of \dAl onto \bsi is
difficult, since in reality, the pool of observed ``dissolved aluminium'' is
operationally defined, i.e., it exists as different kinds of ions as well as
colloids that can go through a filter of, e.g., 0.2 $\mu$m.  Also the adsorbent,
\bsi, exists in various forms and it is, for instance, difficult to even define
the charge density, which is important for its potential to adsorb \dAl.
Usually the surface is covered by natural organic matter
\citep{colloids:lead2007,colloids:filella2007}.
Furthermore, the main scavengers for aluminium in the open ocean also include
Particulate Organic Matter (POM), calcium carbonate (CaCO$_3$) and even
particulates that are not part of PISCES (such as lithogenic particles that do
not participate actively in biogeochemical cycles).
Hence it is difficult to conceive a mechanistic model of ad- and desorption of
``\dAl'' onto the poorly-defined pool of particulates.

%FIXME: in the following: Loucaides (2010) does not take into account the
%natural organic matter which actually makes \bsi positively charged.  On such
%particulates the *negative* (hydroxided) Al ions can adsorp!  In any case, we
%need to make this more consistent.
In our model only \bsi is used as a scavenger of \dAl.  Biogenic silica is likely
to be the most important because (1) it sinks quickly, so that Al has little
chance to significantly desorb from \bsi (ballast effect), and
(2) positively charged aluminium
ions are easily adsorbed onto the negatively charged surface of \bsi particles
\citep{dixit2002,loucaides2010}.
%While particulates are also presumed to be negatively
%charged, observational studies confirm that biogenic silica is a strong
%scavenger of aluminium \citep{stoffyn1982,moran1989}.

Sinking of \bsi and \ads is described as follows.
In the model, \bsi sinks with a speed of $w_\text{ML} = 30$ m/day in the mixed
layer.  Below the mixed layer, it increases linearly with depth, reaching a
value of $w_\text{4km} = 200$ m/day at 4 km below the mixed layer
\citep{aumont2006}.  This means that \ads also sinks with this speed, while it
can desorb following equations \ref{eqn:scav-gehlen} and \ref{eqn:eq-gehlen}.
For every layer in the model, the following equation holds:
\begin{equation}
\frac{\partial}{\partial t} [\ads] \bigg|_\text{sinking} = - \frac{\partial}{\partial z} (w_s \cdot [\ads]) \;,
\end{equation}
where $w_s$ is the sinking speed, given in the mixed layer by a constant
$w_\text{ML}$ and below the mixed layer increasing with depth according to:
\begin{equation}
    w_s = w_\text{ML} + (w_\text{4km} - w_\text{ML}) \cdot
          \frac{z-D_\text{ML}}{4000 \text{m} - D_\text{ML}} \;,
\end{equation}
where $z$ is the model depth.
Once sunk to the ocean floor, we assume that the aluminium is buried permanently.
This is described by the following change of \ads in the bottom layer:
$\frac{\partial}{\partial t} [\ads] \big|_{z=z_\text{sed}} = - [\ads] \cdot w_s / H_\text{sed}$, where
$H_\text{sed}$ is the thickness of the bottom model layer.

Adding advection and mixing, the full equations for [\ads] and [\dAl] away from
the boundaries are as follows:
\begin{align}
    \frac{\text{d}}{\text{d}t} [\ads]
        = \kappa &([\ads^\text{eq}] - [\ads]) - \frac{\partial}{\partial z} (w_s \cdot [\ads])  \notag \\
         &+(\mathcal{A} \nabla_H^2 + \nu_E \frac{\partial^2}{\partial z^2}) [\ads]
    \;,\\
    \frac{\text{d}}{\text{d}t} [\dAl] = -\kappa &([\ads^\text{eq}] - [\ads]) \notag\\
    &+(\mathcal{A} \nabla_H^2 + \nu_E \frac{\partial^2}{\partial z^2}) [\dAl] \;,
    \label{eqn:base_equations}
\end{align}
where the general advection ($\mathbf{v}\cdot\mathbf{\nabla}$) of the tracers is
implicit in the full time derivatives.  $\mathcal{A}$ and $\nu_E$ are the
horizontal and vertical eddy diffusivity coefficients, respectively.
%These equations are discretised in the vertical by replacing the Dirac deltas
%by Kronecker deltas and dividing by the respective layer thicknesses.

The code of the NEMO model, as well as PISCES and the aluminium model, is free
software: software that can be used, studied, and modified without restriction
(except that published modifications must fall under the same license).  It is
available at \url{http://www.nemo-ocean.eu/}.
Version 3.1 of the NEMO model is used, specifically svn revision 1183.  This can
be accessed through \url{http://forge.ipsl.jussieu.fr/igcmg/svn/modipsl/}.
The aluminium specific code is available under the same conditions as an
electronic supplement to this article: \texttt{trcal\_sed.F90} describes
aluminium deposition into the ocean and into the sediment, and
\texttt{trcal\_rem.F90} describes the internal processes of ad- and desorption.

\subsubsection{Input fields}    \label{sec:model_input}
\begin{figure*}
    %\centering      %TODO: maak twee figuren consistent (dus Al XOF alles)
    \subfigure[Dust deposition (g m$^{-2}$ yr$^{-1}$) at the ocean surface]{
        \includegraphics[width=\columnwidth]{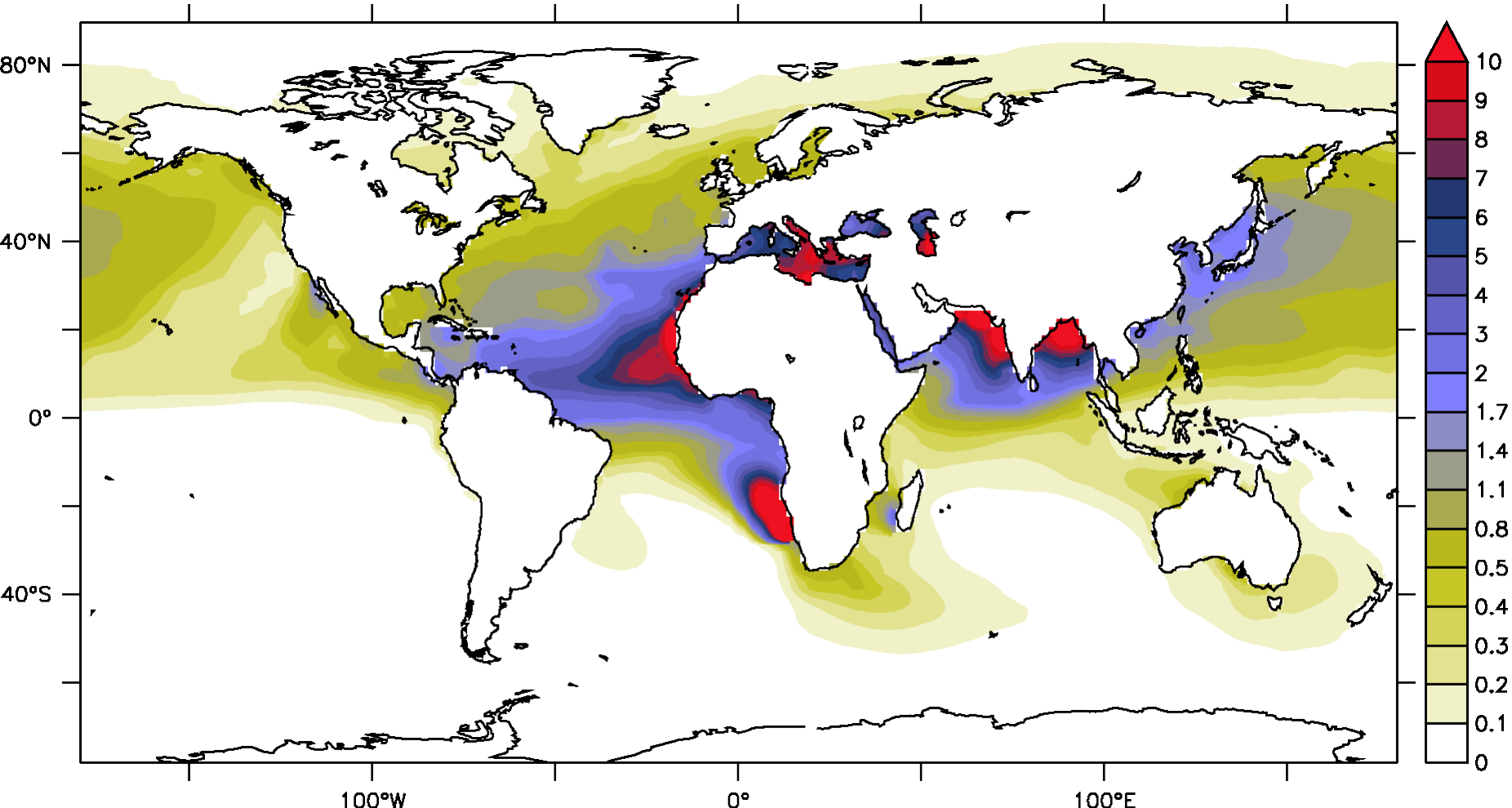}
        \label{fig:dust}
    }
    \subfigure[Sediment sources of Al (arbitrary units) as implemented in
               PISCES for the Fe model]{
        \includegraphics[width=\columnwidth]{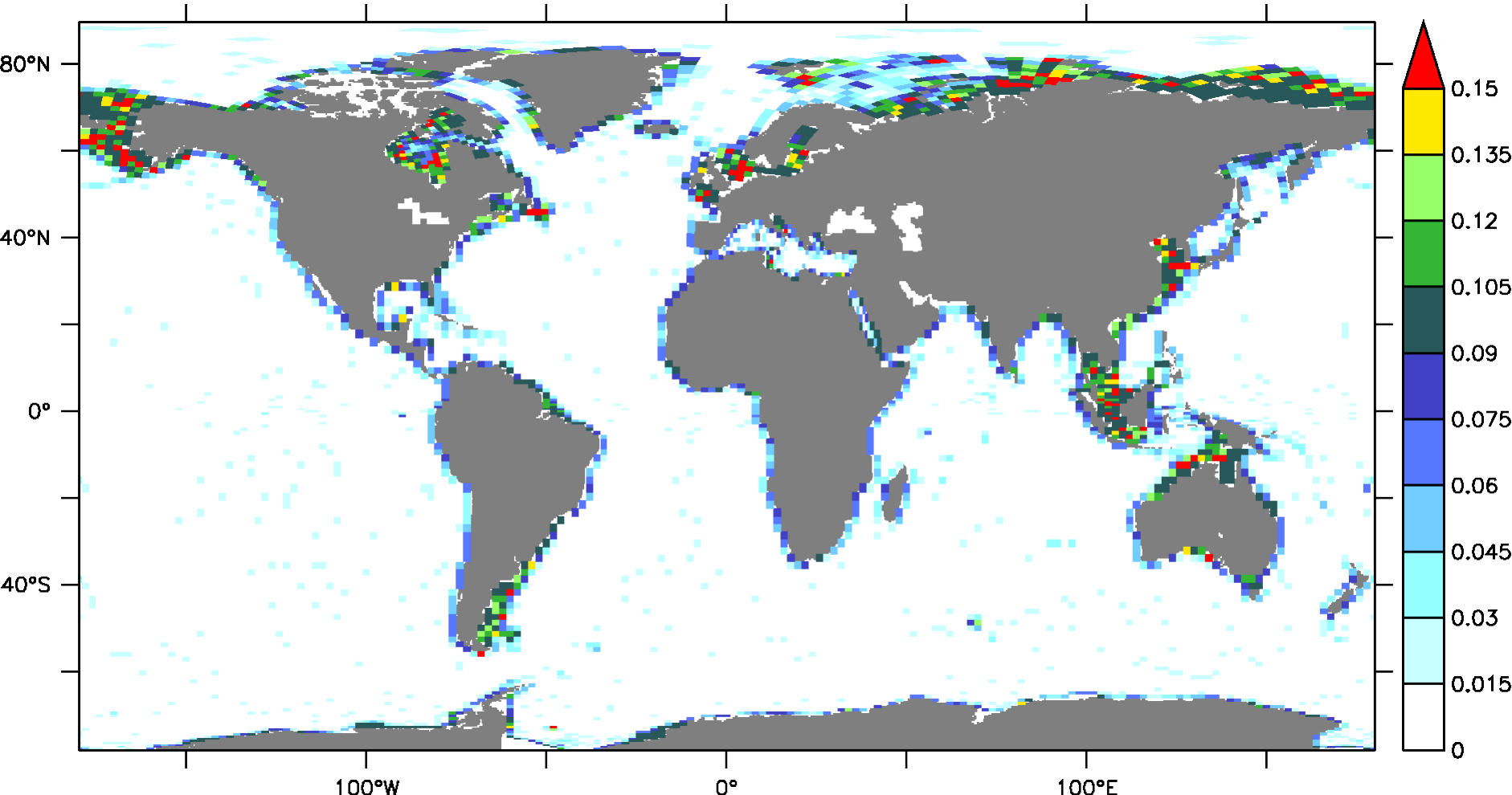}
        \label{fig:margins}
    }
    \caption{Aluminium sources used to force the model, based on the Fe input in
    PISCES, using using an Al\,:\,Fe crustal ratio of 8.1\,:\,3.5.}
    \label{fig:al_input}
\end{figure*}

The dust deposition field was taken from the output of the INCA model, an
atmospheric dust model \citep{whitehead1998}.  INCA is the aerosol module of
the LMDzT atmospheric model \citep{schulz2009}.  The resulting climatology is
used and described by \citet{aumont2008} and evaluated in \citet{textor2006}.
The surface deposition is shown in Fig.~\ref{fig:dust}.
Most dust is deposited in the Atlantic Ocean, west of the Sahara and 
west of the Kalahari Desert.  Another important dust deposition site is found
in the northern Indian Ocean.  Dust deposition east of Asia in the Pacific Ocean
is much smaller, and in the South Pacific Ocean, the Southern Ocean and the
Arctic Ocean there is almost no deposition of dust.  The amount of dust per
basin is listed in Table \ref{tab:dust}.
\begin{table*}
\centering
\begin{tabular}[t]{l|r@{.}lr@{.}l|r@{.}l}
\hline
\multirow{2}{*}{Basin:}          & \multicolumn{4}{l}{\emph{Dust deposition:}} & \multicolumn{2}{|l}{\emph{Sediment sources:}}  \\
%\hline
                & \multicolumn{2}{l}{Absolute (Tg/yr)}  & \multicolumn{2}{l}{Relative (\%)} & \multicolumn{2}{|l}{Relative (\%)} \\

\hline
Atlantic Ocean  & 128&3              & 47&9     &   18&1    \\
Indian Ocean    &  58&3              & 21&8     &    7&4    \\
Pacific	Ocean   &  54&3              & 20&3     &   27&9    \\
Southern Ocean  &   0&3              &  0&12    &   11&4    \\
Arctic Ocean    &   1&5              &  0&55    &   22&8    \\
Mediterranean Sea& 17&8              &  6&6     &    3&6    \\
other seas      &   7&4              &  2&8     &    8&9    \\
\emph{total}    & 267&9              &100&0     &  100&0    \\
\hline
\end{tabular}
\caption{Amount of yearly dust and sediment input in each basin.  The Southern
Ocean is defined as the ocean south of 59\degree S, without overlap with the
Pacific, Atlantic and Indian Oceans.}
% Southern Ocean definition here is rather south, but 60\degree S is the
% definition according to the International Hydrographic Organization (IHO).
\label{tab:dust}
\end{table*}

According to \citet{textor2006} the model that is used to create this dust
deposition field is validated against high-quality observational datasets, but
there are a number of uncertainties, because there are not enough observations
available to sufficiently constrain any dust model globally.  The spatial
resolution is nominally 2\degree, and the temporal resolution is one month
(twelve time steps per year).  In reality, dust deposition is highly sporadic
\citep{aumont2008,rijkenberg2008}.  Dust events are expected to be highly local
and there is sufficient horizontal mixing to dilute the \dAl distribution, such
that there is no significant concentration change.
This has also been observed in the North Atlantic Ocean \citep{rijkenberg2012}.
% TODO?: What about the Al residence time (in the mixed layer)?  If this is much
% higher than the average time between dust events, dust events would be 'local
% in time' as well.
Furthermore, our aim is to look at averages of [\dAl] over at least several
months, and therefore it is acceptable to use monthly averaged dust deposition
fields.
The deposition field (Fig.\ \ref{fig:dust}) is very similar to the one used by
\citet{jickells2005}, except that in our field there is almost no dust deposition
near Argentina and Chili.

In Fig.\ \ref{fig:margins} the sediment source is plotted, which is used for one
of the sensitivity experiments.  It is a function developed for Fe fluxes and
depends on the sea floor topography (which accounts for the degree of
oxygenation of the sediments), as described by \citet{aumont2006}.  The Al\,:\,Fe
mass ratio used is 8.1\,:\,3.5, the same as for the dust deposition field.
%TODO: Check this!  and check dimensions

\subsubsection{Simulations}    \label{sec:model_sensitivity}
Our goal in this study is to better understand how the assumptions made in our
model impact on the distribution of \dAl.  Specifically, we aim to examine the
importance of different Al cycle processes and external \dAl sources. To that
end, five sensitivity experiments are performed (see Table \ref{tab:experiments})
with parameters different from the reference experiment.

For the reference experiment a five percent dissolution of the Al
fraction is used, which is within the known range and close to the values used by
\citet{gehlen2003} (3\%) and \citet{han2008} (5\%).  The partition coefficient
$k_d$ is set at $4\cdot 10^6$ l/kg.  This quantity signifies the amount of Al that
can adsorb onto biogenic silica particulates, which means that if $k_d$ is to be
increased, \dAl is exported in larger quantities by scavenging.  Our value is
four times higher than that of \citet{gehlen2003}, but much smaller than the one
used by K.\ Orians and J.\ McAlister ($5.6\cdot 10^7$ l/kg, used for shelves as
well as open ocean regions, \citet{mcalister2011}).\footnote{%
We did not use the $k_d$ value from \citet{gehlen2003}, because initial
experiments showed that this would leave too much \dAl in the ocean surface,
especially at high latitudes.}
The first order rate constant $\kappa$, which signifies how quickly the
equilibrium between \dAl and \ads is established, is not constrained by the
literature.
All these parameters for the reference experiment are chosen in such a way that
the model simulates the \dAl distribution from the observations of the selected
cruises (see Section \ref{sec:observations}) reasonably well, while all the
parameters remained within literature ranges.

After 600 yr the reference experiment 
is forked into several sensitivity simulations.  These
sensitivity experiments are run, along with the reference experiment, for
another 1300 yr or more to a steady state.  The total model time for the
reference simulation is 2200 yr.
An overview of all key parameters for the simulations is given in Table
\ref{tab:experiments}.
For all experiments we use the OPA physical fields and atmospheric dust
deposition as described in Section \ref{sec:model_input}.  In all simulations,
except for the third which tests the sensitivity of water column dissolution, Al
is dissolved only in the surface ocean.  In the third experiment Al is
dissolved in the water column as well with the fraction $10^{-4}\cdot
e^{-z/z_0}$ of the dust deposition field, where $z$ is the depth in km and $z_0
= 1$ km.  The extra term added to Eq.\ \ref{eqn:base_equations} is
$\widetilde{\alpha} \cdot e^{-z/z_0} \mathcal{D}_\text{Al} / H_k$, where $H_k$
is the thickness of layer $k$ and $\widetilde{\alpha}$ forms with the exponent
the fraction of Al that is dissolved in layer $k$.
%We chose for an implicit parametrisation, instead of adding an extra
%tracer of `dust' to the ocean water column, because such an addition would
%introduce several new underdetermined parameters in the equation for the
%dissolution speed of Al from the aluminosilicates, which would require several
%computationally heavy reverse modelling experiments.  It would be interesting to
%model such a tracer as a range of sensitivity experiments, but this is
%for future work.

% Table with sensitivity experiments
% Version 2.0
% History:
%   1.1 initial separate table
%   1.2 removed 'first submission' entries
%   2.0 submission
%\documentclass[a4paper,14pt]{article}
%\usepackage[english]{babel}
%\usepackage{amsmath,amssymb}
%\begin{document}

\begin{table*}
\begin{tabular}{l|lllll}
\hline
Experiment:          & surf.diss.\ (\%) & subsurf.diss.\!   & $k_d$ ($10^6$ l/kg) & $\kappa$ (yr$^{-1}$) & sediments \\
\hline
Gehlen et al.\ (2003) & 3                &   no              & 1                   & 10000               & no \\
Han et al.\ (2008)   & 5                &   no              & -                   & -                   & no \\
%                                                   (not based on equil.) FIXME: Han (2008) contains diatom incorporation, thus not comparable!
\\
Reference experiment & 5                &   no              & 4                   & 10000               & no \\
Doubled dissolution  & \textbf{10}      &   no              & 4                   & 10000               & no \\
With subsurface diss.& 5                &   \textbf{yes}    & 4                   & 10000               & no\\
With ocean margins   & 5                &   no              & 4                   & 10000               &\textbf{yes}\\
Halved part.\ coeff.\ $k_d$ & 5                &   no              & \textbf{2}          & 10000	             & no \\
Slow equilibration $\kappa$   & 5                &   no              & 4                   & \textbf{100}         & no \\
\hline
\end{tabular}
\caption{Overview of reference and sensitivity experiments. Deviations from the
reference experiment are in bold.}
\label{tab:experiments}
\end{table*}

%\end{document}

The sensitivity simulations can be divided into two types of sensitivity
experiments: the first three test the sensitivity to Al sources (amount of dust
dissolution into the surface ocean and into the ocean water column, and the
inclusion of ocean sediment dissolution) and the last two test the
sensitivity to the internal Al cycling (lower partition coefficient $k_d$ and
lower first order rate constant $\kappa$).

\subsection{Observations}   \label{sec:observations}
\begin{figure}
    \centering
    \includegraphics[width=\columnwidth]{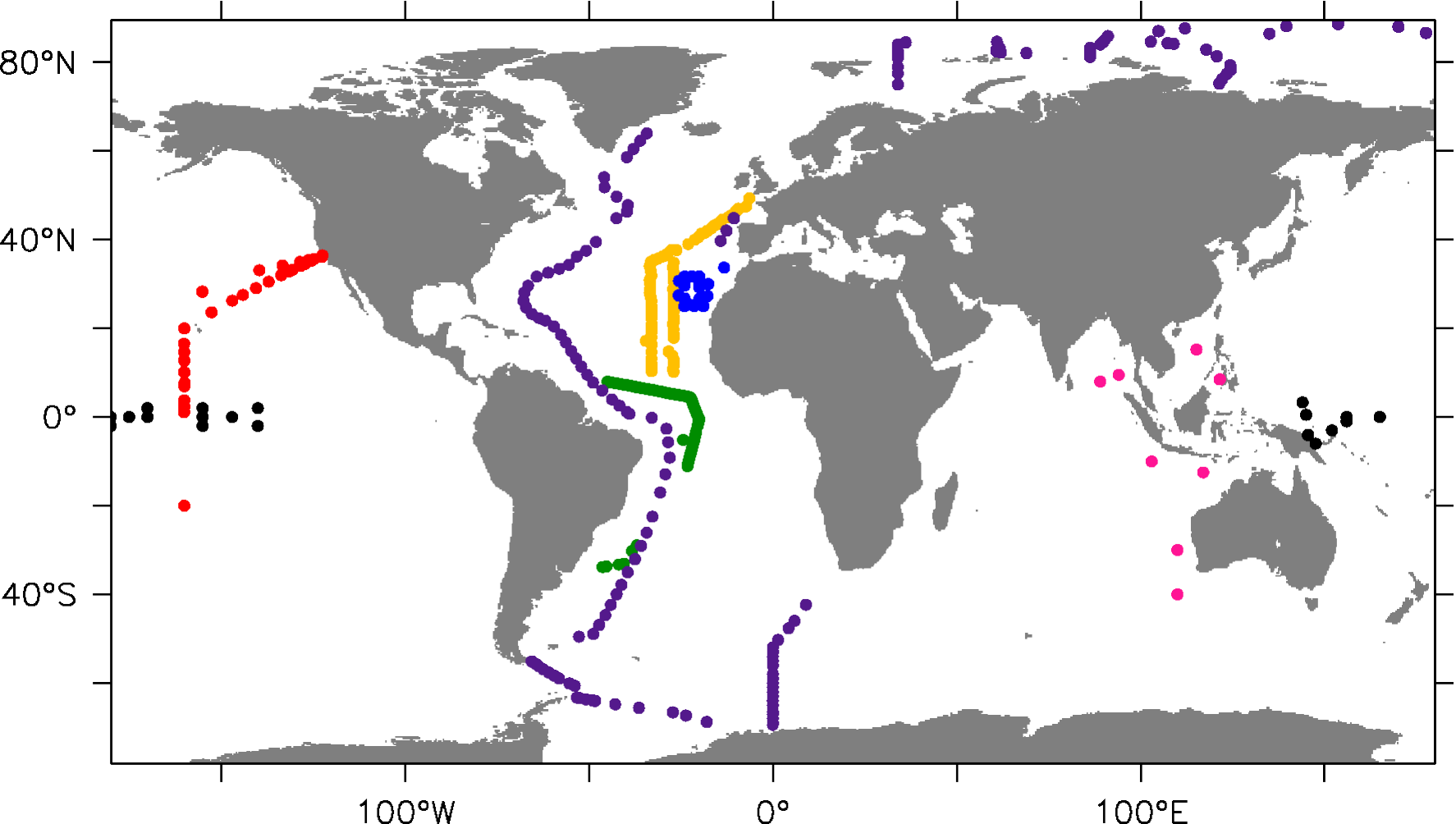}
    \caption{Stations of the cruises:
         M 60 [yellow]; 
         MC-80 and VERTEX [red]; 
         IOC96 [green]; 
         IRONAGES III [blue];
         KH-98-3 [pink];
         EUCFe [black];
         Geotraces-NL [purple].
         See Table \ref{tab:data} for an overview with references
         and the number of observations.
    }
    \label{fig:cruises}
\end{figure}

The recent IPY-Geotraces observations in the Arctic \citep{middag2009},
North-east Atlantic \citep[][Chapter 5]{middag2010}, West Atlantic \citep{middag_prep}
Oceans and the Atlantic section of the Antarctic Ocean
\citep{middag2011} are used for a detailed comparison and optimisation of the
model parameters.  See the upper part of Table \ref{tab:data} for these datasets
and Fig.\ \ref{fig:cruises} for the station positions.  These datasets comprise
overall 3455 individual data values for dissolved Al. All values have been
verified versus international reference samples and their consensus values of
the SAFe and Geotraces programmes, see supplementary material S-1. Moreover,
during cruise 64 PE 321 (2010) excellent agreement was obtained for a complete
vertical profile of dissolved Al at the BATS station between a previous
occupation and dissolved Al data in 2008 by the US Geotraces group, see
supplementary material S-2.
This large amount of observations give the possibility to compare the model with
deep ocean observations of [\dAl], allowing us to validate the model in the deep
ocean and to study the global Al cycle in more detail.

For a worldwide global ocean comparison one has to rely on data that was
collected in the era before the reference samples of SAFe and Geotraces were
available. Inevitably the definition of criteria for selecting such previously
published datasets is less rigorous, see Electronic supplement S-3 for the
criteria used for each of the selected datasets. The selected datasets are
listed in Table \ref{tab:data}, lower part, and positions are shown in Fig.\
\ref{fig:cruises}.

\begin{table*}[t]
\noindent\makebox[\textwidth]{%
\begin{tabular}{l|l|l|l|l|r}
\hline
Cruise      &   Research vessel  & Year    & Location               &   Source              & Nr.\ of data \\
\hline
ARK XXII/2  &   Polarstern       & 2007    & Arctic                 & \citet{middag2009}     & 1080 \\
%ANT XXIV/3  &   Polarstern       & 2008    & Southern Ocean         & Middag et al.\ in press &  919 \\
ANT XXIV/3  &   Polarstern       & 2008    & Southern Ocean         & \citet{middag2011,middag2012}&  919 \\
64 PE 267   &   Pelagia          & 2007    & North-east Atlantic    & Middag et al.         &  137 \\
64 PE 319   &   Pelagia          & 2010    & North-west Atlantic    & Middag et al.         &  383 \\
64 PE 321   &   Pelagia          & 2010    & Centre-west Atlantic   & Middag et al.         &  504 \\
JC057       &   James Clark Ross & 2011    & South-west Atlantic    & Middag et al.         &  432 \\
\hline
\multicolumn{5}{l|}{Subtotal used primarily for detailed comparison and optimisation of the model} & 3455 \\
\hline
\multicolumn{6}{l}{}    \\
\hline
\multicolumn{6}{l}{Compilation of selected other observations for global ocean comparison:} \\
\hline
%M 12 (IOC90) &   Meteor          & 1990    & East Atlantic          & \citet{measures1995}   &   76 \\
IOC96       &   Knorr            & 1996    & Central-south Atlantic & \citet{vink2001}       & 1049 \\
M 60        &   Meteor           & 1982    & North-east Atlantic    & \citet{kremling1985}   &   91 \\
IRONAGES III &  Pelagia          & 2002    & North-east Atlantic    & \citet{kramer2004}     &  181 \\
EUCFe       &   Kilo Moana       & 2006    & Equatorial Pacific     & \citet{slemons2010}    &  195 \\
MC-80       &   Thompson         & 1980    & Pacific                & \citet{orians1986}     &   92 \\
VERTEX-4    &   Wecoma           & 1983    & North Pacific          & \citet{orians1986}     &   54 \\
VERTEX-5    &   Thompson         & 1984    & North Pacific          & \citet{orians1986}     &   59 \\
KH-98-3     &   Hakuho-Maru      & 1996    & East Indian            & \citet{obata2007}      &  152 \\
%Discovery   &   Discovery        & 1988-93 & Mediterranean          & \citet{chou1997}       &  149 \\
%                                             Mediterranean            Hydes et al.
%Cruise     &   Research vessel  & Year    & Location               &   Source           & Nr.\ of data points \\
\hline
\multicolumn{5}{l|}{Subtotal compilation of selected other observations for global comparison} & 1873 \\
\multicolumn{5}{l|}{Grand total of all dissolved Al values used in this study} & 5328 \\
\hline
\end{tabular}
}
\caption{\hbox{Observational data used for comparison with model.  For positions see Fig.~\ref{fig:cruises}.}}
\label{tab:data}
\end{table*}

% TODO: make plot okay for colourblind people

% Results and discussion, version 2.0
% 
\section{Results}    \label{sec:results}
\subsection{Reference simulation}   \label{sec:reference}
\begin{figure}
    \centering
    \includegraphics[width=\columnwidth]{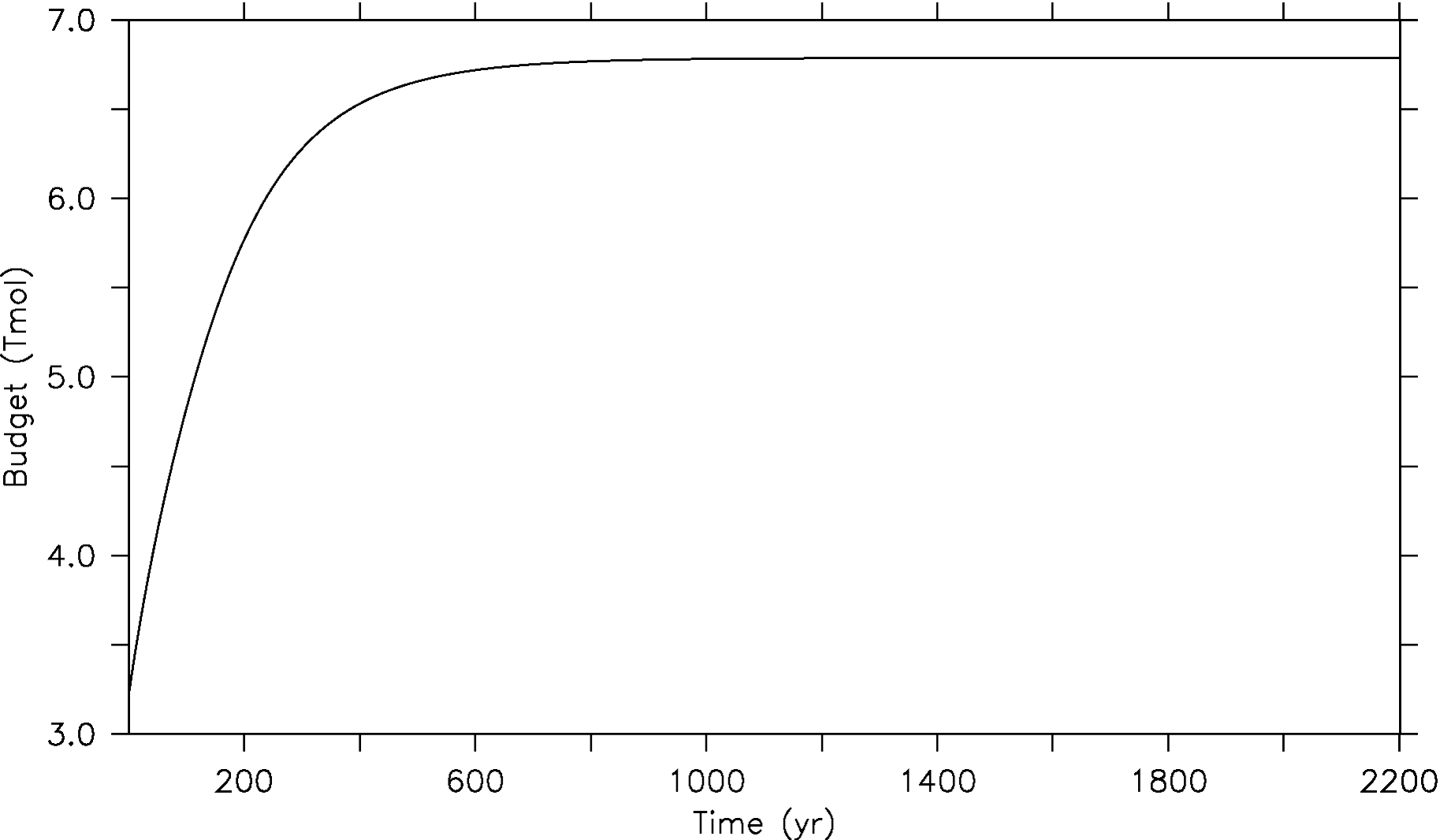}
    \caption{Total Al budget ([\ads]+[\dAl]) (Tmol) in the world ocean, plotted
    against the model years}
    \label{fig:budget_ocean_refexp}
\end{figure}

Our reference experiment is performed for a spin-up period of 2200 yr.  The
resulting total ocean Al budget in the reference simulation is around 7 Tmol (1
Tmol = $10^{12} $ mol).  Already after 600 yr the total Al distribution is more
or less in a steady state as shown by Fig.~\ref{fig:budget_ocean_refexp}, where
the total integrated aluminium (dissolved and adsorbed) is plotted against
time.  Therefore, from 600 yr our sensitivity experiments (see Section
\ref{sec:sensitivity}) are forked.%
\footnote{The relevant tracers of the raw model output can be found at
\url{http://data.zkonet.nl/index.php?page=Project_view&id=2916&tab=Datasets}.}
% N.B. La kalkulo kvanto[i=@sum,j=@sum,k=@sum]/1e9 donas Tmol (M/1000 * volumo)!
%   refrun: P1600-3800     2200 yr (=600+1600)
%   sensit: P2201-3500     1300 yr (or more)

%%% Model results: surface %%%
\begin{figure}
    \centering
    \includegraphics[width=\columnwidth]{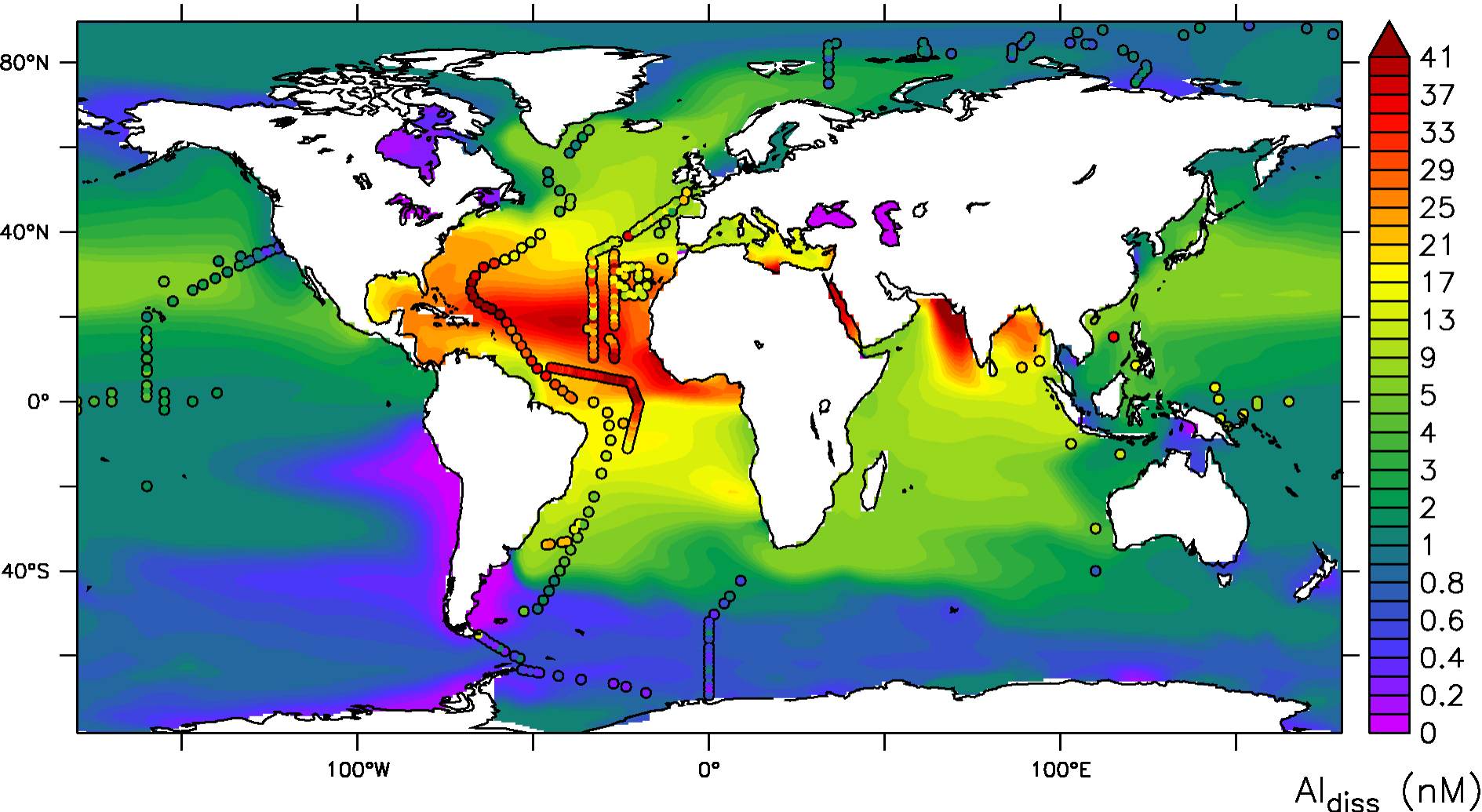}
    \caption{Dissolved aluminium concentration (nM) from the reference
    experiment at the surface ocean, with respective observations plotted on top of
    it.  Note that the scale is expanded for low concentrations.}
    \label{fig:refexp_surface}
\end{figure}
\begin{figure}
    \centering
    \includegraphics[width=\columnwidth]{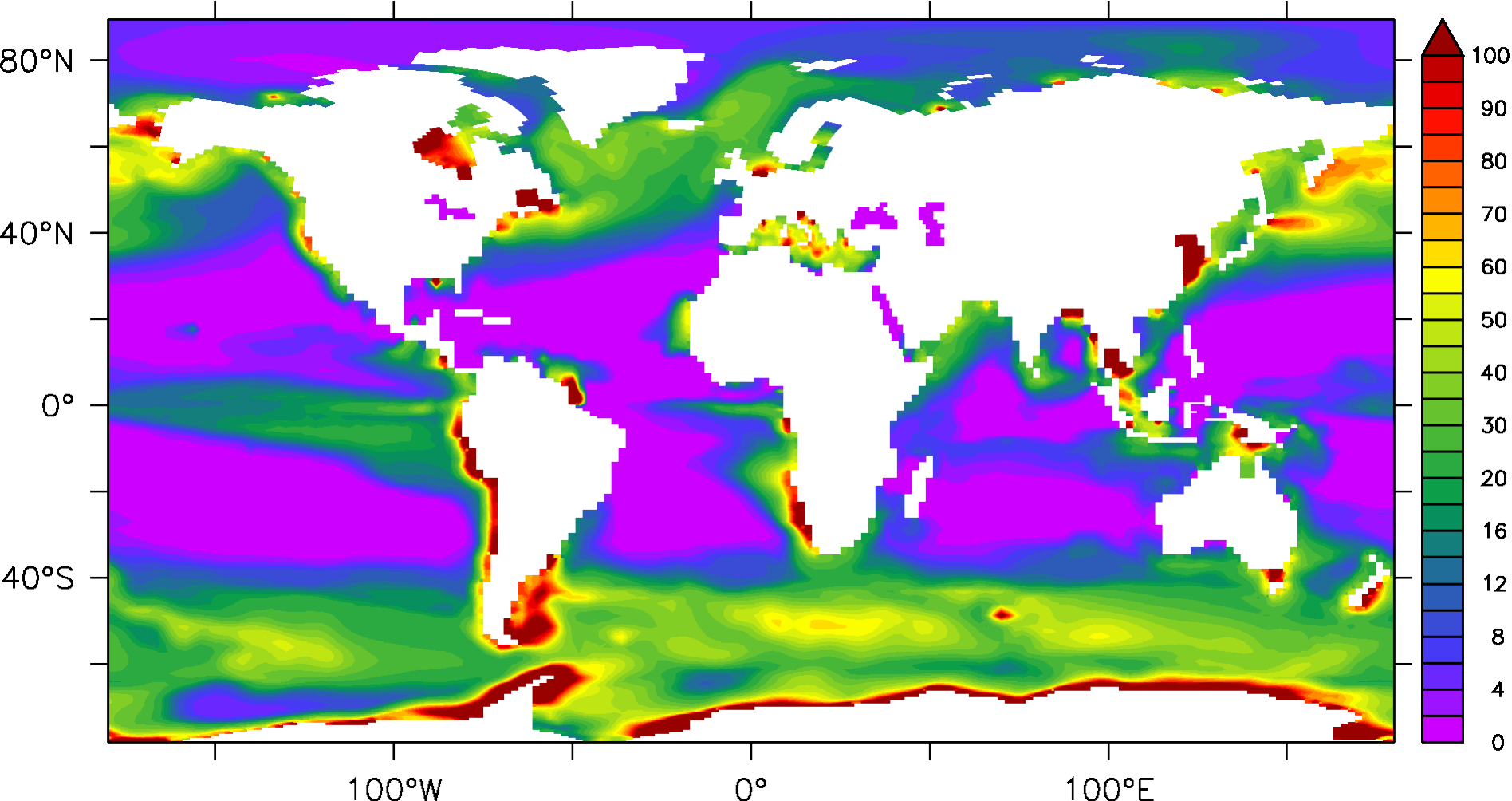}
    \caption{Biogenic Si concentration (nM) in the PISCES model, at the surface,
    yearly average.}
    \label{fig:BSi_surface}
\end{figure}

Fig.~\ref{fig:refexp_surface} shows the \dAl surface concentration of the
reference experiment.  As in all subsequent plots, yearly averages are shown.
There is a seasonal cycle in [\dAl] in our model, but it is small relative to
the spatial trends.  The observed \dAl concentrations from our Al data
compilation (see Section \ref{sec:observations}) are plotted as circles over
the model \dAl concentration in this and later figures.  A comparison of the
model with the observations will be done in Section \ref{sec:comp.observation}.
Consistent with the dust deposition field (Fig.\ \ref{fig:dust}), the largest
modelled \dAl surface concentrations are in the central Atlantic Ocean, with
values between 30 and 40 nM (1 nM = $10^{-9}$ mol/l) near 20\degree N,
decreasing northward to values in the order of several nM north of 60\degree N.
The concentration of \dAl between 40\degree N and 60\degree N is larger than
expected based on the dust deposition field (Fig.\ \ref{fig:dust}).  This is
because the Gulf Stream and the North Atlantic Current transport \dAl northward
before most of it is scavenged.  As will be shown in Section \ref{sec:timescale}
and \ref{sec:moc}, most \dAl is scavenged before it passes Iceland.  The result
is that [\dAl] in the surface waters of the Arctic Ocean is in the order of 1
nM.  On the other side of the globe, south of 40\degree S, the Al concentration
is even lower, less than 1 nM.  This is to be expected on the basis of [\bsi]
(Fig.\ \ref{fig:BSi_surface}) and the dust deposition flux (Fig.\
\ref{fig:dust}).  In the Pacific Ocean the concentration is relatively low as
well, except near 20--35\degree N where a wide band of high [\dAl] can be
found.
This maximum arises from the large dust deposition along the Asian continental
margin between 20\degree N and 60\degree N.  North of 40\degree N, the
concentration of \bsi is high, as can be seen in Fig.\ \ref{fig:BSi_surface}.
Therefore, in that region the dissolved Al is quickly depleted by the high \bsi
concentration in the surface, while south of 40\degree N the North Pacific
Current transports it eastward into the central North Pacific Ocean.

%%% Model results: depth %%%
\begin{figure*}
    \centering
    \includegraphics[width=\textwidth]{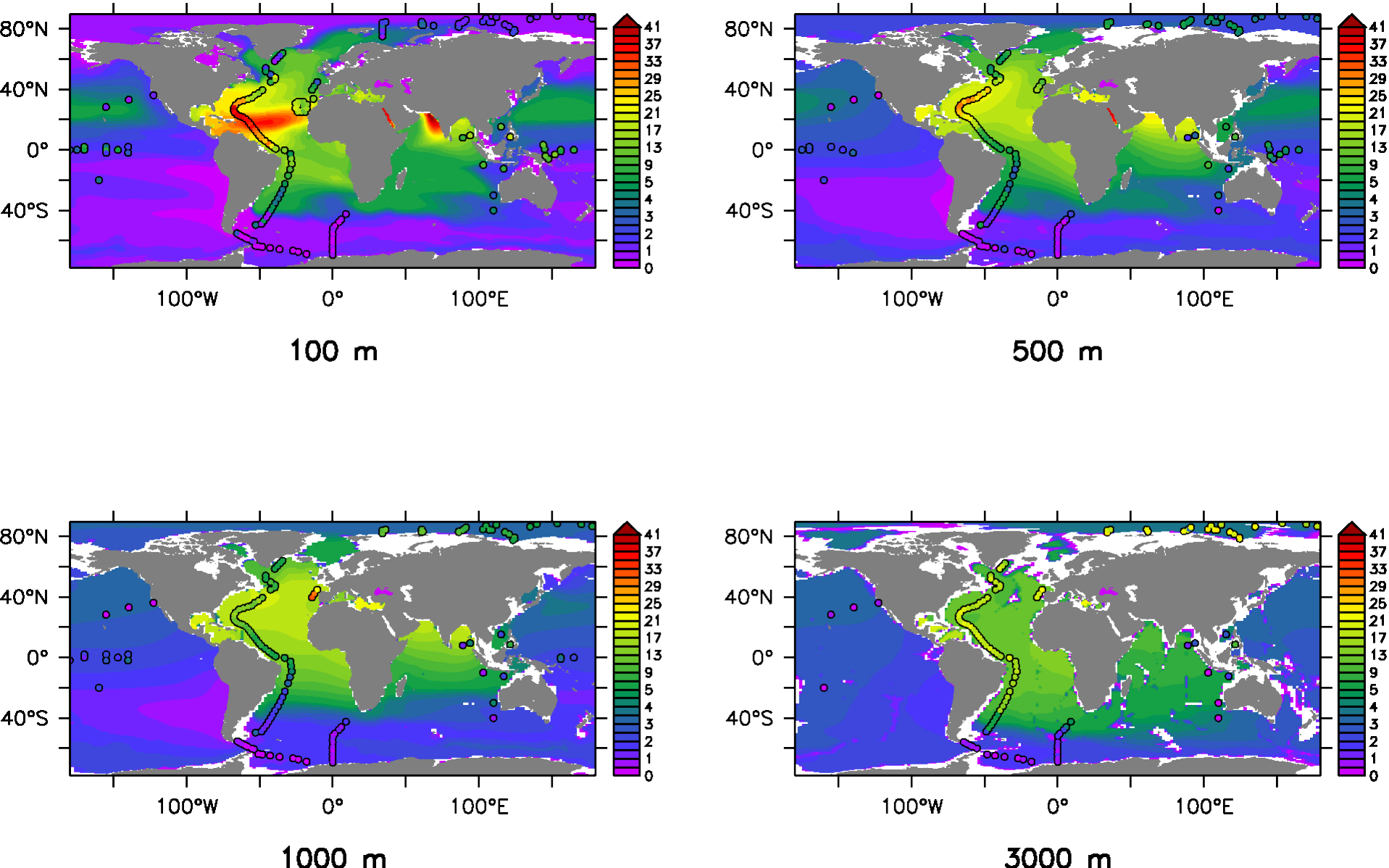}
    \caption{Dissolved aluminium concentration (nM) from the reference
    experiment at several depths, with respective observations plotted on top of
    it.}
    \label{fig:refexp_layers}
\end{figure*}
\begin{figure}
    \centering
    \includegraphics[width=\columnwidth]{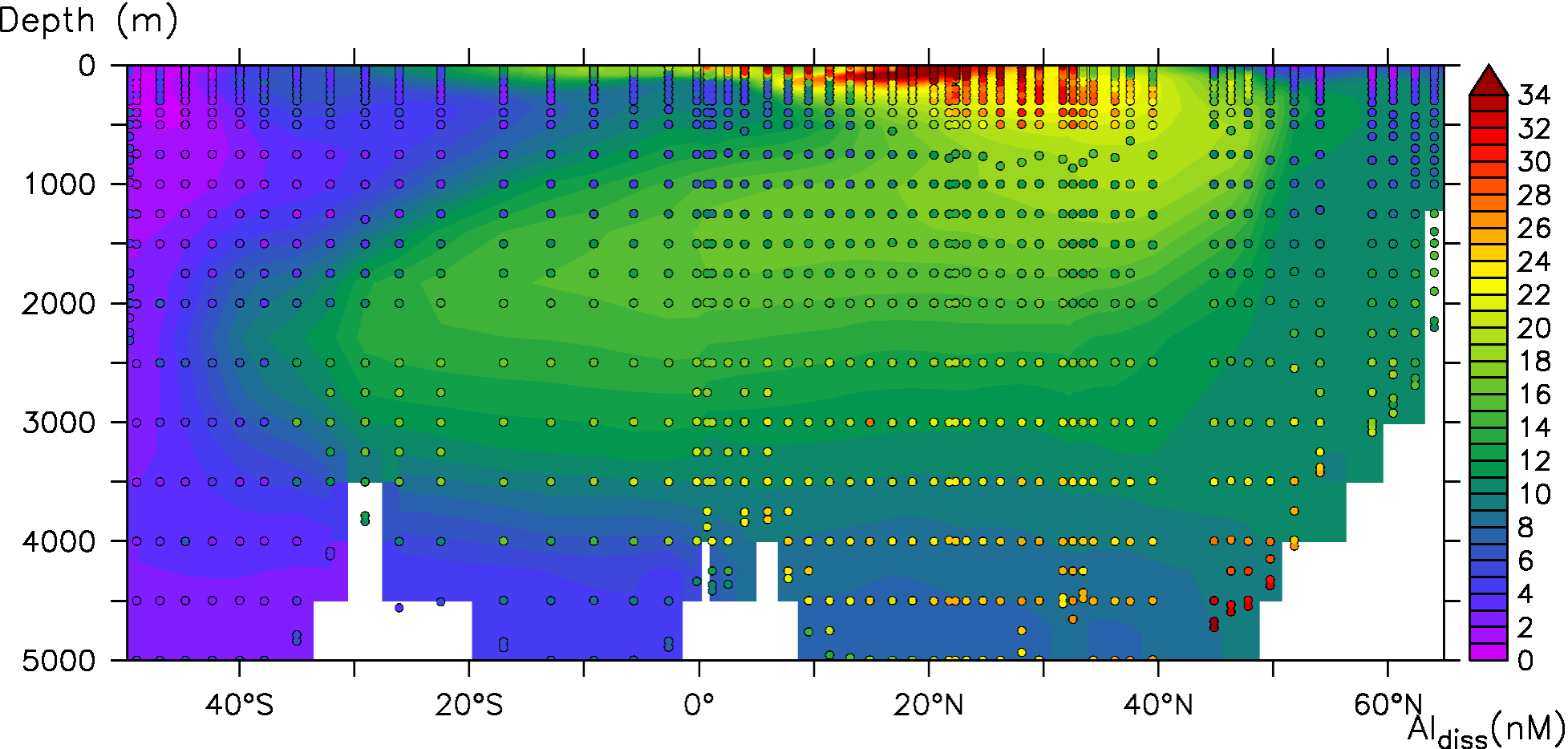}
    \caption{Dissolved aluminium (nM) from reference experiment along the
    West Atlantic Geotraces cruise track.  Observations are plotted on top.}
    \label{fig:refexp_section}
\end{figure}
In Fig.\ \ref{fig:refexp_layers} the modelled \dAl concentration of the
reference experiment is shown at selected depths (100\,m, 500\,m, 1000\,m
and 3000\,m) and in Fig.\ \ref{fig:refexp_section} the Geotraces cruise
section in the West Atlantic Ocean.  This section was calculated from the
three-dimensional model data by converting the ORCA2 gridded model data to a
rectilinear mapping and after this interpolating the rectilinear data onto the
cruise track coordinates.
In the Atlantic Ocean there is a maximum of [\dAl] from about 30\degree S at 2
km depth, getting higher northward (Fig.~\ref{fig:refexp_section}).  According
to the model, deep \dAl concentrations are low in the Southern Ocean, the
Arctic Ocean and the Pacific Ocean (Fig.~\ref{fig:refexp_layers}).

\subsubsection{Comparison with observations}   \label{sec:comp.observation}
In Figs.\ \ref{fig:refexp_surface}, \ref{fig:refexp_layers} and
\ref{fig:refexp_section} observed \dAl concentrations (see Section
\ref{sec:observations}) are plotted as circles over the modelled results.
The precision of the measurements performed during the Geotraces cruises at a 1
nM level and lower (polar oceans) was about 4\% and at a 20 nM level (West
Atlantic Ocean) about 2\% \citep[][p.~25]{middag2010}.  These observations have
a standard deviation that is smaller than can be distinguished based on the
colour bar.  Older observations generally have lesser precision, but are of
sufficient quality (see Section \ref{sec:observations}).  Therefore the model
results can in principle be compared with the observations without regard of
measurement errors.

%%% Compared with observations: surface %%%
Fig.\ \ref{fig:refexp_surface} shows that the model and the observations reveal
similar patterns in the surface ocean: high concentrations in the central
Atlantic Ocean and low concentrations in the polar oceans and the Pacific
Ocean.  However, details in the modelled \dAl distribution do not agree with
the observations.  For instance, the large concentration in the Pacific Ocean
between 20 and 35\degree N is not visible in the observations of
\citet{orians1986}.
Also between 45 and 65\degree N in the West Atlantic Ocean the concentration of
\dAl in the model is overestimated compared with the Geotraces observations
near Greenland \citep{middag_prep}.  This suggests that Al is
not sufficiently scavenged in this area.  This is further discussed in Section
\ref{sec:adv_vs_scav}.
The model underestimates [\dAl] near Brazil.  This can be related to errors in
ocean currents (Fig.\ \ref{fig:opa}), as the transport of low [\dAl] water
might be too high in the Brazil current.  In reality this current is narrower
than the 2\degree\ resolution of the model.  Another possibility is that the
dust deposition field is not realistic in this area.
Furthermore, the observations generally have more spatial variability than the
modelled [\dAl].  Many of these differences depend on several model boundary
conditions (dust deposition distribution, velocity field, particle sinking
speed) as well as temporal variability, because the model results plotted are
yearly averages, while observations are done throughout the year in precise
locations.

\begin{figure}
    \centering
    \includegraphics[width=\columnwidth]{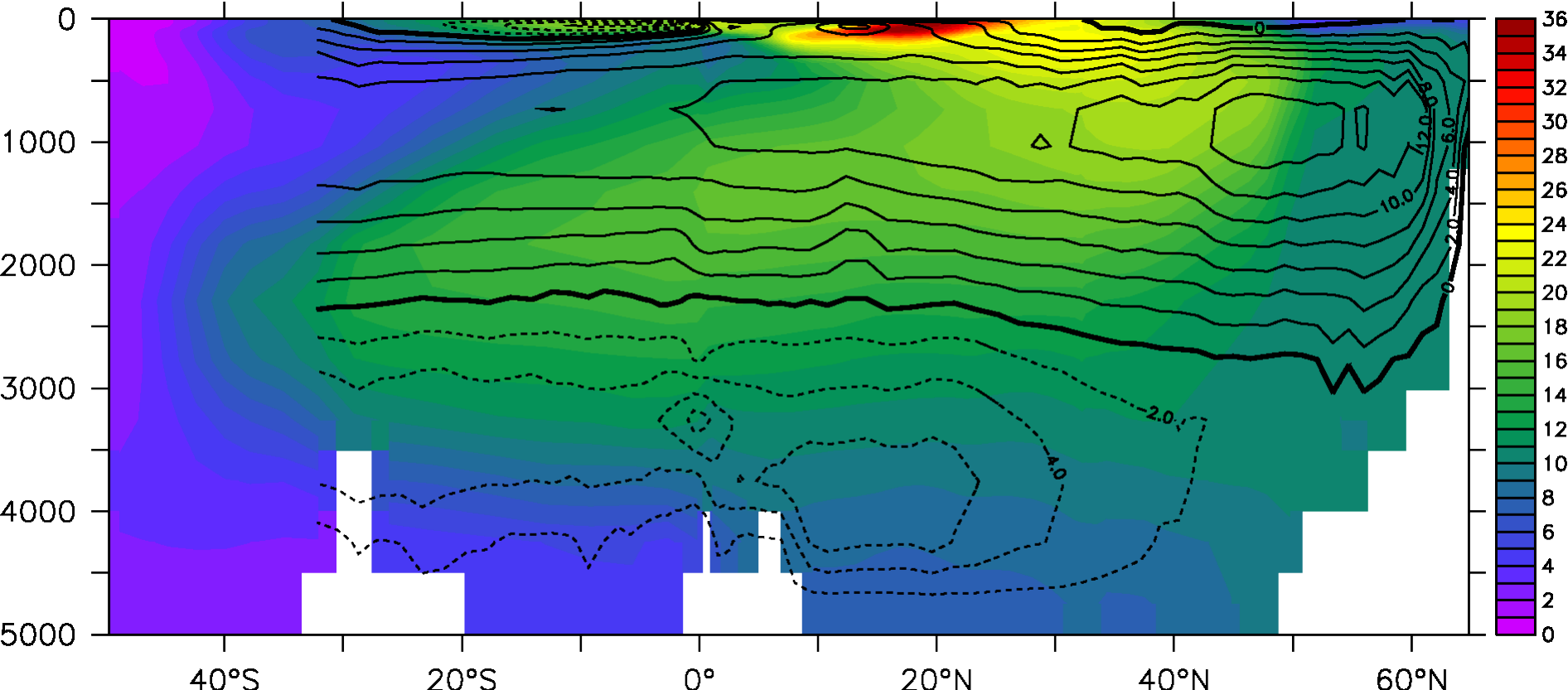}
    \caption{The quantity in the coloured area is the model [\dAl] (nM)
    at the Geotraces West Atlantic cruise section.
    The contours represent the model Atlantic Overturning Stream Function (Sv).}
    \label{fig:contour_on_model}
\end{figure}

%%% Compared with observations: depth %%%
Fig.\ \ref{fig:refexp_section} shows that similar patterns are present in the
West Atlantic Ocean in the model and the observations.  There is a clear
pattern of North Atlantic Deep Water (NADW), which is mainly visible between
40\degree S and 20\degree N in both the model and the observations.  In Fig.\
\ref{fig:contour_on_model} the Atlantic Overturning Stream Function (OSF) is
plotted over the model results, showing that the patterns of the OSF and [\dAl]
coincide.  In the Southern Ocean, low concentrations of \dAl in the upper km
are clearly penetrating from the Southern Ocean northward to at least 20\degree
S around 500 m depth.  This is Antarctic Intermediate Water (AAIW)
and Subantarctic Mode Water (SAAMW).  From the Southern Ocean near the bottom,
there is also Antarctic Bottom Water (AABW) flowing northward, visible in the
observations and the model.

The similarity between the model and the observations lessens in the deeper
North Atlantic Ocean, where according to the observations [\dAl] increases
with depth (for depths below 800 m), while in the model there is a decrease
with depth below 1.5 km.  Besides this general pattern of increase of
[\dAl] with depth in the observations, a very high concentration of \dAl is
present between 45 and 50\degree N near the sediment, which enhances the
dissimilarity between the model and the observations.
This problem will be discussed further in Section \ref{sec:moc}.
%FIXME: This problem is far from solved (in sec:sloweq or any other section).

Furthermore, in the southern hemisphere in the model the maximum of [\dAl] is at
about 2 km depth, indicating a southward flow around 2 km depth.  In the
observations this maximum is at 3 km depth, indicating a southward flow at 3 km
depth.  Indeed, according to the hydrographical analysis of the West
Atlantic Geotraces cruise (\citet{vanaken2011}, cf.~\citet{krauss1996}) the flow is at 3 km depth.  As can
be seen in Fig.\ \ref{fig:contour_on_model} the southward flow is too high in
the water column compared to the observations.

\begin{figure}[h]
    \centering
    \includegraphics[width=\columnwidth]{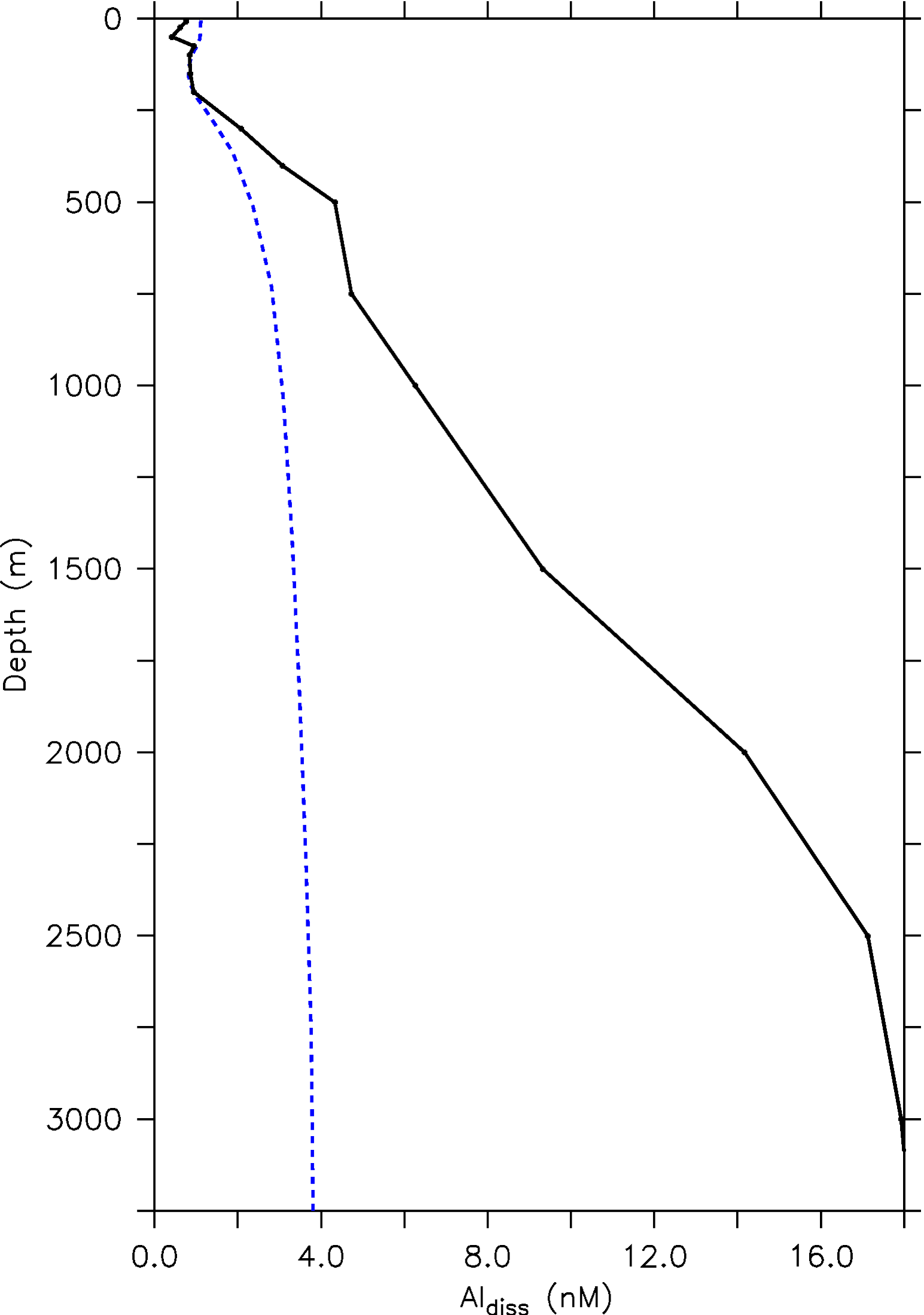}
    \caption{[\dAl] (nM) profile in the eastern Arctic of reference experiment
    (blue dashed line) and the observations of IPY-Geotraces-NL observations (black
    solid line) at 213.6\degree E, 87.03\degree N}
    \label{fig:East_Arctic_profile}
\end{figure}

Another feature which is not captured by the model is the aluminium
profile in the eastern Arctic Ocean, which is shown in
Fig.\ \ref{fig:East_Arctic_profile}.
It shows that the model [\dAl] increases with depth from a minimum near 150 m
to a maximum near the bottom by a factor of 3 (and is in the order of 2 nM,
see Fig.~\ref{fig:East_Arctic_profile}).  The observations show an increase as
well (as also presented in Fig.~\ref{fig:refexp_layers}), but here it is far
more pronounced with a small minimum at the surface (less than 1 nM) to a large
maximum near the abyss (around 18 nM) \citep{middag2009}.  Thus in the Arctic
Ocean the vertical change in [\dAl] in the model is by far not as pronounced as
in the observations.
% NOTE: There actually is a linear relation for certain points in the Arctic,
% but it is so far difficult to relate this to the resutls of Rob!  I think
% that scavenging can also play a role in such a linear relation, since
% remineralisation of detrital silica not only results in \dSi but also in
% \dAl; just loko at the equations.
The absence of sediment sources of Al might play a role in this.  The results
of a simulation with sediment sources will be discussed in Section
\ref{sec:sensitivity}.
%These parameters basically increase the absolute value at all depths, so that
%the strong silicon-like profile is not replicated.  Furthermore, the linear
%relation between [\dSi] and [\dAl] in \citet{middag2009} is not reproduced by
%the model (not plotted).  Thus we may need another process in our model, like
%biological incorporation and subsequent remineralisation at depth along with
%the diatom frustules.

\subsection{Sensitivity simulations}    \label{sec:sensitivity}
Section \ref{sec:model_sensitivity} described how our sensitivity simulations
are set up.  The reasons why we performed these specific experiments are
closely related to the problems we encountered while interpreting the results
from the reference experiment (Section \ref{sec:reference}).  After a brief
summary of two of these problems and associated sensitivity simulations, 
the results of these simulations will be presented.%
\footnote{The relevant tracers of the raw model output can be found at
\url{ftp://zkoclient:zko@dmgftp.nioz.nl/zko_public/00056}.}

How can it be that in the Atlantic Ocean from 800 m depth to the bottom of the
ocean there is a decrease in the \dAl concentration in the model, while
according to the observations there is an increase (see
Fig.~\ref{fig:refexp_section})?  The only transformation mechanism in our model
is reversible scavenging, which should give a profile where [\dAl] increases
with depth, and the only source is dust.
However, the model biogenic silica concentration, its sinking speed, advection
and mixing and the first order rate constant $\kappa$ are all relevant for the
vertical distribution of \dAl as well.  Since the biogenic silica concentration
and the physical fields can both be assumed to be reasonably realistic (see
Section \ref{sec:model}), no sensitivity simulations concerning these fields
are described in this paper.  We do perform one simulation with a different
first order rate constant (Section \ref{sec:sloweq}) to investigate where in
the ocean this parameter influences the \dAl distribution.  The value of
$\kappa$ is significantly decreased in this simulation (this parameter is very
poorly constrained by observations).

Concerning the strongly depth-increasing profile of \dAl as observed in the
eastern Arctic Ocean (Fig.\ \ref{fig:East_Arctic_profile}), we perform a
simulation where sediment ocean margins are added as a source of \dAl,
following the approach of \citet{aumont2006} and assuming an Al\,:\,Fe fraction
of 8.1\,:\,3.5 in the sediments (thus we assume sediment Al supply is coupled
to Fe).  The result of this simulation is discussed in Section
\ref{sec:sediment}.  Furthermore, an experiment is performed where the
partition coefficient is decreased from $4\cdot 10^6$ l/kg to $2\cdot 10^6$
l/kg, which is a way to evaluate the importance of relative partitioning
between the dissolved and adsorbed fraction.  This experiment will be discussed
in Section \ref{sec:deckd}.

\subsubsection{Increased dust dissolution fraction}  \label{sec:incdis}
\begin{figure}
    \includegraphics[width=\columnwidth]{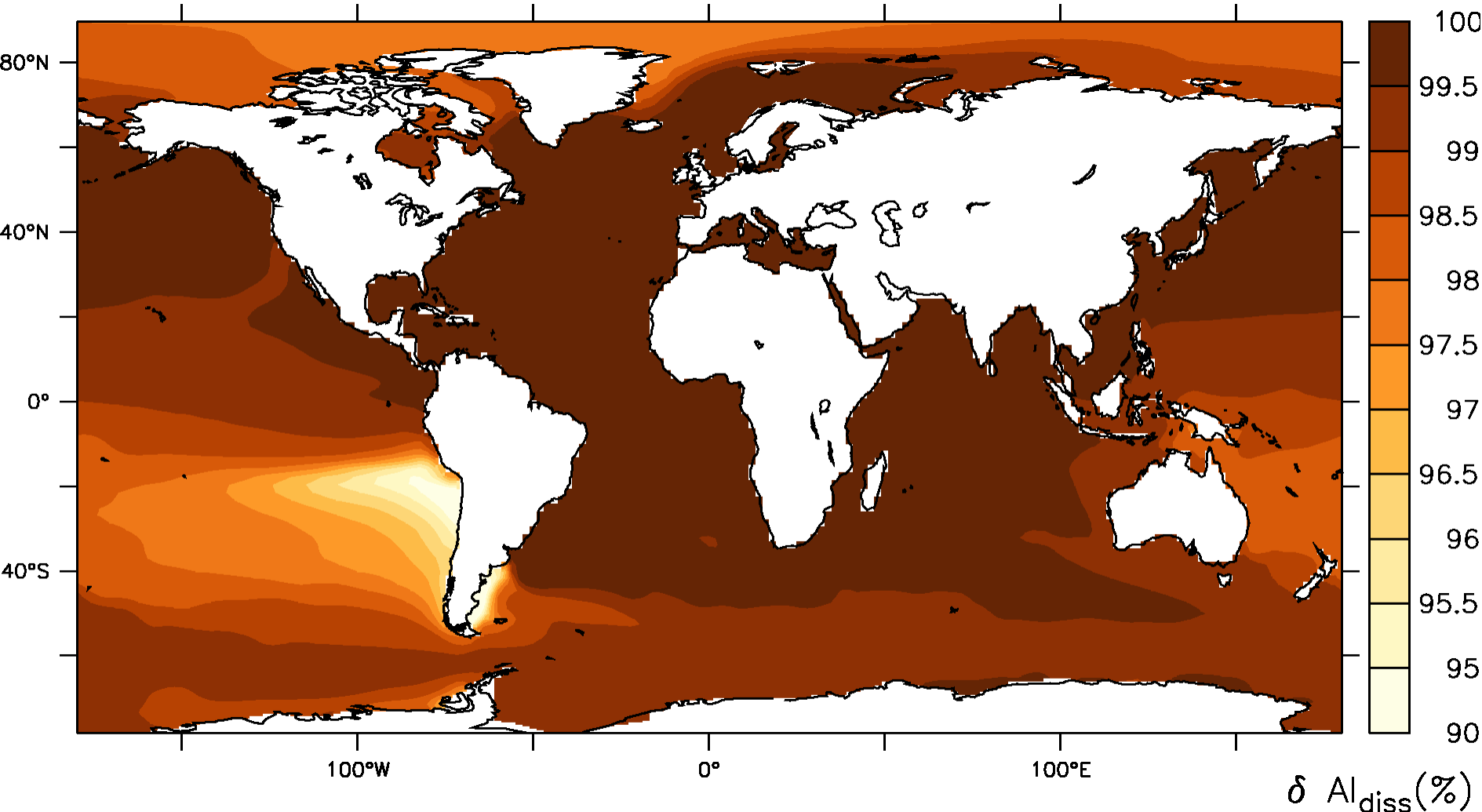}
    \caption{The relative surface difference of the dissolved aluminium concentration
    between the experiment with an increased dissolution (10\%) and
    the reference experiment (5\% aluminium dissolution).
    The scale is in percentages, of which the largest part from 95\% to 100\%.
    }
    \label{fig:incdis_SENS}
\end{figure}

To examine the impact of greater dust dissolution on the cycling and
distribution of \dAl, we increase the surface dissolution percentage of Al from
5 to 10\% (`doubled dissolution').
In Fig.~\ref{fig:incdis_SENS} the resulting change in [\dAl] in the surface
ocean.  In the West Atlantic Ocean along the Geotraces cruise track the increase
in [\dAl] is everywhere between 99 and 100\% (not plotted).
The [\dAl] is increased with a factor of two in most locations in
the ocean.

The doubling can be explained when the model equations (Eqs.\
\ref{eqn:scav-gehlen} and \ref{eqn:eq-gehlen}) are considered.
Consider a one-box model in which equilibration is instantaneous ($\kappa =
\infty$).  If we increase dust dissolution everywhere with a factor of two and
wait until steady state, we also get a doubled particulate Al burial.  This
means that [\ads] must be doubled in the ocean since the sinking/burial speed
does not change.  Since in steady state [\ads] is proportional to [\dAl] (Eq.\
\ref{eqn:eq-gehlen}), [\dAl] must be doubled as well.  Hence a doubled [\dAl]
is expected.  In other words, the total Al budget is linear with the input.
(See for instance \citet{broecker1982} for background.)
This means that the amount of dissolution in the surface ocean has an effect
everywhere in the ocean, i.e.\ the effect is global and not restricted to dust
deposition sites.

\begin{figure}
    \centering
    \includegraphics[width=\columnwidth]{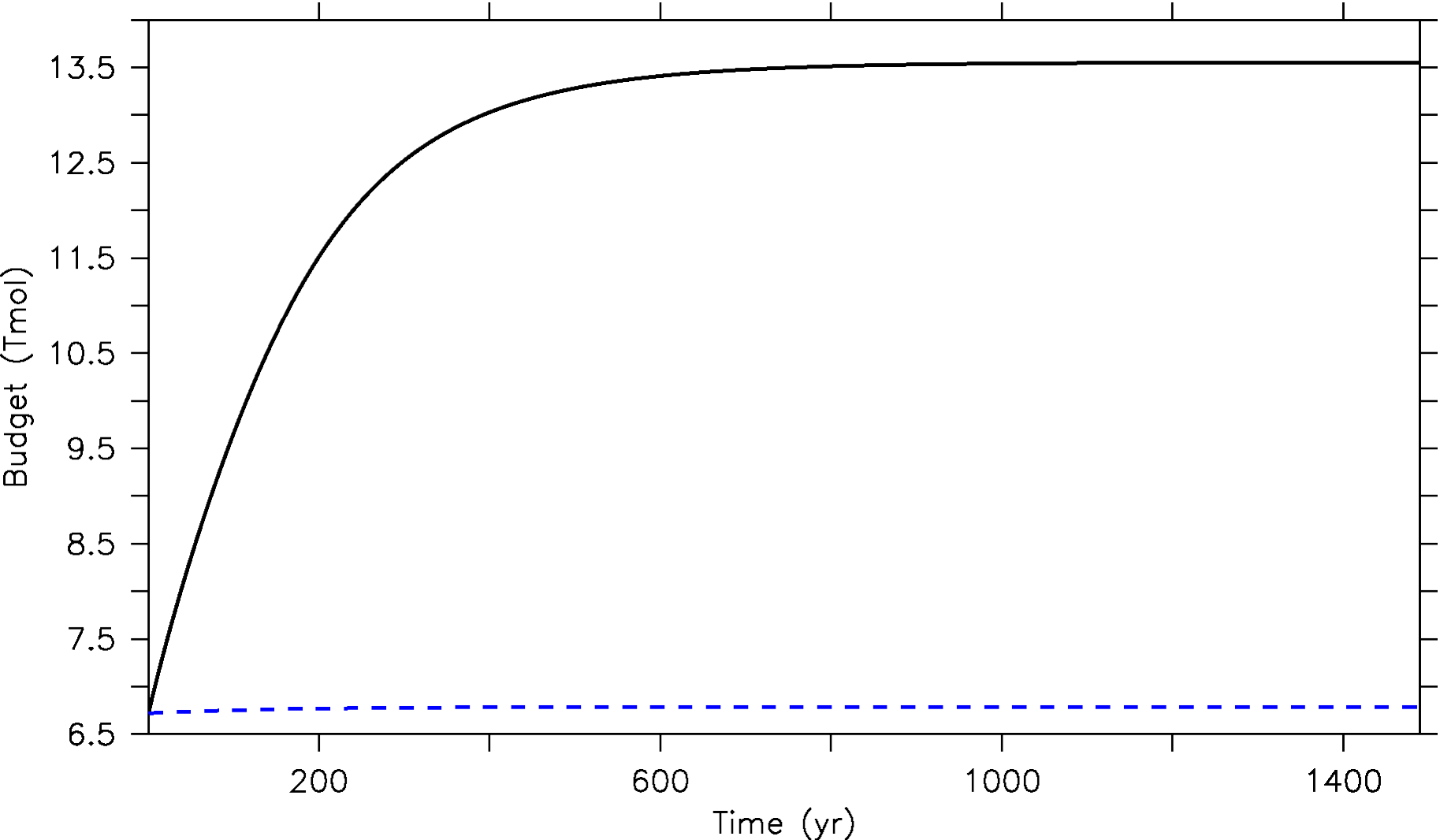}
    \caption{Total Al budget (Tmol) in world ocean during spin-up of 1490 years
    for increased dissolution sensitivity simulation (solid black line; forked
    from the reference simulation at year 600).  The dashed blue line signifies
    the Al budget of the reference simulation.}
    \label{fig:budget_ocean_incdis}
\end{figure}

The Al budget is not doubled everywhere.  Fig.\
\ref{fig:incdis_SENS} shows that in the coastal upwelling region
near Chili the increase is only around 95\%.  This is because the model is not
completely spun up, as can be seen in Fig.\ \ref{fig:budget_ocean_incdis}, so
that the increased dissolution of Al from dust has not reached the deep Pacific
Ocean yet.  Therefore the percentages in Fig.\ \ref{fig:incdis_SENS} indicate
how close the subsequent sensitivity simulations are to steady state in
different regions in the ocean.

\subsubsection{Water column dust dissolution}  \label{sec:noss}
For the reference experiment we included dust dissolution only in the surface
ocean. %, analogously to iron \citep{aumont2006}.  A priori there is no reason
to assume that Al from dust only dissolves in the surface ocean.  During
sinking of dust more Al might be dissolved.  To test the effect of water column
dissolution we performed a sensitivity simulation with water column dissolution
included.

\citet{gehlen2003} have included \dAl input at the surface only, but they do
compare different dust deposition distributions, as do \citet{han2008}.  The
latter also dissolve dust below the surface layer, but no sensitivity
experiments have been presented for water column dissolution.  None of our
sensitivity experiments, as part of an OGCM, have been published before.

\begin{figure}
    \subfigure[World ocean surface difference (nM)]{
        \includegraphics[width=\columnwidth]{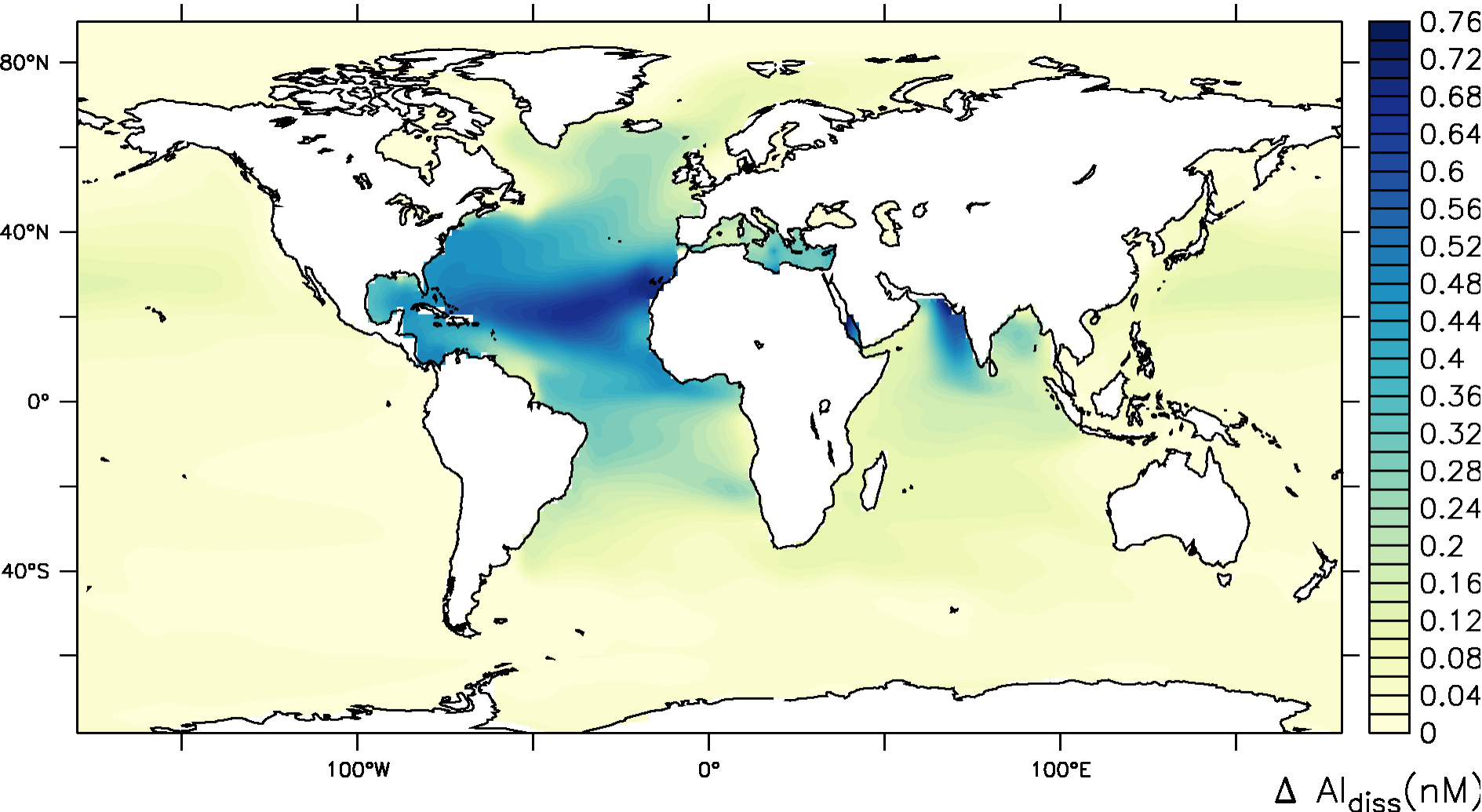}
    }
    \subfigure[West Atlantic section difference (nM)]{
        \includegraphics[width=\columnwidth]{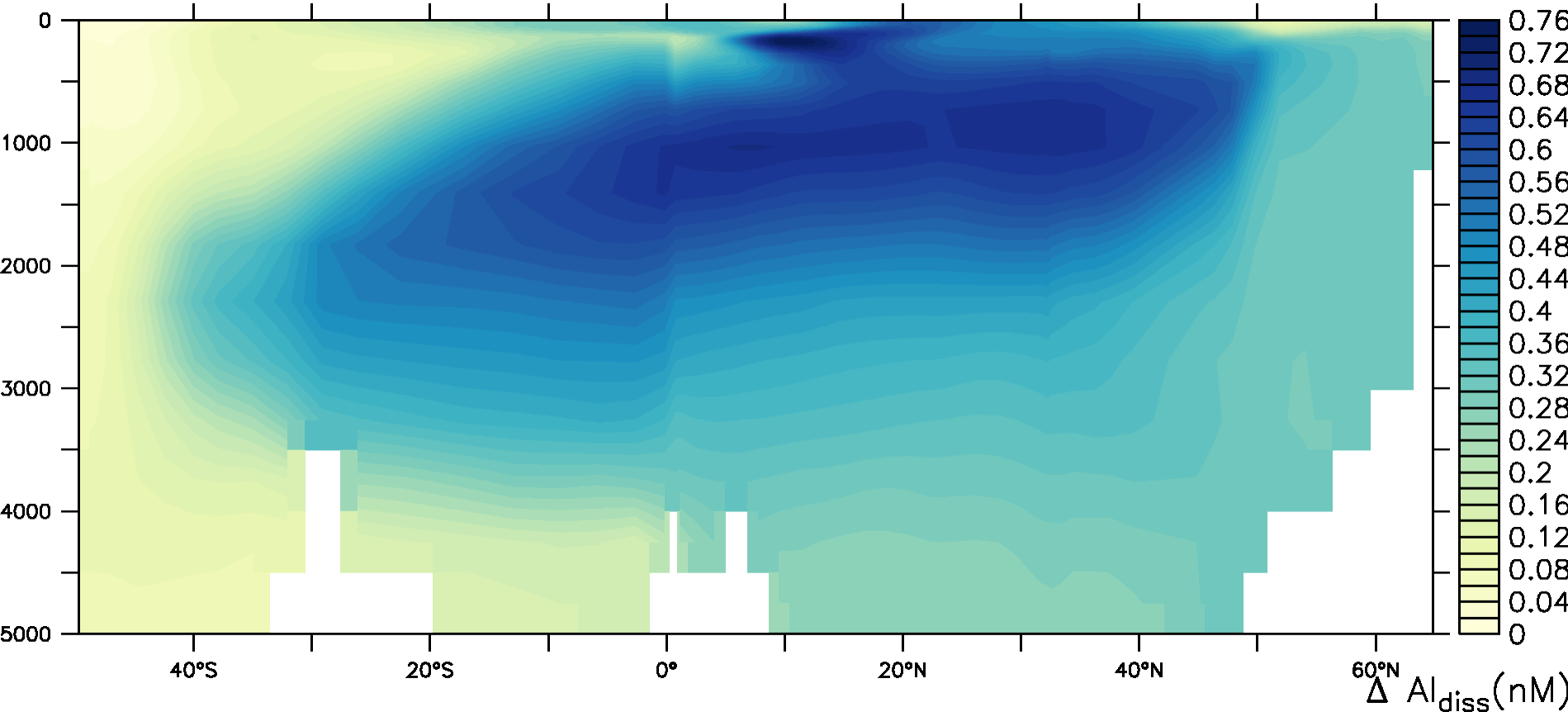}
    }
    \caption{Difference in [\dAl] (nM) between the simulation with water column
    dissolution with the reference simulation (without water column dissolution)}
    \label{fig:noss_section}
\end{figure}

Our simulation with water column dissolution resulted in slightly higher values
of [\dAl] of up to 0.7 nM more in some places.  Fig.\ \ref{fig:noss_section}
compares [\dAl] in the water column dissolution experiment with the reference
experiment (Figs.\ \ref{fig:refexp_surface}, \ref{fig:refexp_section}).

On first sight, north of the equator, water column dissolution moves the
modelled [\dAl] away from the observations.  This is clear if the decrease in
[\dAl] with depth is considered in the reference simulation, while in the
observations [\dAl] increases with depth.  The simulation with water column
dissolution makes the modelled decrease even more pronounced.  Also the high
[\dAl] in the upper 500 m is only slightly improved, and not at precisely the
right location: in the upper 500\,m the model especially underestimates [\dAl]
between 20 and 35\degree N.  Therefore, water column dissolution does not
improve the model results.

\subsubsection{Ocean sediments source} \label{sec:sediment}
\begin{figure*}
    \subfigure[World ocean surface concentration (nM)]{
        \includegraphics[width=\columnwidth]{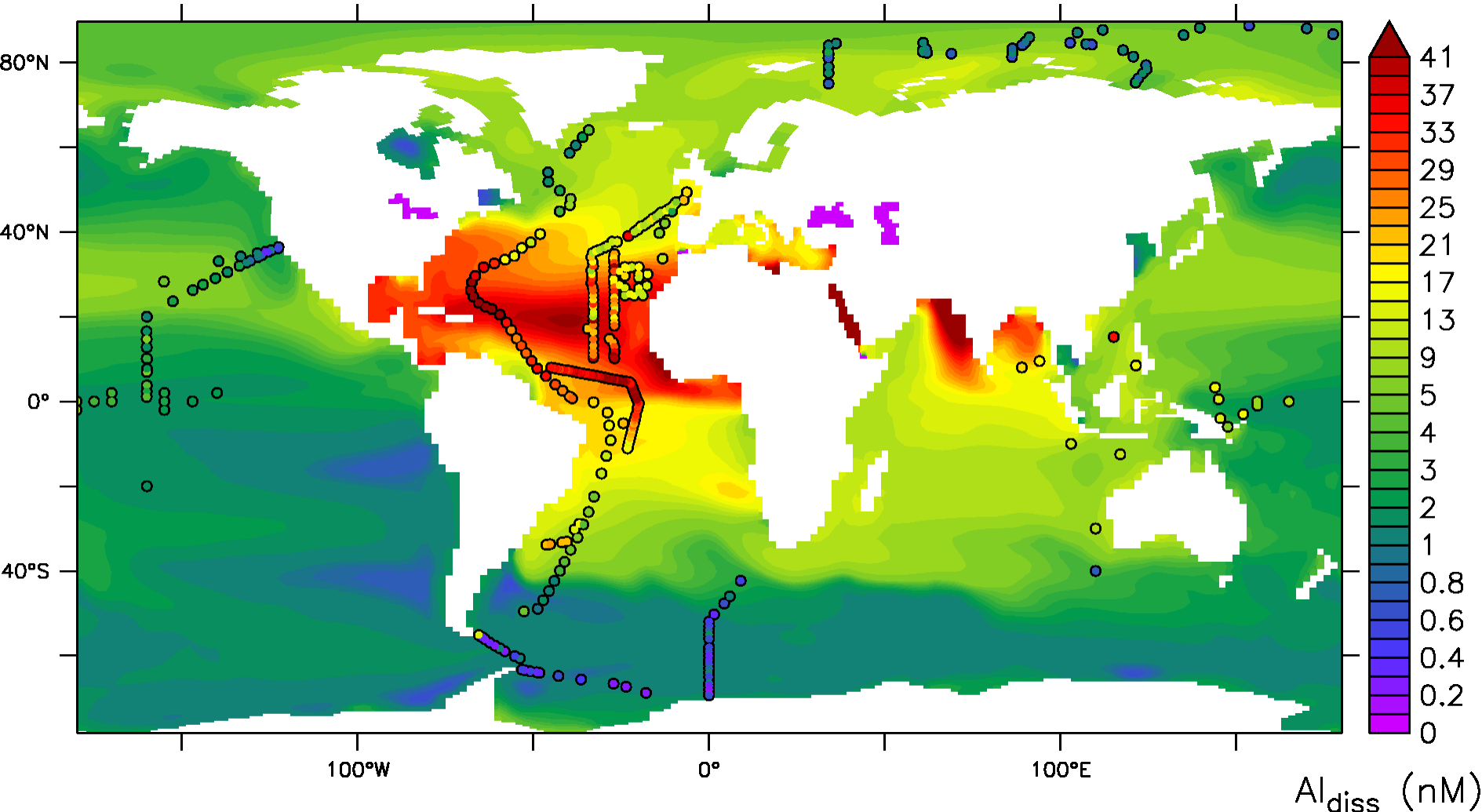}
        \label{fig:marg_surface}
    }
    \subfigure[West Atlantic section concentration (nM)]{
        \includegraphics[width=1.1\columnwidth]{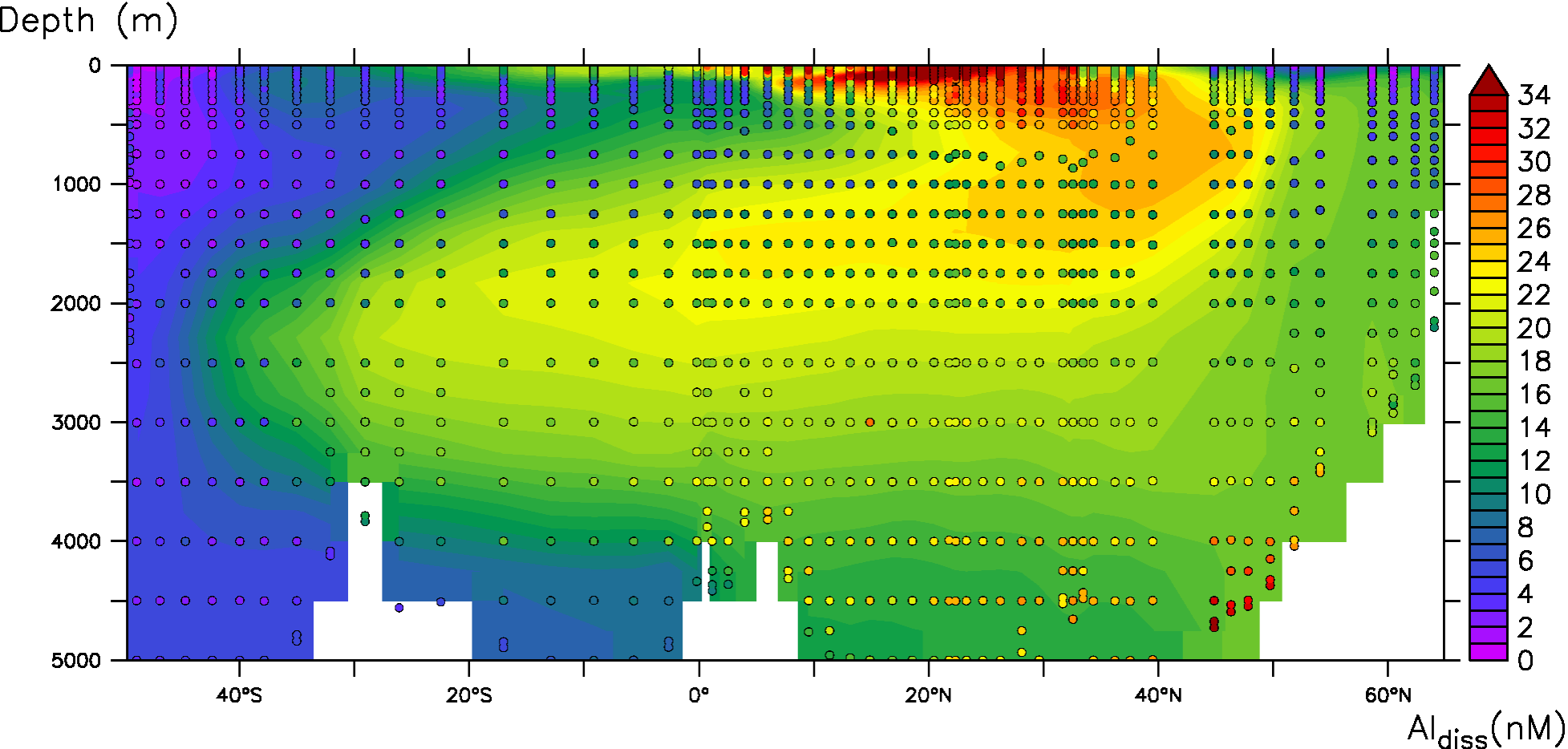}
        \label{fig:marg_section}
    }
    \subfigure[World ocean surface difference (nM)]{
        \includegraphics[width=\columnwidth]{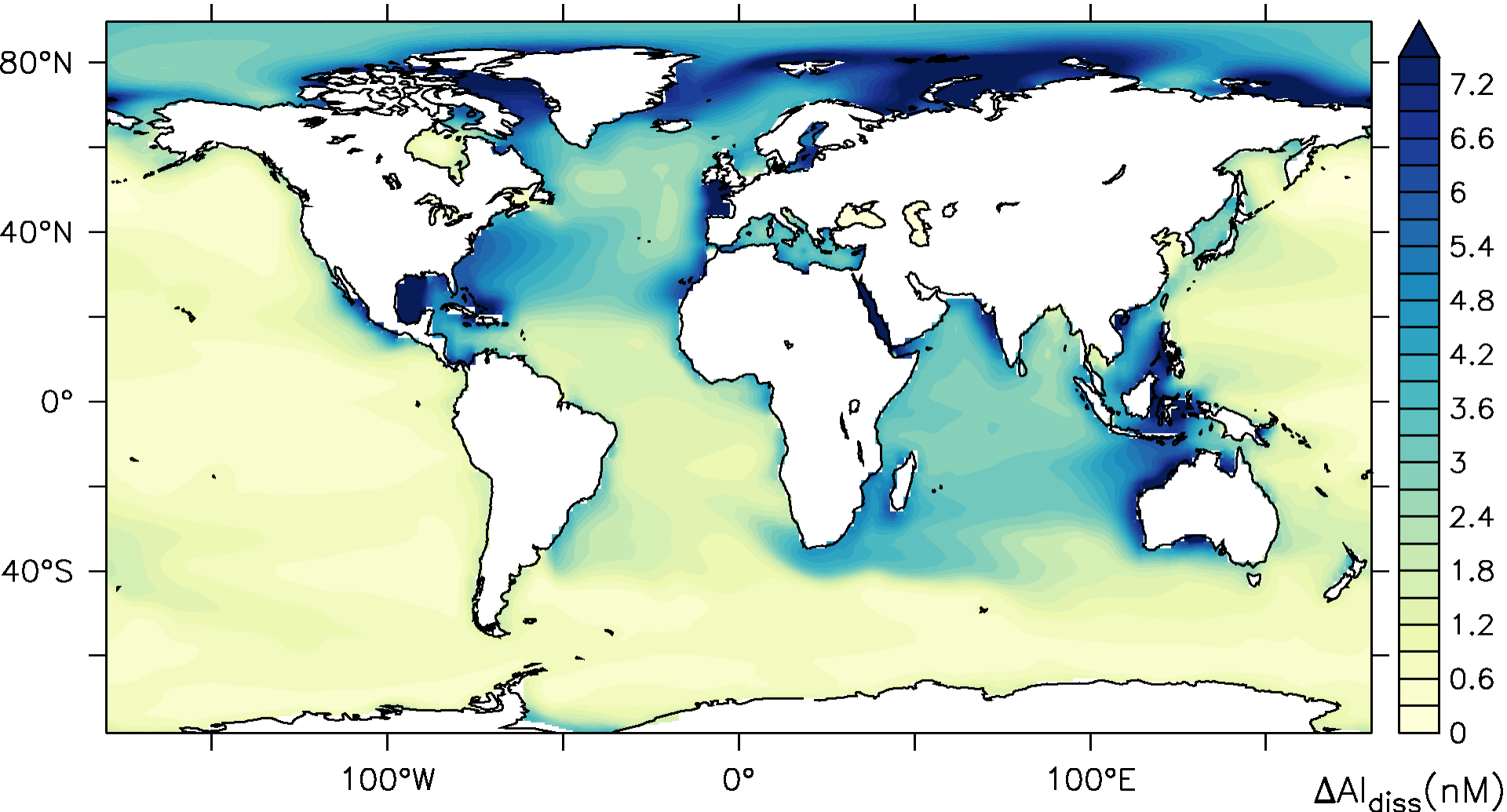}
        \label{fig:marg_surface_abs}
    }
    \subfigure[West Atlantic section difference (nM)]{
        \includegraphics[width=1.1\columnwidth]{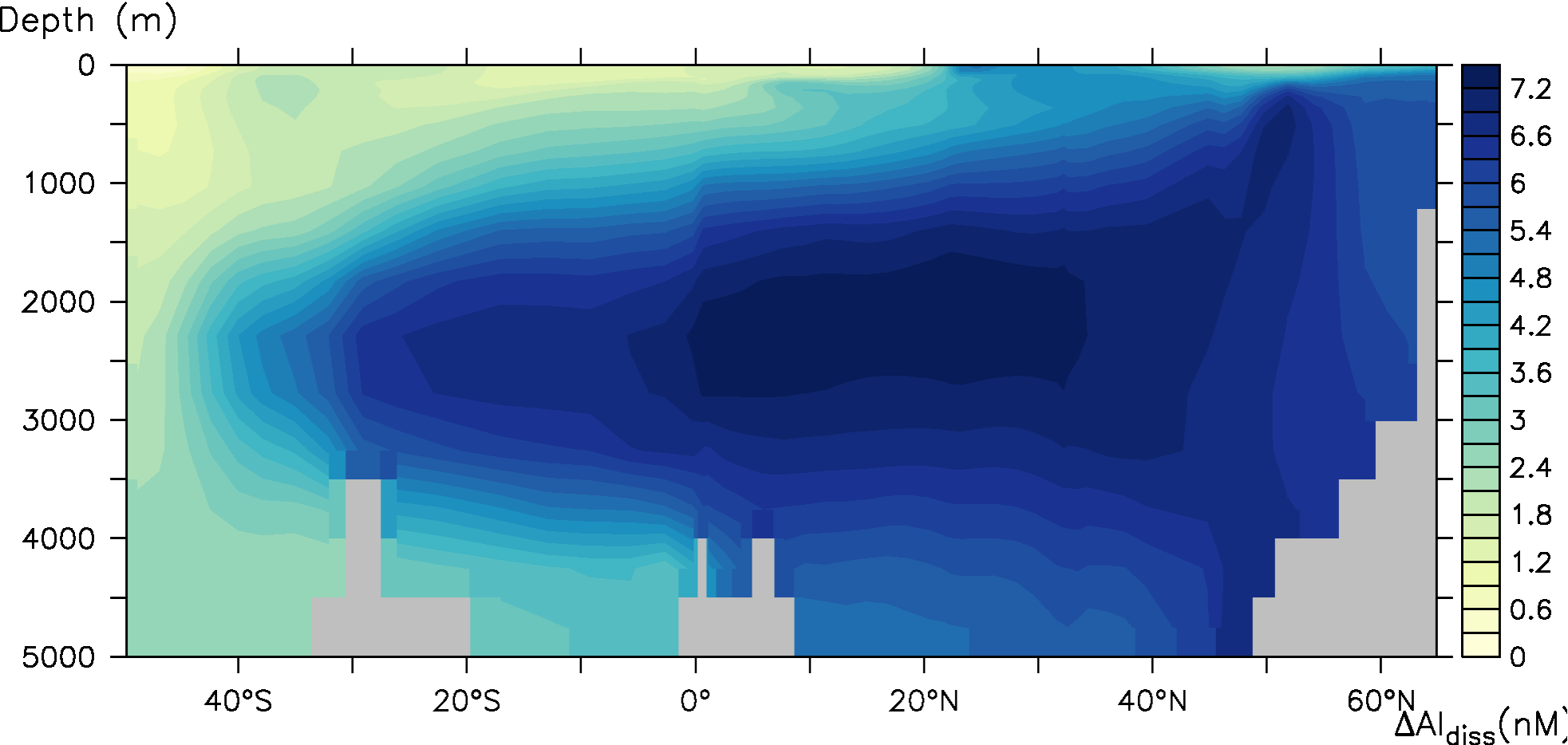}
        \label{fig:marg_section_abs}
    }
    \caption{Experiment with sediment input: [\dAl] (a,b) and its
    difference with the reference run (c,d)}
    \label{fig:marg}
\end{figure*}

It is possible that dissolution of dust is not the only way in which aluminium
enters the ocean.  River input and hydrothermal vents are excluded as a
significant sources of Al (see Section \ref{sec:introduction}).  But one other
source that might be significant, and might also help our model to produce more
realistic [\dAl] (e.g.\ in the Arctic Ocean), is redissolution of Al from
sediment resuspension.
Therefore, in one of the simulations, we have included a rough approximation of
Al margin sources, as described in Section \ref{sec:model_input}.  The
resulting [\dAl] and its difference with the reference experiment is plotted in
Fig.~\ref{fig:marg}.
The influence of shelf sediment sources on the aluminium distribution can
clearly be seen.  Especially in the Arctic Ocean, the Indonesian Archipelago
and near the east coast of North America, dissolved aluminium is increased
substantially when sediment input is added to the model.  This is to be
expected, since the largest sediment source is on the shelf areas (see Fig.\
\ref{fig:margins} for the added Al source).

Let's now look at the West Atlantic cruise track.  Fig.\
\ref{fig:marg_surface_abs} represents the difference between the sediment
source experiment and the reference simulations at the ocean surface.  There is
a clear increase at and near the surface between 20 and 40\degree N in the West
Atlantic Ocean.
The cruise section difference plot in Fig.\ \ref{fig:marg_section_abs} shows a
large increase in the NADW.  So it seems that high [\dAl] can be explained by
not only dust deposition but also sediment sources.  This is consistent with
the explanation of \citet{middag_prep}, even though their arguments
mostly concern deep sediment resuspension and to a lesser extend margin
sediments.  The sediment source in our model is mostly from near-shore sediment
resuspension (Fig.\ \ref{fig:margins}), in line with the findings of
\citet{mackin1986}.  Of course, a deep sediment source might also explain the
high [\dAl] near the sediment at 45--50\degree N, and it may therefore be
important to include such a source in a future model study, since our current
results suggest that margin sediments do not effectively reproduce this
maximum.

\begin{figure}[h]
    \includegraphics[width=\columnwidth]{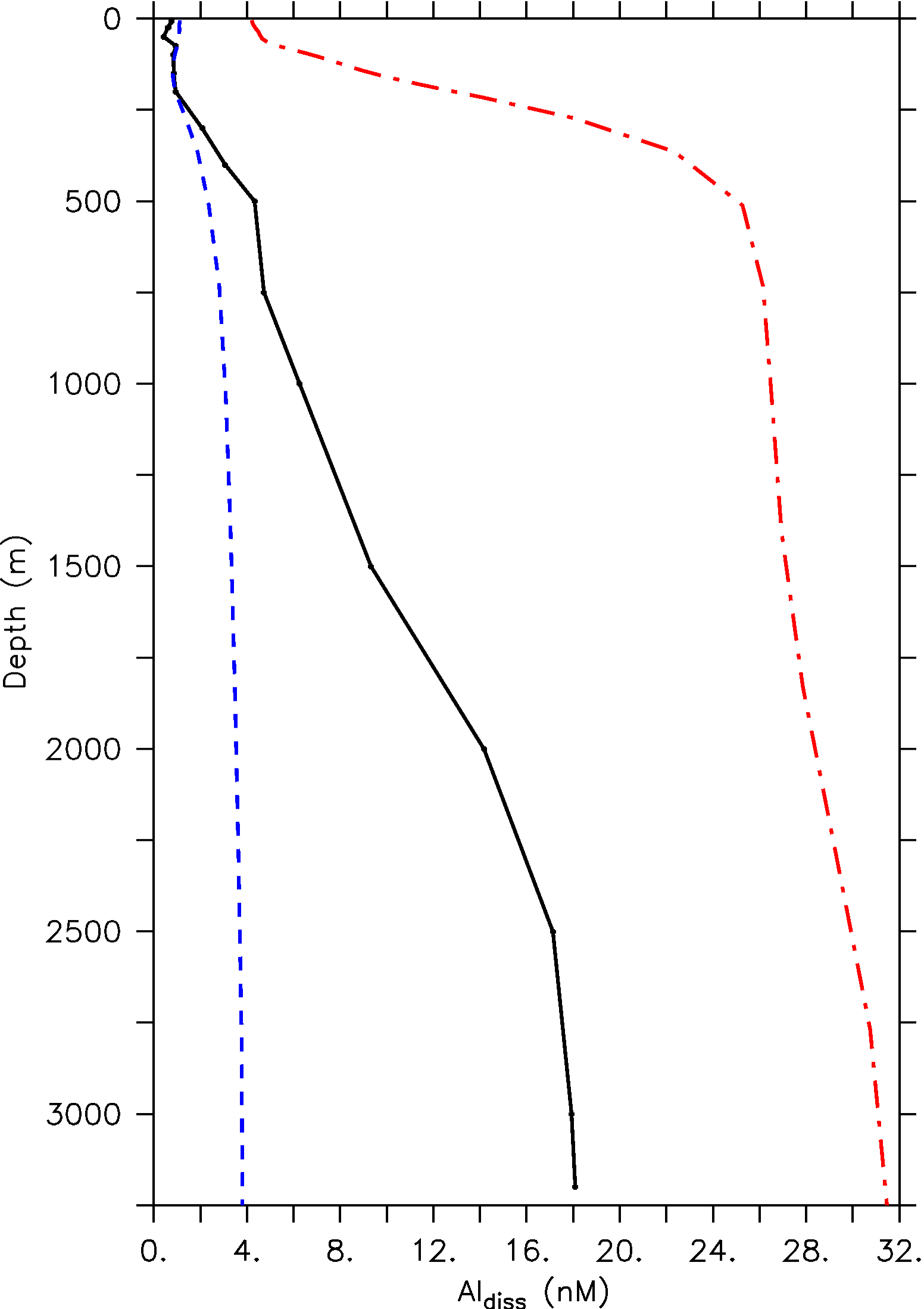}
    \caption{[\dAl] (nM) profile from a station in the eastern Arctic
    (ARKXXII/2): observations (black solid line), the reference simulation (blue
    dashed line), and sediment sensitivity simulation (red dash-dotted
    line), all at 213.6\degree E, 87.03\degree N}
    \label{fig:East_Arctic_profile_marg}
\end{figure}

The concentration in the model gets significantly higher in the Arctic Ocean
which is bad for depths between the surface and a few km, but better for deeper
waters, where a much higher concentration of aluminium was observed.  This is
shown in the profile plots of Fig.~\ref{fig:East_Arctic_profile_marg}.
Sedimentary input has as an effect that [\dAl] increases everywhere in the
Arctic Ocean and especially with depth, giving a profile that has the same
\emph{form} as the observations, suggesting the potential importance of margin
\dAl sources herein.
%FIXME: This is not true!  TODO: Check profile of spun-up exp.
%This is a silicon-like profile, illustrating the link between Si and Al (see
%\citet{middag2009}). -- FIXME: not enough evidence for that (Marion)
%
However, the modelled [\dAl] is too large compared with the observations.
Potential reasons for this are described in Section \ref{sec:discuss_marg}.

\subsubsection{Decreased partition coefficient} \label{sec:deckd}
One of the key parameters in our model is the partition coefficient $k_d$.  It
regulates the amount of Al that can be adsorbed by \bsi.  To see what effect
this parameter has in our model setup, we performed an experiment in which 
$k_d$ is decreased by a factor of two.

\begin{figure}
    %\subfigure[World ocean surface concentration (nM)]{
    %    \includegraphics[width=\columnwidth.3]{kd_surface_v2_pix}
    %    \label{fig:kd_surface}
    %}
    %\subfigure[West Atlantic section concentration (nM)]{
    %    \includegraphics[width=\columnwidth.36]{kd_section_v2_pix}
    %    \label{fig:kd_section}
    %}
    \subfigure[World ocean surface difference (\%)]{
        \includegraphics[width=\columnwidth]{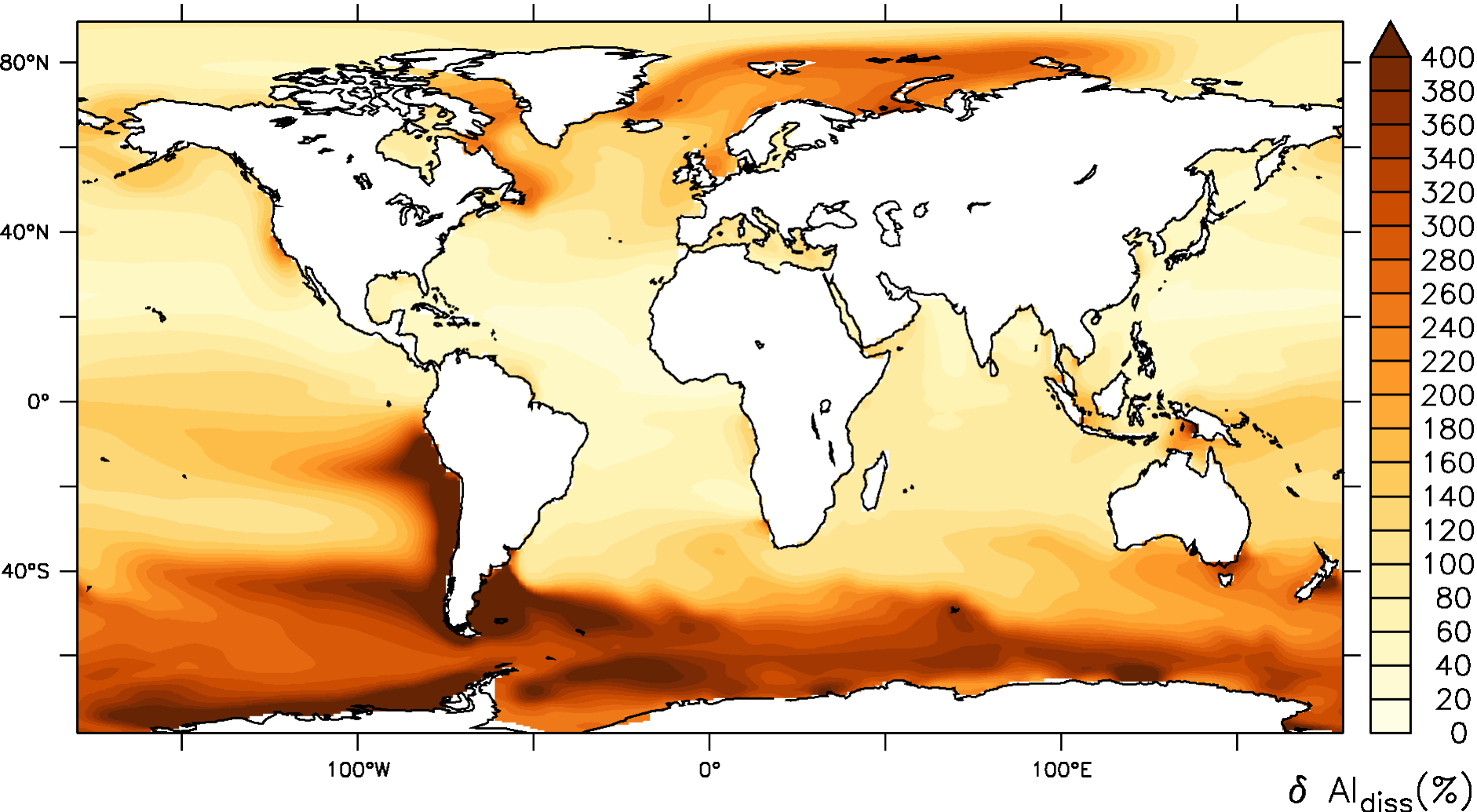}
        \label{fig:kd_SENS_surface}
    }
    \subfigure[West Atlantic section difference (\%)]{
        \includegraphics[width=1.1\columnwidth]{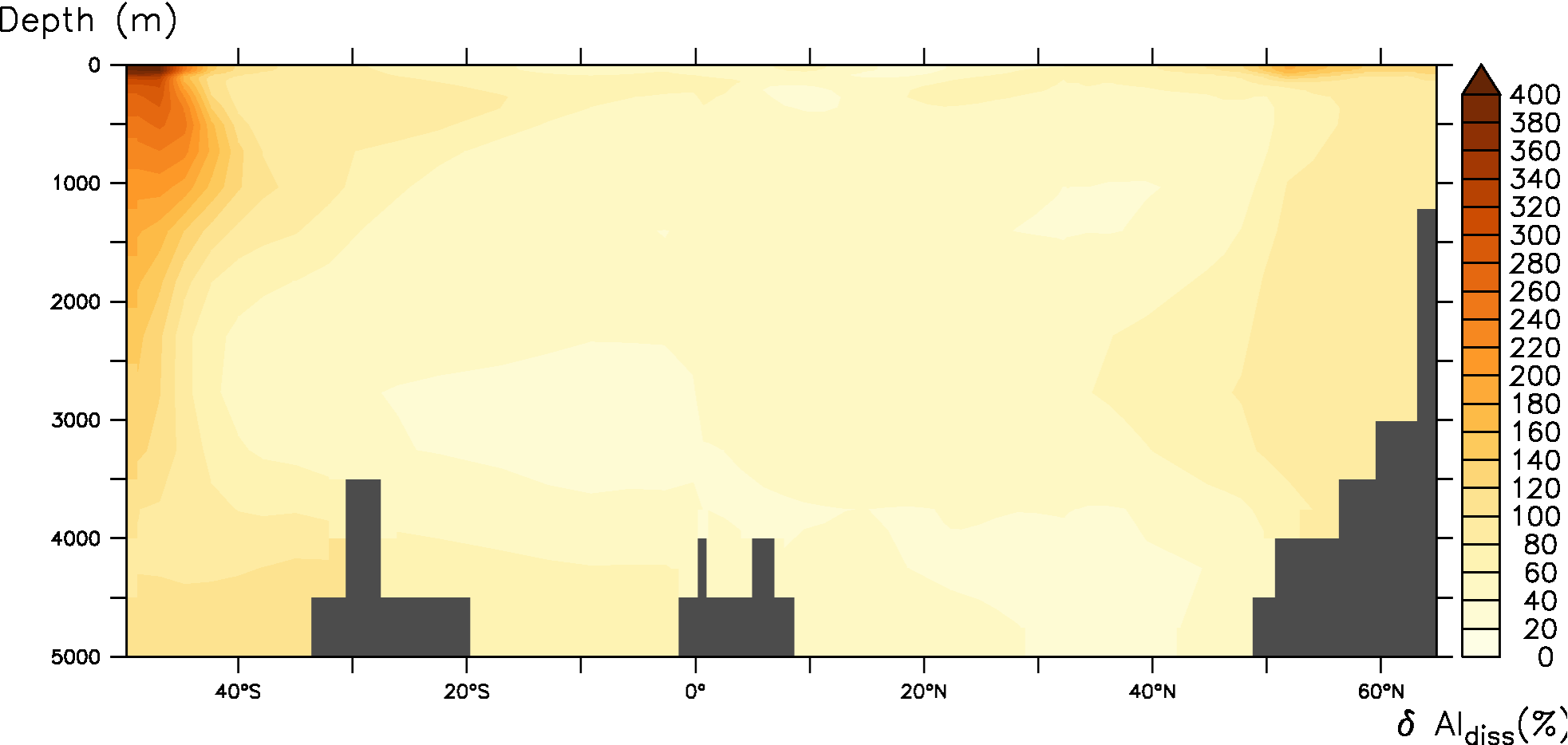}
        \label{fig:kd_SENS_section}
    }
    \caption{%
    The difference of the experiment with a decreased $k_d$ relative to
    the reference experiment.} \label{fig:deckd_rel}
\end{figure}
Fig.~\ref{fig:deckd_rel} shows that a decrease of $k_d$ leads to the most
significant relative increase of [\dAl] in the Southern Ocean.
The reason for the large increase in [\dAl] in the Southern Ocean, and also a
significant amount in the northern seas, is that these are locations of high
diatom production, which results in large [\bsi].  As can be seen from Eq.\
\ref{eqn:eq-gehlen}, the equilibrium concentration of \ads is proportional to
$k_d$ and [\bsi], hence $k_d$ is important in regions with high [\bsi].
This will be further explained by looking at the Al budget and timescales in
Section \ref{sec:discuss_kd}.

\subsubsection{Decreased first order rate constant}   \label{sec:sloweq}
The second most important scavenging parameter is the first order rate constant
$\kappa$.  This parameter should have a dynamic effect on the distribution of
\dAl, since it describes how quickly \dAl and \ads equilibrate (see Eq.\
\ref{eqn:eq-gehlen}).  To test this in our model, we decreased $\kappa$ from
$10^4$ to $10^2$ yr$^{-1}$.

\begin{figure*}
    \subfigure[World ocean surface concentration (nM)]{
        \includegraphics[width=\columnwidth]{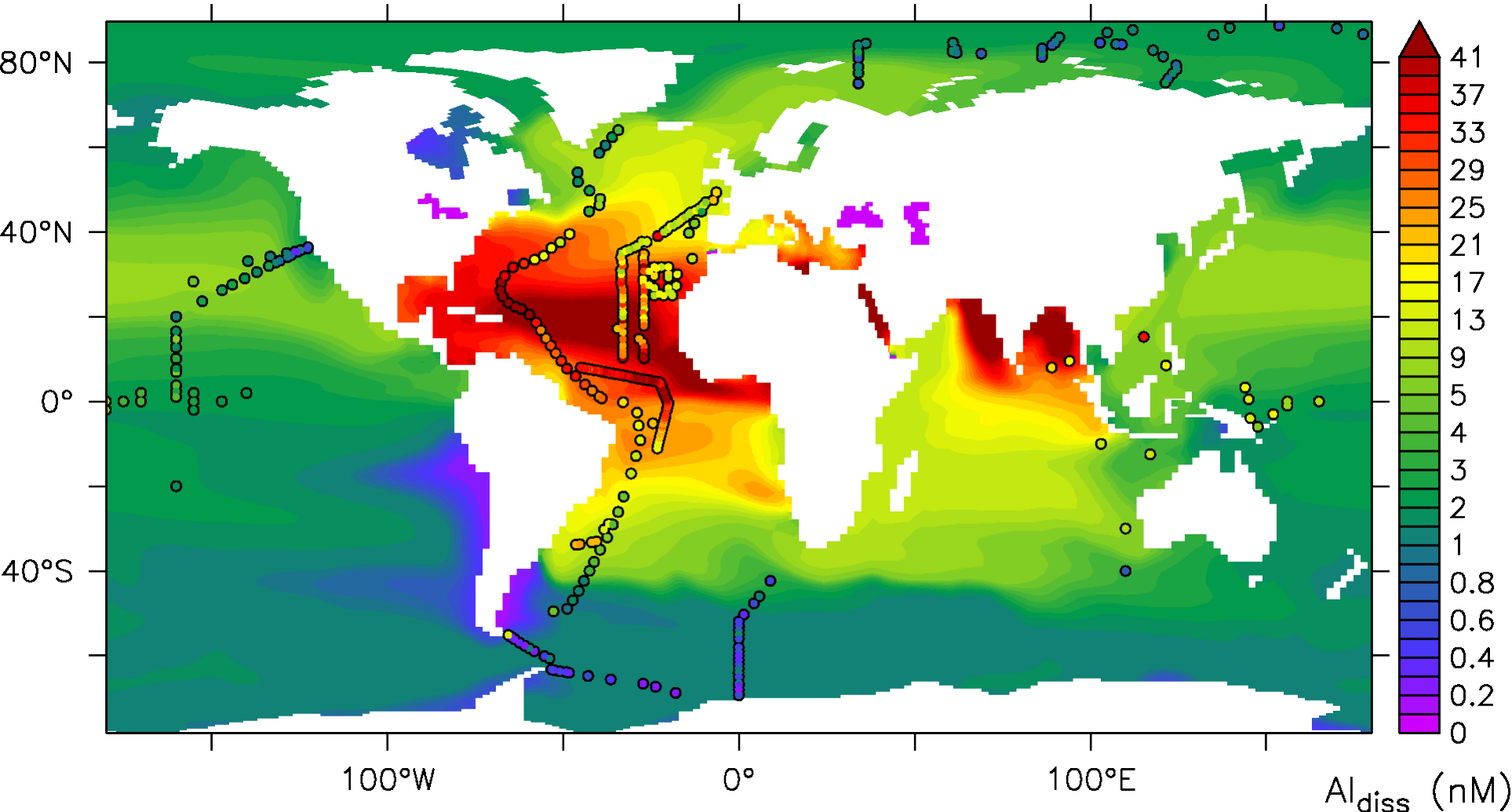}
        \label{fig:slow_surface}
    }
    \subfigure[West Atlantic section concentration (nM)]{
        \includegraphics[width=1.1\columnwidth]{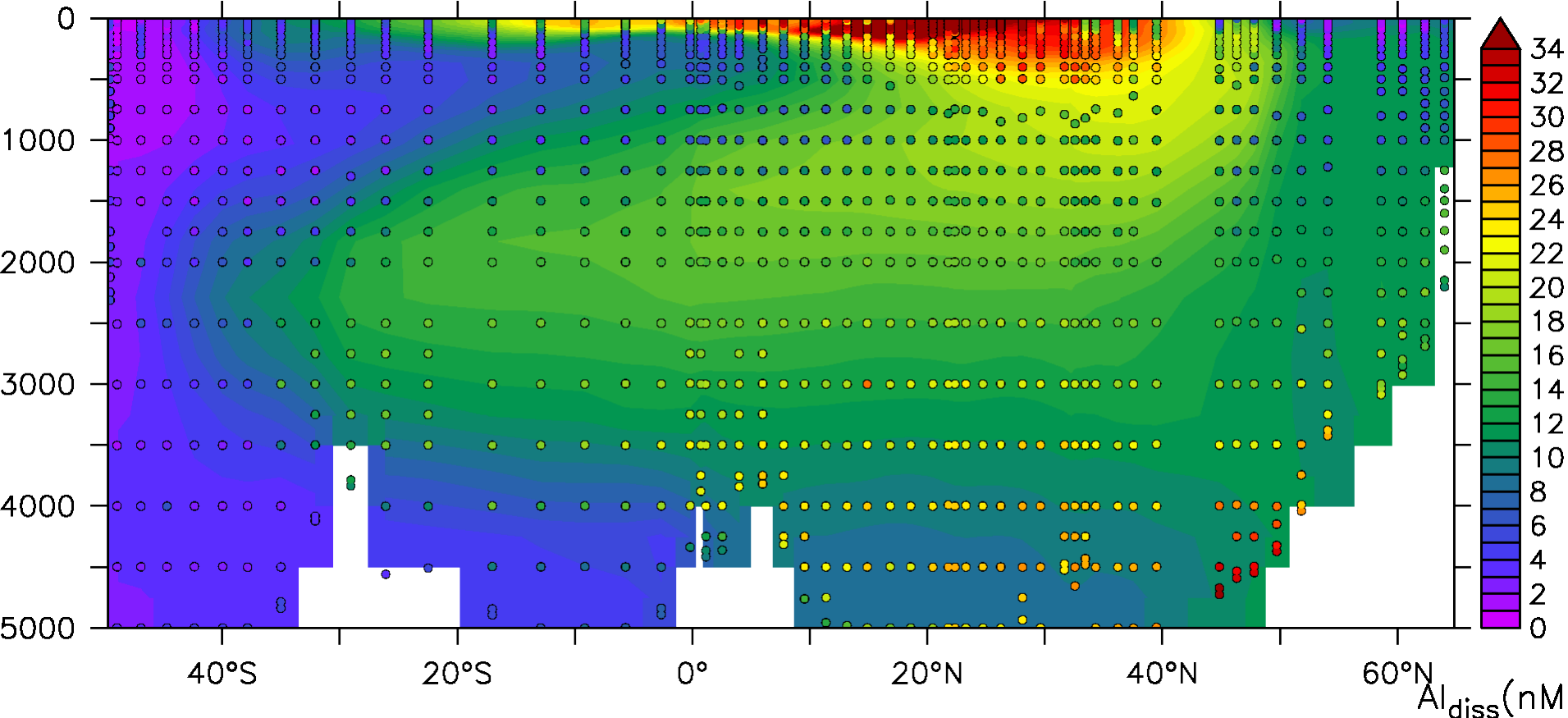}
        \label{fig:slow_section}
    }
    \subfigure[World ocean surface difference (\%)]{
        \includegraphics[width=\columnwidth]{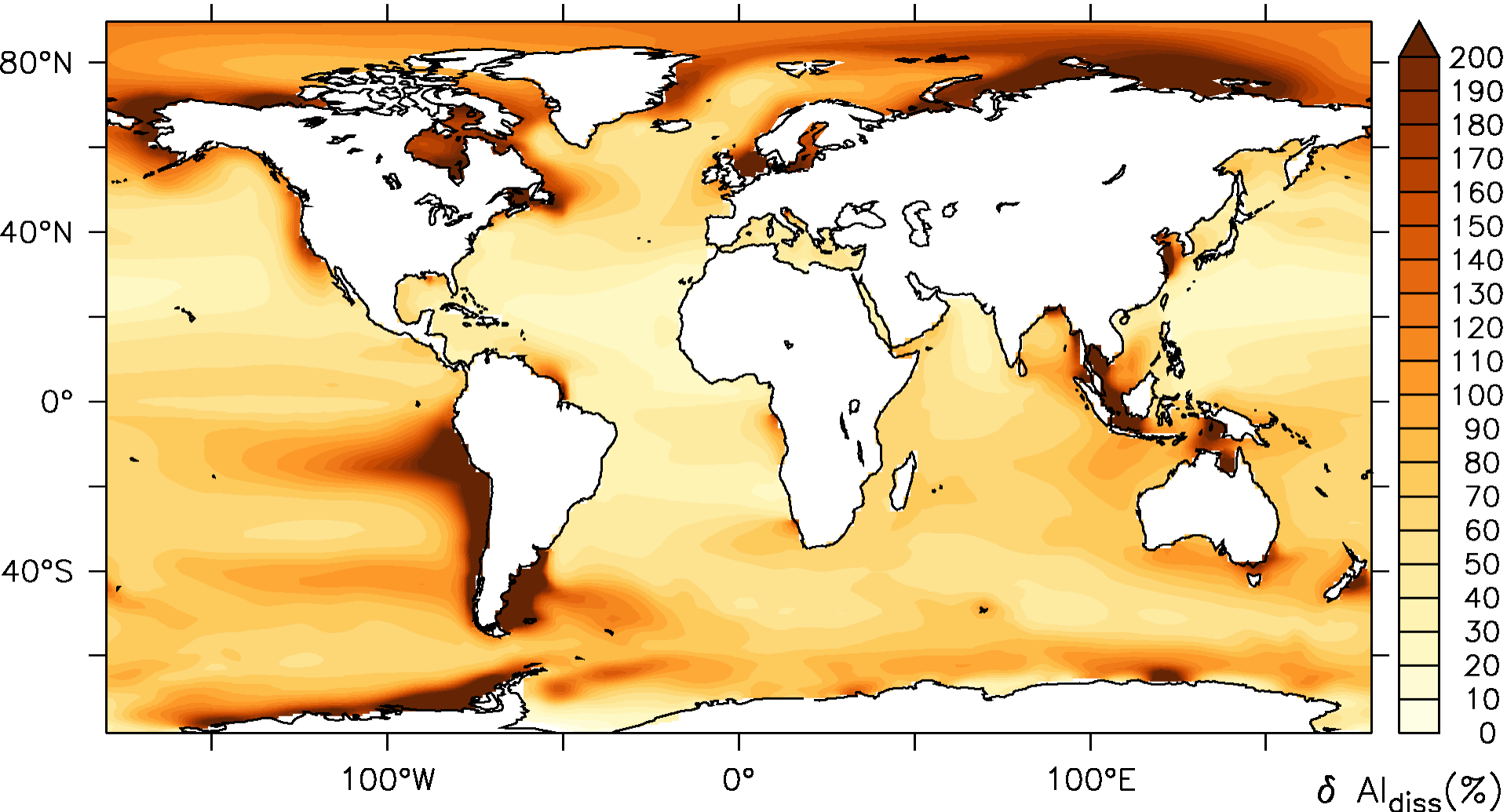}
        \label{fig:slow_surface_rel}
    }
    \subfigure[West Atlantic section difference (\%)]{
        \includegraphics[width=1.1\columnwidth]{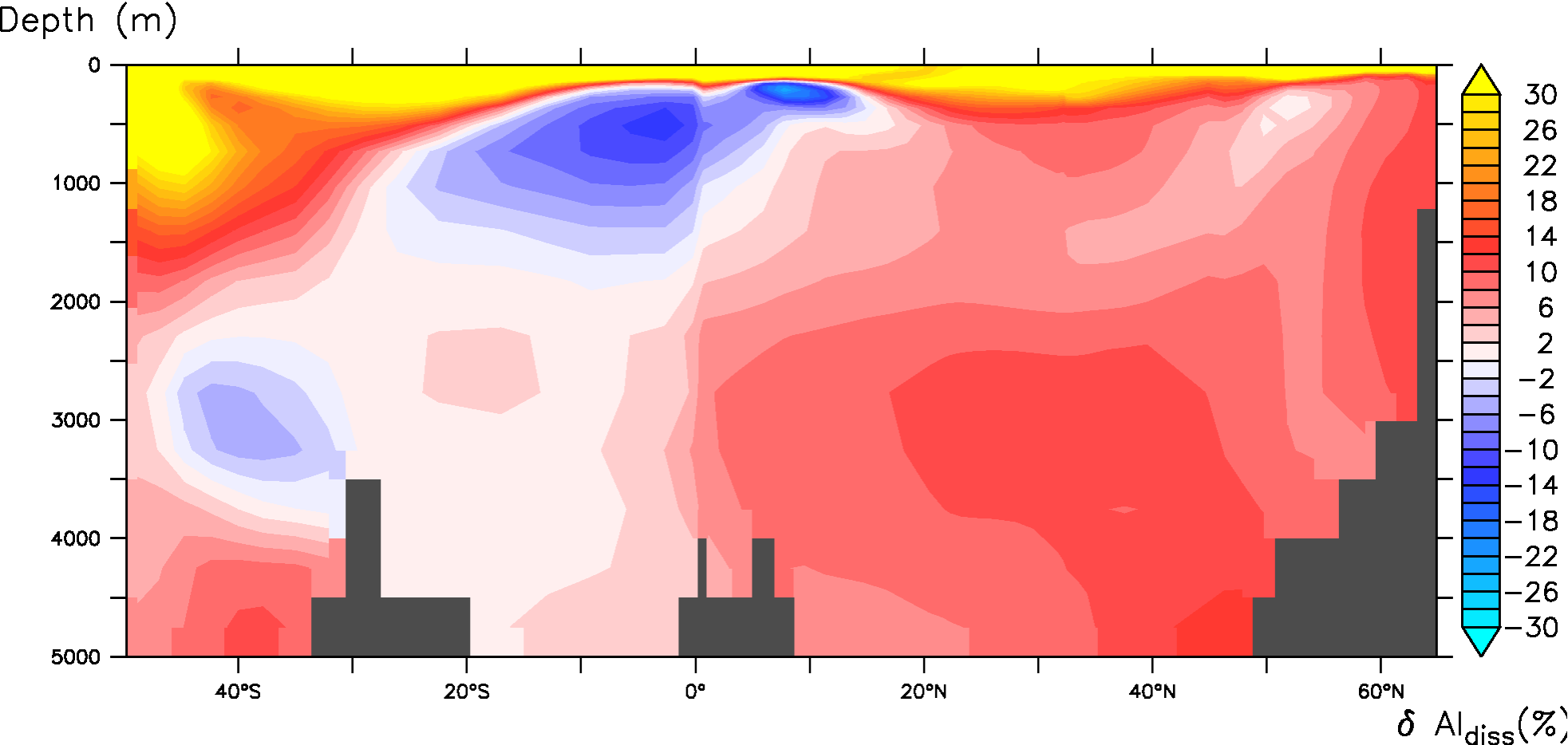}
        \label{fig:slow_section_rel}
    }
    \caption{[Al$_\text{diss}$] (nM) in the slow
    equilibration experiment ($\kappa = 100 $ yr$^{-1}$) and the relative
    difference with the reference experiment ($\kappa = 10^4 $ yr$^{-1}$).}
    \label{fig:slow}
\end{figure*}

As can be seen in Fig.\ \ref{fig:slow}, surface [\dAl] increases significantly
compared with the reference experiment (compare Fig.\ \ref{fig:refexp_surface}
with Fig.~\ref{fig:slow_surface}).  Fig.~\ref{fig:slow_section_rel} shows the
relative change in [\dAl] in the West Atlantic Geotraces section.  From the
equator northward below 2 km depth an increase of aluminium is visible.  There
is also an increase near the deep sediment between 45 to 50\degree N, but it is
too small to explain the elevated concentration in this area found in the West
Atlantic Geotraces observations (see bottom right in Fig.\
\ref{fig:refexp_section}).

\section{Discussion}     \label{sec:discussion}
\subsection{Comparison with Gehlen et al.\ (2003)}
Since our model is very similar to the model of \citet{gehlen2003}, a good
agreement between our results and theirs is expected.  Even though several
features in the two model results are the same, there are also noteworthy
differences. These can arise because of differences in the ocean physics, the
dust deposition field, as well as the biogeochemical model.

For the velocities the output of the NEMO model on the ORCA2 configuration is
used in our model.  On the other hand, in the model of \citet{gehlen2003} a
climatology from the Hamburg Large-Scale Geostrophic (LSG) OGCM is used
\citep{maier1993}.  The basic characteristics of the physics of the two models
are the same, but a few features, like the Mixed Layer Depth (MLD) and the
Atlantic MOC, are significantly different.  Even though LSG is a low resolution
model, the depth of the meridional southward return flow of the Atlantic MOC is
almost 3 km deep, which is realistic.  On the other hand, in the OPA model
(with the ORCA2 configuration) this return flow is much more shallow (a bit
over 2 km deep), as shown in Fig.\ \ref{fig:contour_on_model}.  A shallow
return flow is typical for low resolution models (e.g.\ \citet{dutay2002}).
Furthermore, the AABW in LSG goes to 15\degree N, while in our model it goes
all the way to 40\degree N.  Overall, the physics of the two models are
comparable, but we should keep in mind that the overturning in the ORCA2
configuration of OPA is too shallow and AABW goes too far north, compared with
the LSG model.

%%% Compared with Gehlen: surface %%%
In both the work of \citet{gehlen2003} and in our work, the highest
concentration of \dAl is found in the surface model layer
(Fig.~\ref{fig:refexp_surface}) in the Atlantic Ocean around 20\degree N.  In
both models the \dAl concentration is low in the Southern Ocean and in the
Pacific Ocean.  However, there are distinct differences of [\dAl] in the
surface ocean in the models.  Several features visible in the observations are
better captured by the model of \citet{gehlen2003}, while other features are
simulated better by our model.
These differences are likely to be due to the different dust deposition fields
used in the model of \citet{gehlen2003} compared to the one we used.

In the eastern Atlantic Ocean around 20\degree N near the coast of Africa,
\citet{gehlen2003} simulate a \dAl concentration of over 300 nM, or around 50 nM
for a different dust deposition field.  The concentrations in our model and the
observations, however, are generally around 30 nM.

In and near the Indonesian Archipelago [\dAl] ranges from 1 to 20 nM in both
models, while the observations of \citet{obata2007} and \citet{slemons2010} are
much more homogeneous, about 10 nM.  This might be because the ocean currents
in both models are not realistic because of a too low resolution of our model
for this region.  Possibly other effects in the real ocean play a role, like
sedimentary input from the Indonesian Archipelago and east of Indonesia (see
Fig.\ \ref{fig:margins} for the relative importance).

%%% Compared with Gehlen: depth %%%
The vertical meridional section in Fig.~\ref{fig:refexp_section} through the
West Atlantic Ocean shows a reasonable correspondence between the two models.
The concentration at the surface is high in both models, it roughly
decreases with depth in the Atlantic Ocean.  In the Southern Ocean aluminium
concentrations are low in the whole column, but especially in the surface
(Fig.~\ref{fig:refexp_section}, very left).  However, in our model between
10\degree S and 45\degree N there is a significant decrease of [\dAl] with
depth, which is not visible in the model results of \citet{gehlen2003} where
below 2 km north of the equator [\dAl] is rather homogeneous in both dimensions
along the cross-section.
The model of \citet{gehlen2003} shows North Atlantic Deep Water (NADW) in the
\dAl distribution, spreading southward at a depth of almost 3 km, as also shown
by the observations (Fig.\ \ref{fig:refexp_section}, but also Van Aken, in
preparation).  Our model also captures NADW, but this is at only 2 km depth, as
explained before.
%This will be explained in the Section \ref{sec:moc}.

\subsection{Advection versus scavenging}    \label{sec:adv_vs_scav}
%The model used in this study to simulate the Al concentrations is a General
%Circulation Model (GCM) with scavenging of \dAl.  We have used a GCM, because
%the \dAl distribution is not purely controlled by dust deposition and
%scavenging.  Certain areas of the ocean might be regulated mostly by scavenging,
%while other areas might be mostly controlled by advection.  
While Al is entering the ocean through the surface by dust deposition and
leaving it at the bottom through burial, it is also transported by the currents.
In this section we try to find out whether scavenging or advection is more
important in setting the modelled Al concentration at different regions of the
ocean.

In our model scavenging is reversible, which we base on a priori arguments
\citep{bacon1982,anderson2006} and the general increase of [\dAl] with depth in
observations.  Therefore it is unexpected that in the Atlantic Ocean a depth
decreasing profile of \dAl is simulated (see e.g.\ Fig.\
\ref{fig:slow_section}).
One possible reason for this is that the \bsi distribution is not realistic.
However, the PISCES model with the Si cycle is used by several modelling studies
\citep{aumont2006,dutay2009,arsouze2009,dutay2009,tagliabue2010} and has the
same spatial patterns as known \bsi export fields (e.g.\ \citet{sarmiento2006}).
% FIXME/TODO: This text does not really suffice -- a comparison with e.g.
% Phoebe's data should be done!

It should be emphasised that we only scavenge Al by biogenic silica.  Of course
CaCO$_3$ and POC might also scavenge in reality.  Looking back to our
comparison with the observations of \citet{middag_prep}, we found that near
Greenland [\dAl] was overestimated by the model.  This would be an example
where the addition of other scavengers, especially CaCO$_3$ can help in
decreasing the modelled [\dAl] near Greenland, and also in the western South
Atlantic Ocean between 30 and 40\degree S.  Refer to e.g.\
\citet{dittert2005,sarmiento2006} for export of \bsi, CaCO$_3$ and POC; or
\citet{lam2011} for concentrations.

The importance of scavenging is visible from several features in our model
data.
Firstly, Fig.\ \ref{fig:refexp_layers} shows that on a global scale the pattern
of [\dAl] has roughly the same form at different depths, which is a result of
reversible scavenging, i.e., at the surface where [\bsi] is high, Al is mainly
adsorbed onto \bsi, while at depth Al is mostly desorbed from \bsi.
Specifically, [\dAl] is high in the Atlantic and Indian Oceans north of
30\degree S, and much lower in other regions of the ocean.
Secondly, the fact that the concentration in the Pacific Ocean is low at all
depths, is because \dAl is removed by scavenging before it reaches the Pacific
Ocean.
These are just a few of the features in the \dAl distribution that suggest that
on a large scale scavenging is more important than advection.

% Importance of advection:
However, the observations and the model show that advective transport by ocean
currents is important as well.
For instance, Fig.\ \ref{fig:dust} shows that large quantities of dust are
deposited just west of the Sahara.  However, large concentrations of \dAl are
not only visible directly below the dust deposition site, but even higher
concentrations are found further to the west (Fig.\ \ref{fig:refexp_surface}).
Large amounts of dissolved Al must be advected from the dust deposition site
toward Central America.  Hereafter it seems to be advected northward
until Iceland, where [\dAl] is high despite very little dust deposition.
Deeper in the North Atlantic Ocean the original dust signal decreases, and at 2
km depth in the model (or 3 km depth in the observations) \dAl is transported
into the southern hemisphere by the NADW.  This southward transport
can be seen clearly in Fig.\ \ref{fig:refexp_section}, but no further than
40\degree S.  South of this latitude the \dAl concentration is low.
%
%In the Pacific Ocean there is not so much dust deposition and therefore the
%primary way in which [\dAl] can increase in our model is by means of advection
%from the Atlantic (and possibly Indian) Ocean.
% ^^ DOES not really SEEM consistent with scavenging argument ^^
These examples indicate that also advection plays an important role in the
redistribution of Al.

\subsubsection{Timescales}     \label{sec:timescale}
While the analysis of [\dAl] distributions points to the combined importance
of scavenging and advection in setting its overall pattern, it does not allow to
quantify their relative importance. In order to assess the relevance of these
different processes, associated timescales are defined based on our model equations.

% ttrans and tscav:
A priori the relative relevance of scavenging can partly be derived from the
model equations.
Substituting Eq.\ \ref{eqn:eq-gehlen} into Eq.\ \ref{eqn:scav-gehlen} yields:
\begin{align}
    \frac{\textrm{d}}{\textrm{d}t} [\ads] &= \kappa \cdot ([\text{Al}^\text{eq}_\text{ads}] - [\ads]) \notag\\
                                          &= \frac{1}{\tau_\text{ads}} [\dAl] - \frac{1}{\tau_\text{eq}} [\ads] \;,
    \label{eqn:hom}
\end{align}
where $\tau_\text{ads} = 1/(\kappa k_d [\bsi])$ is the typical adsorption
timescale.  This is the typical time necessary for Al to transform from \dAl to
\ads (or vice versa).  This process depends on the amount of available \bsi and
the amount of \dAl which can adsorb onto \bsi, giving the actual rate of
conversion between \dAl and \ads.
The addition of \ads is equal to the removal of \dAl, so $\tau_\text{ads}$ is
either the time necessary to add \ads, or the time to remove \dAl.
The second timescale in Eq.~\ref{eqn:hom}, $\tau_\text{eq} = 1/\kappa$, is the
time it takes for \ads to equilibrate to $\text{Al}^\text{eq}_\text{ads}$.
This process provides a stabilising feedback on the growth of [\ads] (or
decrease of [\dAl]).

Scavenging is the process of adsorption and sinking, so we need to know how fast
particles sink.  This is used together with the adsorption timescale to
define a scavenging timescale.
Since the surface ocean is analysed, the typical sinking speed might be defined as
sinking through the first 10 m, the thickness of the upper model gridbox.
However, after exporting \ads out of a gridbox by sinking, the particles sink
through the mixed layer (which is generally thicker than 10 m).  During sinking,
the particles will partly be mixed back into the upper gridbox.  Therefore to
speak of a removal, it is a necessary condition for Al to sink out of (at least)
the mixed layer.
In our model, the sinking velocity of \bsi and \ads is defined to be constant in
the mixed layer, namely $w_s = 30 $ m/day.  Now it is easy to define the
typical sinking timescale: $ \tau_\text{sink} = D_\text{ML} / w_s $,
%    \label{eqn:tsink}
where $D_\text{ML}$ is the mixed layer depth.
The scavenging timescale can be defined as the maximum of the adsorption and
the sinking timescales:
\begin{equation}
    \tau_\text{scav} = \text{max}( \tau_\text{ads},\tau_\text{sink} )\;.
    \label{eqn:tscav}
\end{equation}
This signifies how fast Al is exported from the mixed layer.  Except for a few
places where the mixed layer is very deep, like the centre of the Labrador Sea,
the time it takes to sink out of the mixed layer is generally very small
compared to the time it takes the Al to adsorb onto a particle
($\tau_\text{sink} \ll \tau_\text{ads}$) and thus almost everywhere
$\tau_\text{scav} = \tau_\text{ads}$.

\begin{figure}
    \centering
    \includegraphics[width=\columnwidth]{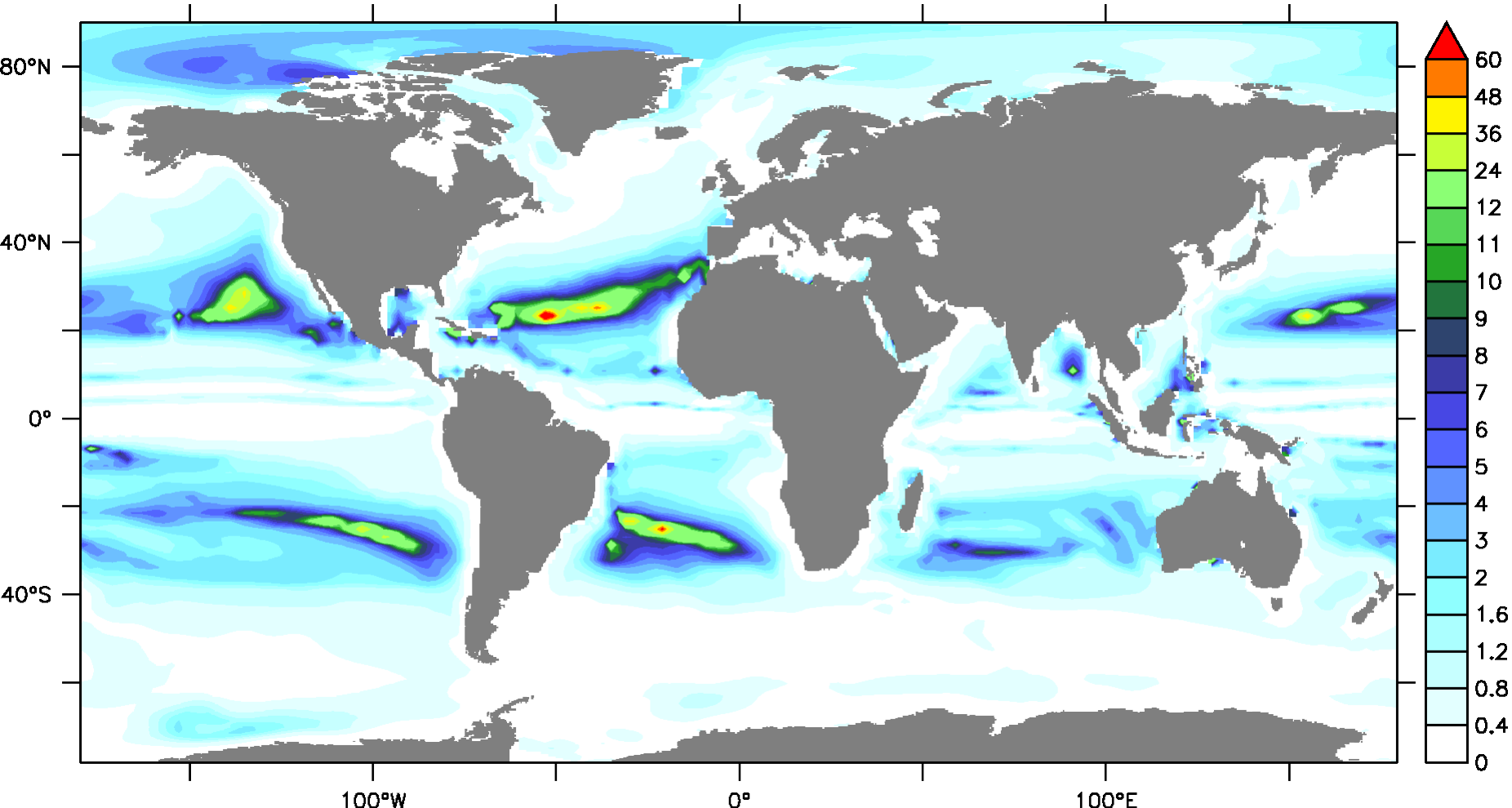}
    \caption{Mean local residence time (in months) of dissolved Al in the
    surface mixed layer based on scavenging and advection processes.}
    \label{fig:tres}
\end{figure}

% tadv:
A typical advection timescale is defined as follows:
\begin{equation}
    \tau_\text{adv} = \text{min}\left( \frac{L}{V}, \frac{D_\text{ML}}{W} \right) \;.
    \label{eqn:tadv}
\end{equation}
Here $L$ is the typical length scale, defined as the horizontal diameter of a
gridbox, and $V$ is the horizontal speed.  Since the vertical velocity
component is very small ($W \ll (D_\text{ML}/L) U$), at most locations only the
horizontal components need to be considered, but for correctness the vertical
advection $W$ is included in the calculation as well.  The meaning of this
timescale is that within a time $\tau_\text{adv}$ dissolved Al (or
any other non-buoyant tracer) is advected out of a gridbox (if it is not
scavenged before).

% tres:
The residence time can be defined as the minimum of the scavenging and the
advection timescales ($\tau_\text{adv} = \text{min}\left( \frac{L}{V},
\frac{D_\text{ML}}{W} \right)$).  It is the typical time that \dAl stays within
a volume box of horizontal grid resolution (about $2\degree \times 2\degree$)
times $D_\text{ML}$.  This quantity is presented in Fig.\ \ref{fig:tres}.
The oligotrophic gyres in the Atlantic, Pacific and Indian Oceans are clearly
visible.  In the centre of these gyres it takes much longer than one year
before Al is exported out of a volume box, either by advection or by
scavenging.  The modelled \dAl distribution of Fig.\ \ref{fig:refexp_surface}
shows in the Atlantic Ocean the largest values between 10 and 30\degree N.
This can be partly explained by the large dust input between 5 and 20\degree N,
where \dAl is removed relatively fast by advection but dust input keeps [\dAl]
large, and partly by the large residence time between 20 and 30\degree N, where
there is no large dust input but \dAl simply stays there for a long time, since
there is no \bsi present for scavenging and advection is very small.  A similar
argument can be given for the high [\dAl] near the North Pacific gyre.

% Upsilon:
\begin{figure*}
    \centering
    \includegraphics[width=\textwidth]{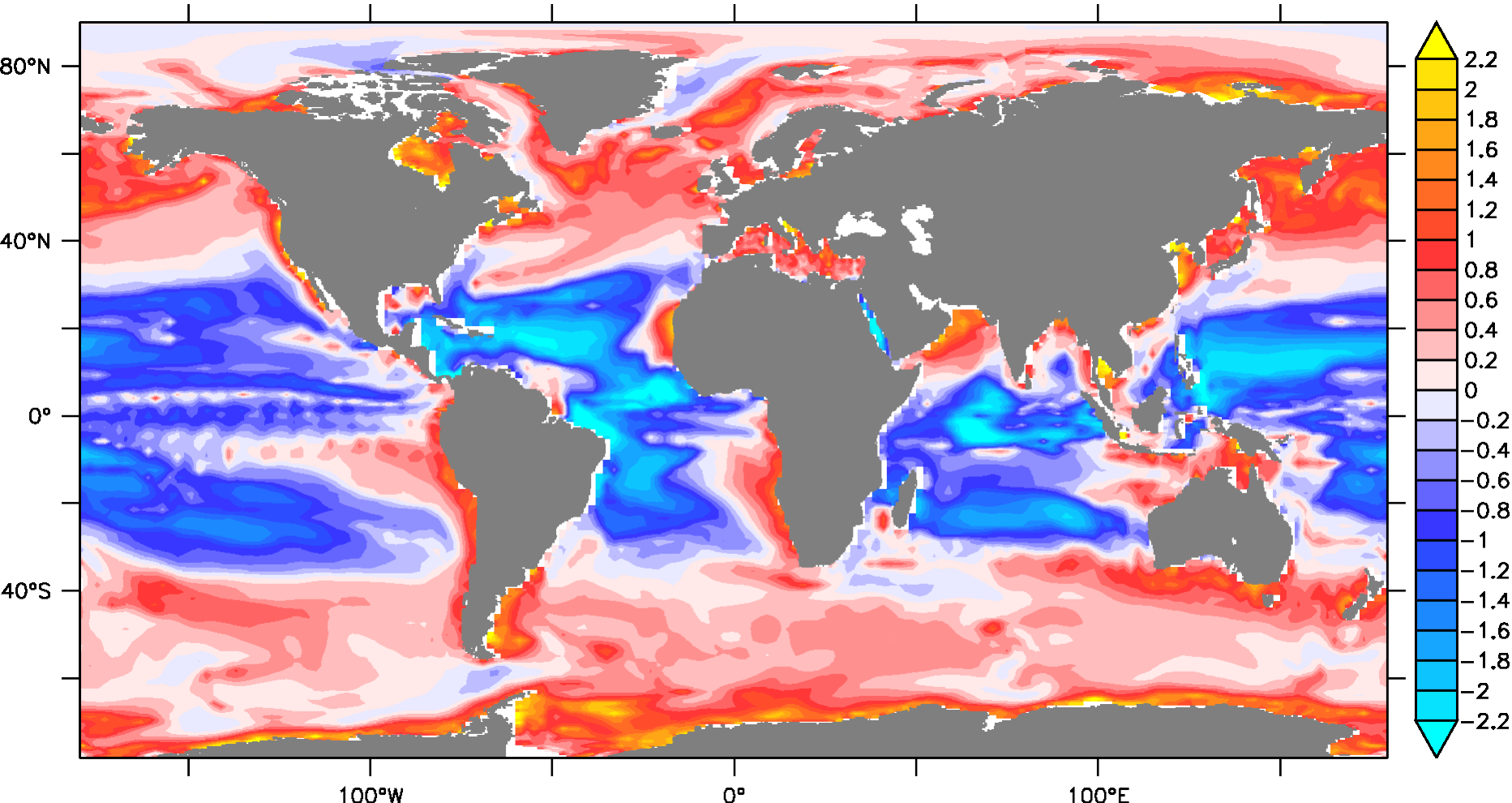}
    \caption{log$_{10}(\Upsilon)$, when
    $\kappa = 10^4$ yr$^{-1}$ and $k_d = 4\cdot 10^6$ l/kg.  For the advection time
    $\tau_\text{adv}$ the definition of Eq.\ \ref{eqn:tadv} is used,
    and the scavenging time $\tau_\text{scav}$ is the maximum of
    $\tau_\text{ads}$ and $\tau_\text{sink}$.}
    \label{fig:adv_vs_scav}
\end{figure*}
If we want to know which of the timescales are more important, a relative
relation between $\tau_\text{scav}$ and $\tau_\text{adv}$ must be defined.  The
number for relative importance of scavenging versus advection for Al export is
the following:
\begin{equation}
    \Upsilon = \frac{\tau_\text{adv}}{\tau_\text{scav}} \;.
    \label{eqn:adv_vs_scav}
\end{equation}
The logarithm of this quantity is plotted in Fig.\ \ref{fig:adv_vs_scav}.
In regions where $\Upsilon \gg 1$, like north of 40\degree N in the Atlantic
and Pacific Oceans and in the south of the Southern Ocean, scavenging is more
important than advection.  On large scales the large $\Upsilon$ regions
coincide very well with the large \bsi regions (Fig.\ \ref{fig:BSi_surface}).
Looking more in detail, we can see that $\Upsilon$ is less homogeneous than
[\bsi].  For instance, in the Drake Passage and in the Pacific sector of the
Southern Ocean at 150\degree W the timescale fraction is smaller then one would
expect based on the \bsi distribution.  This is because of the strong velocity
of the Antarctic Circumpolar Current (see Fig.\ \ref{fig:opa}).

In regions where $\Upsilon \ll 1$, in the low latitudes, advection is more
important than scavenging.
One can say that if $\Upsilon \ll 1$, [\dAl] is advection driven, while for
$\Upsilon \gg 1$, [\dAl] is scavenging driven.  In the high $\Upsilon$
regions a one-dimensional model would be a reasonable approximation, provided
that there is sufficient Al input to keep [\dAl] in steady state.  One region
where this should perfectly work is the Mediterranean -- there $\Upsilon$ is
very large and the dust flux is high.
Several other regions like the northern seas, near Antarctica and some other
coastal areas would also be good candidates for a one-dimensional model.  Other
regions of the ocean cannot be well described by a one-dimensional model.
     % FIXME: HOWEVER, \Upsilon SHOULD BE LARGE THEN AT ALL DEPTHS !!!
Specifically the Arctic Ocean would not be a good candidate.  Even though it is
a semi-closed basin, internal circulation has an effect of the same order as
scavenging.
%This has likely to do with the %relatively large currents FIXME: NO, THESE ARE
%SMALL !
%
These arguments must be taken into consideration when using one-dimensional
models.

%Adsorbed aluminium must sink sufficiently fast for this argument to hold.
%According to the model bathymetry, the average ocean depth is 3726 m and the
%sinking speed linearly increases from 30 to 200 m/day.  Then the time for \ads
%(and \bsi) to sink out of the model domain is about $\tau_\text{sink} = 45$
%days.  This time scale is much shorter than the \dAl to \ads conversion time
%scale $\tau_\text{trans}$ (as well as the advection time scale).  Therefore the
%typical scavenging time scale is equal to the conversion time scale
%($\tau_\text{sink} \ll \tau_\text{trans}$, and hence $\tau_\text{scav} :=
%\text{max}(\tau_\text{trans}, \tau_\text{sink}) = \tau_\text{trans}$).

%{\bf\large N.B.: Missing ni the analysis so far: scavenging vs. advection at depth!
%THis is also part of the reason why the explanations in next subsubsection are
%incomplete!}

%\subsection{Sensitivity simulations}
%We will now discuss further three of our sensitivity experiments.  Potential
%reasons for the too large [\dAl] when adding ocean margin sediments as a source
%of Al, will be discussed.

\subsection{Ocean sediments source}  \label{sec:discuss_marg}
Including margin sediment input of Al results in overestimated values for [\dAl]
compared to the observations in the Arctic Ocean.  Potential reasons for this
will now be discussed.

As described in Section \ref{sec:model_input} the Al flux is
proportional to the Fe flux which depends on the degree of oxygenation of the
sediments.  This is probably not a reasonable assumption, since generally Al
does not flow out of undisturbed sediments, but enters the water column through
resuspension of the sediments (e.g.\ \citet{mackin1986control,vanbeusekom1997}).
Since there is not enough data on resuspension rates, we chose this simplistic
parametrisation for our sensitivity experiment.
A priori we therefore have no reason to expect an improvement in the
model.  The sensitivity experiment can only be used to get an idea of the first
order effects of an ocean sediment source of Al, henceforth referred to
as ``sediment input''.

Since the sediment input in the Arctic Ocean is treated in the same way as the
rest of the ocean, it is surprising that [\dAl] gets much too high (Fig.\
\ref{fig:marg}),
certainly when considering the improvement in the West Atlantic as described
below.  There is a lack of knowledge of boundary exchange processes (e.g.\
\citet{arsouze2009}), which keeps open the possibility that the large sediment
source from the margins is compensated by boundary exchange processes.
In our model only \bsi was used as a scavenger, while near the margins, other
scavengers like POC and CaCO$_3$ might play an important role.
Also scavengers not present in the PISCES model, like clay minerals
\citep{walker1988} might be important for reducing [\dAl].
The possibility of this \emph{boundary scavenging}
\citep{bacon1988,arsouze2009} is convincingly shown to be present in
observations (e.g.\ \citet{brown2010aluminium}).

Since most of the data used here is from the open ocean and not from coastal
areas, and our model is too coarse for an analysis of the coastal aluminium
concentrations, it is not possible to make a good a priori estimate of the
amount of sediment input needed for a better simulation.  Since dust input and
internal processes contain parameters which are, just like sediment input, not
completely constrained, a change in sediment input could result in a simulation
which predicts the open ocean aluminium concentration better, but does a bad job
in other areas.
Nevertheless, consistent with work by \citet{moran1991,moran1992,middag2011},
there are clearly some areas where sediment supply of \dAl appears important and
a better understanding of the processes governing its supply is needed.

\subsection{Internal coefficients}  \label{sec:discuss_kd}  \label{sec:discuss_slow}
For the experiments where we changed two internal
parameters ($k_d$ and $\kappa$) the simulations can now be analysed in a more
sophisticated manner by using the above timescale approach.

\begin{figure}
    \centering  %TODO: update in January
    \includegraphics[width=\columnwidth]{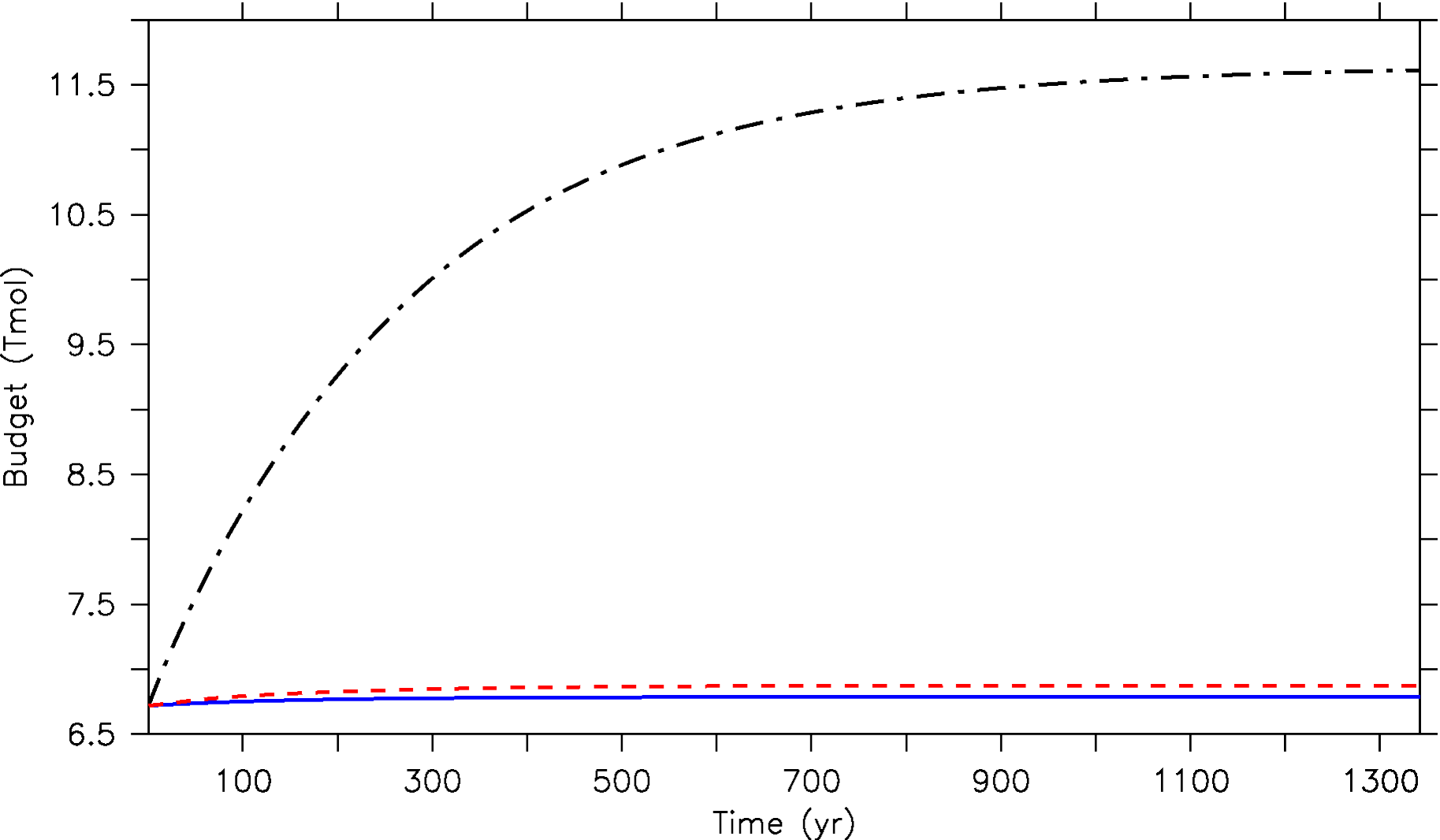}
    \caption{Total Al budget (Tmol) in the world ocean after a partial spin-up
    of 600 yr of the reference simulation.  From this point the reference
    simulation together with the sensitivity experiments is run for another
    1300 yr or more.  The blue solid line is the budget of the reference
    simulation, the red dashed line of the slow-equilibration simulation and
    the black dash-dotted line of the simulation with a decreased partition
    coefficient.}
    \label{fig:budget_ocean_kd_and_slow}
\end{figure}

%\subsubsection{Decreased partition coefficient}    \label{sec:discuss_kd}
As can be seen from the black dash-dotted line in Fig.\
\ref{fig:budget_ocean_kd_and_slow}, the effect of a halved partition coefficient
$k_d$ on the whole ocean is an almost doubling of the total Al budget.
According to Eq.\ \ref{eqn:eq-gehlen} indeed a halved $k_d$ means half as much
$[\ads^\text{eq}]$.  In all experiments equilibration is very fast, except the
one where $\kappa$ is decreased significantly.  Because of the fast
equilibration, with halved $k_d$, export will be half as fast.  In a steady
state, this would result in a doubled \dAl content, if \bsi would be distributed
homogeneously.  Since it is not, this does not hold for our model.  At the dust
deposition site in the Atlantic
Ocean, Fig.\ \ref{fig:kd_SENS_surface} shows an increase of [\dAl] of less
than 100\% in the Atlantic Ocean near the dust deposition site.  Because of
the reduced $k_d$ more \dAl can be transported northward.  When it reaches large
[\bsi], it is scavenged with twice the speed compared to the reference
simulation.  Since this extra \dAl was able to reach the high [\bsi] site, the
effect of the reduced $k_d$ is dampened and therefore the ocean Al budget is
less than doubled.

\begin{figure}
    \centering
    \subfigure[decreased $k_d$ ($k_d = 2\cdot 10^6$ l/kg)]{
        \includegraphics[width=\columnwidth]{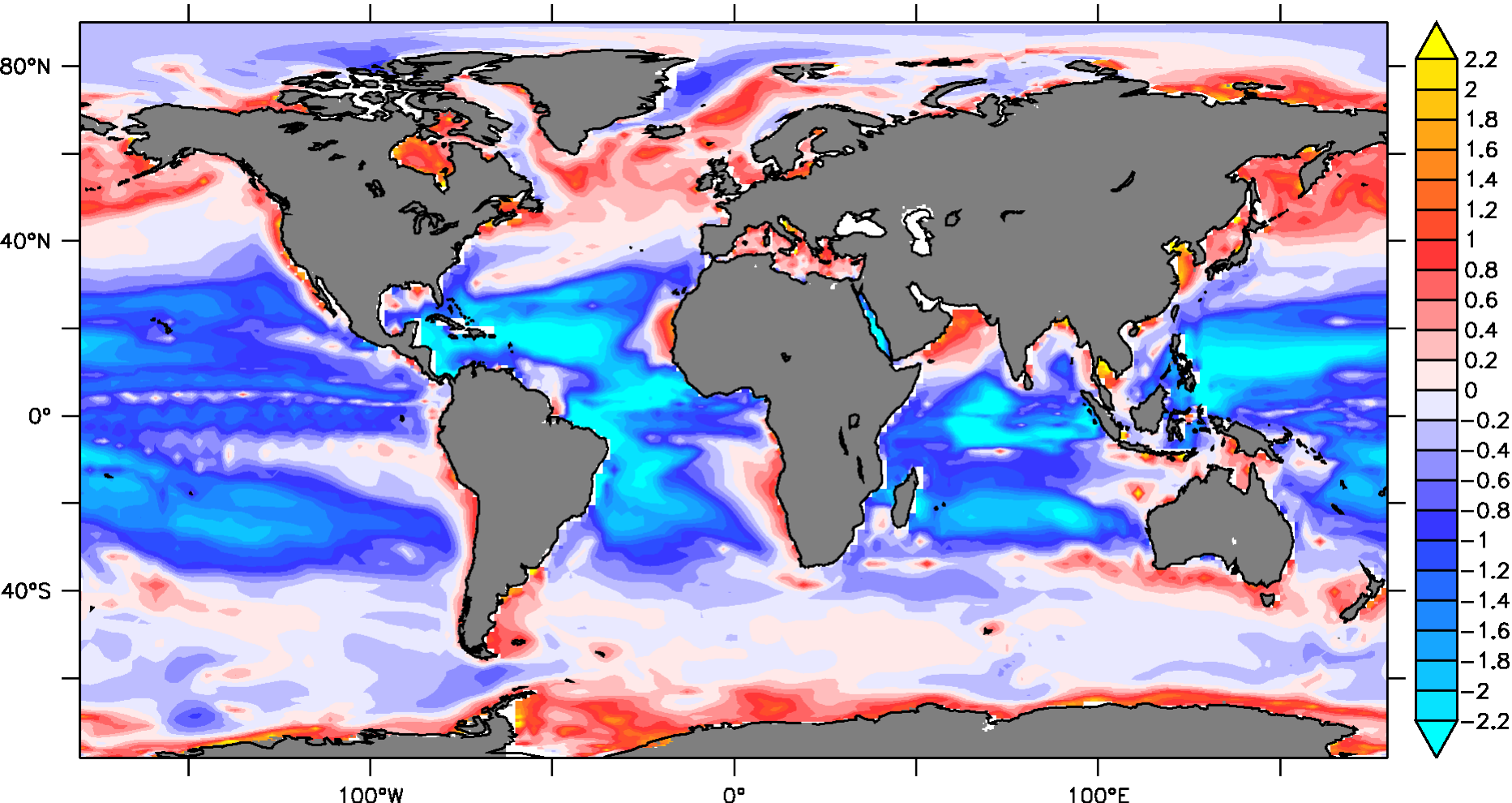}
        \label{fig:adv_vs_scav_kd}
    }
    \subfigure[decreased $\kappa$ ($\kappa = 100$ yr$^{-1}$)]{
        \includegraphics[width=\columnwidth]{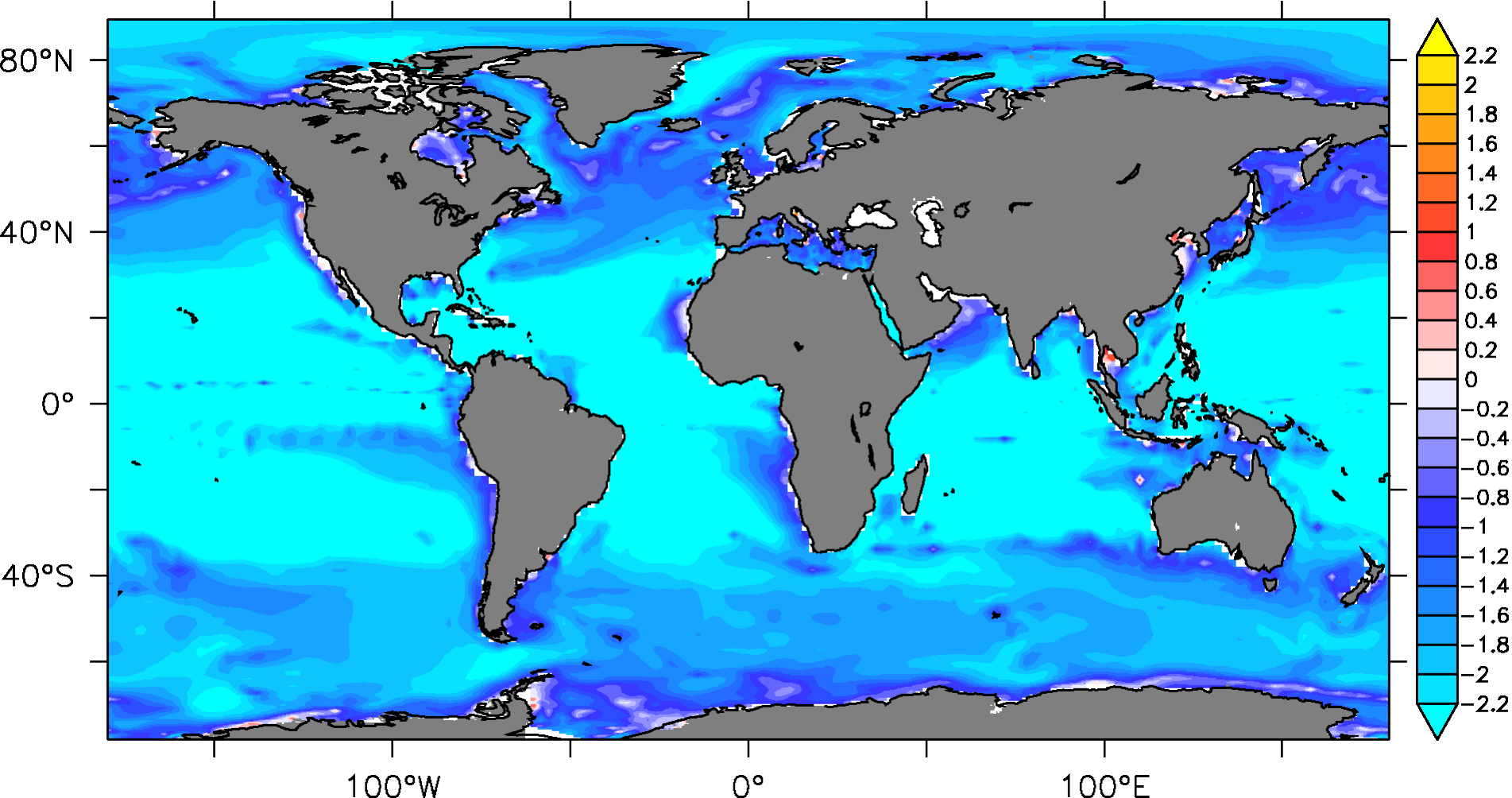}
        \label{fig:adv_vs_scav_slow}
    }
    \caption{log$_{10}(\Upsilon)$, for different adsorption parameters.
    For the advection time $\tau_\text{adv}$ the definition of Eq.\
    \ref{eqn:tadv} is used, and the scavenging time $\tau_\text{scav}$ is the
    maximum of $\tau_\text{ads}$ and $\tau_\text{sink}$.}
\end{figure}

Since the relative effect of advection compared to scavenging can be different
in this simulation, $\Upsilon$ should be analysed.  This timescale ratio is
proportional to $k_d$ and [\bsi] (see Eq.\ \ref{eqn:adv_vs_scav}).
Halving $k_d$ results in only a small change in importance of scavenging relative
to advection (compare Fig.\ \ref{fig:adv_vs_scav} with Fig.\
\ref{fig:adv_vs_scav_kd}).

%\subsubsection{Decreased equilibration speed}     \label{sec:discuss_slow}
A decrease in the first order rate constant $\kappa$ means that equilibration
goes slower (Eq.\ \ref{eqn:scav-gehlen}).  If dust is dissolved in the surface
ocean, its conversion to \ads is slower, so that \dAl is scavenged slower.
Because of the resulting higher [\dAl], more aluminium is transported northward
by the North Atlantic Current.  Fig.\ \ref{fig:BSi_surface} shows the biogenic
silica (\bsi) concentration according to the model at the surface.  In the
Atlantic Ocean are high concentrations of \bsi north of 40\degree N.
The dissolved Al which arrives in this area, is scavenged by this high [\bsi].
This results in high concentrations at all depths around 40\degree N.  However,
because of slow equilibration, \dAl can go further north before it is actually
scavenged.  This is the reason why a significant increase in [\dAl] is visible
around 60\degree N at all depths, and also further south at a depth around 2 km
because this extra Al is advected southward via the MOC
(Fig.~\ref{fig:slow_section_rel}).
Also desorption goes slower, which means that \dAl that is adsorbed at the
surface around 40\degree N, will desorb slowly while sinking, such that compared
to the reference simulation a relatively large quantity of \dAl will appear in
the deep Atlantic Ocean.  This can be seen in the lower right of Fig.\
\ref{fig:slow_section_rel}.

According to Fig.\ \ref{fig:slow_section_rel} there is a decrease of [\dAl]
at low latitudes around 500 m and 1 km depth.  Less dissolved Al from dust in
the surface ocean is scavenged and therefore less adsorbed aluminium is present
in the deeper regions to desorb at depth.
So in the deep ocean there are regions of increased [\dAl] and regions of
decreased [\dAl].

If a lot of particulate aluminium sinks where there is almost no biogenic
silica, it will desorb and the resulting \dAl will stay there.  According to the
equations, with a very large $\kappa$
this will happen instantaneously and the dissolved aluminium will stay there
indefinitely (in the limit of zero biogenic silica and zero advection), i.e.,
$\tau_\text{res} = \infty$.  This really means, as discussed in Section
\ref{sec:timescale}, that in that case advection is much more important than
scavenging.  If $\kappa$ is decreased, the particulate aluminium has enough time
to fall out of this low biogenic silica domain before it desorbs.

In Fig.\ \ref{fig:adv_vs_scav_slow} the logarithm of the relative importance of
scavenging, $\Upsilon$, is plotted for the slow equilibration experiment.  It
shows that everywhere advection is more important than scavenging.
As a consequence, the \dAl distribution obtained with a low $\kappa$ is more
homogeneous than in the reference experiment.
As can be seen from the dashed red line in Fig.\
\ref{fig:budget_ocean_kd_and_slow}, the relative increase of the Al budget is
very small.  The only strong increase is in the surface ocean.  In the
rest of the ocean, [\dAl] is just homogenised.

\subsection{Al versus Si in the MOC}  \label{sec:moc}
% --- TODO: MOVE THIS TO GENERAL DISCUSSION ---
%Our modelling results in the Atlantic Ocean show that around 2 km depth [\dAl]
%decreases from north to south, while [Si$_\text{diss}$] increases, all the way
%into the Pacific Ocean, following the 
%In Fig.\ \ref{fig:contour_on_model} the Atlantic Overturning Stream Function
%is contoured onto the modelled [\dAl].  The depth of the maximum [\dAl] is at
%about 2 km, consistent with the physics used in the model.
% -^- TODO: MOVE THIS TO GENERAL DISCUSSION -^-

\begin{figure}
    \includegraphics[width=\columnwidth]{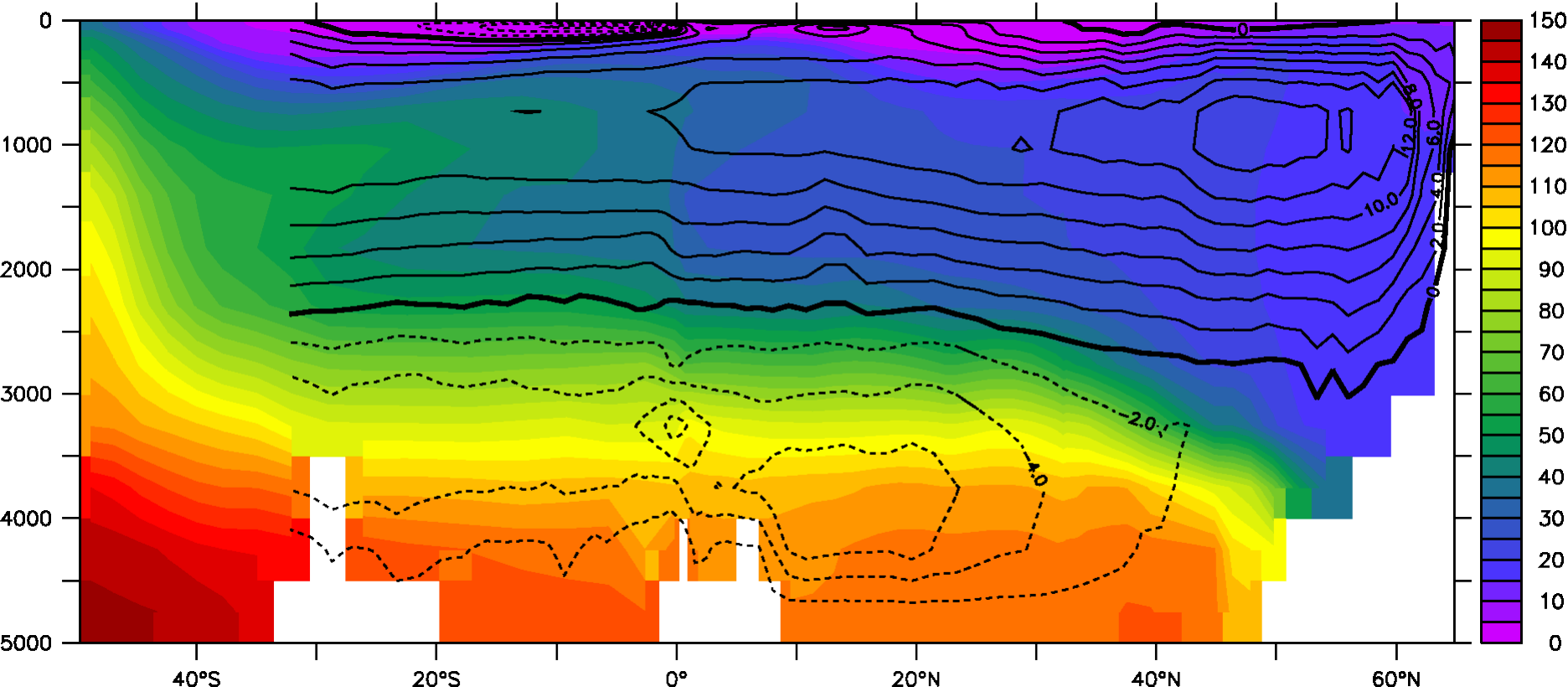}
    \label{fig:moc_on_si}
    \caption{The quantity in the coloured area is
    [\dSi] ($\mu$M) at the Geotraces West Atlantic cruise section.
    The contours represent the Atlantic Overturning Stream Function (Sv).}
\end{figure}

Our modelling results in the Atlantic Ocean show that at around 2 km depth
[\dAl] decreases from north to south, while the concentration of dissolved
silicon (\dSi) increases, all the way into the Pacific Ocean, both following the
Meridional Overturning Circulation (MOC).  This is shown in Figs.\
\ref{fig:moc_on_si} and \ref{fig:contour_on_model}, representing the
concentrations of \dSi and \dAl respectively in the West Atlantic Ocean.  On top
of these figures a contour of the Atlantic Overturning Stream Function (OSF) in
Sv (1 Sv = $10^6$ m$^3 \text{s}^{-1}$) of the physical forcing is plotted.

% TODO?: Possibly remove this alinea (suggestion reviewer #2):
%        But it is a good explanation and synthesis.
The patterns for both [\dAl] and [\dSi] are in first order easily
interpreted by looking at the source of the tracer and the general currents in
the Atlantic Ocean.  The source of \dAl is dust dissolution in the surface of
the central Atlantic Ocean.  It is transported northward by the Gulf Stream and
the North Atlantic Current.  Before the \dAl reaches locations where NADW is
formed, it is scavenged by biogenic silica from 40\degree N northward (Fig.\
\ref{fig:BSi_surface}) and based on our timescale analysis (Fig.\
\ref{fig:adv_vs_scav}) possibly already near 30\degree N.  Because of
remineralisation [\dAl] increases in the NADW.  This is consistent
with the fact that in the model \dAl does not nicely follow the Atlantic
MOC in the very north of the Atlantic Ocean, but rather
tends to sink into the NADW indeed near 30\degree N.  After this the
\dAl is taken along with the NADW, reaching 40\degree S at around 2 km depth.
To complete the picture of the modelled \dAl distribution, low [\dAl] SAAMW/AAIW
enters just below the surface of the Atlantic Ocean from the Southern Ocean, and
low [\dAl] AABW into the Atlantic Ocean below 4 km depth.

The source of \dSi is the large amount of diatoms that sink and remineralise south
of 50\degree S, after which it is advected as AABW as clearly shown in Fig.\
\ref{fig:moc_on_si}.  Other features visible are the SAAMW/AAIW moving below the
surface of the Atlantic Ocean and low [\dSi] NADW diluting the
dissolved Si concentration in the Atlantic Ocean.
The pattern of [\dSi] is similar to the observations described by
\citet{middag_prep}.  The main difference is that the model
overestimates [\dSi] in the deep North Atlantic Ocean, and is a
little bit too low in the deep South Atlantic Ocean.  This is consistent with
the overestimated AABW inflow in the model velocity field.
According to observations, in the lower cell of the Atlantic MOC, the OSF
reaches a maximum of 1 or 2 Sv, while according to our physical model forcing
its maximum is about 6 Sv.  Furthermore, the AABW does not go as far north as in
our model.
Because of this the transport of any tracer following the pathway of AABW will
be too large.  Therefore [\dSi] is overestimated in the deep North Atlantic
Ocean.
Coming back to the discussion in Section \ref{sec:comp.observation}, because of
the same reasons [\dAl] is too quickly diluted in the deep North Atlantic Ocean.

The underestimation of [\dAl] in the deep North Atlantic Ocean is more
significant than the overestimation of [\dSi] in this same area.  This suggests
there might be another process playing a role.  A likely candidate is an
adjusted depth-depending sinking velocity of \bsi and \ads which is
depth-increasing in our current model.  Recent observations
\citep{mcdonnell2010,mcdonnell2011} suggest that sinking speed is not strictly
depth-increasing, but other functions with depth should be considered.  Also
more sophisticated aggregation methods should be studied, as for instance done in models by
\citet{kriest1999,gehlen2006,burd2009}, even though none of these specific efforts give results
consistent with the study by \citet{mcdonnell2010}.  Another possibility is that
the model needs sediment sources more sophisticated than in our sensitivity
experiment, as also shown by observations by \citet{moran1992,middag_prep}
and references therein.

%% TODO: remove next paragraph, since our residence time is incommensurable with
%%       their definition.
%\citet{gehlen2003} and \citet{han2008} calculated the surface residence times
%without taking advection into account, even though in some regions advection is
%quite important.  We get a similar distribution of the residence time as they
%do, and its global average is in the order of about 1 month to 10 yr.
%% This is much shorter than earlier estimates by \citet{orians1985}.
%%       N.B.: Orians spreekt over global ocean, met tres = 100--200 yr

% Conclusions, version 2.0
% 
\section{Conclusions and outlook} \label{sec:conclusion}
The objective of this study is to come to a better understanding of the
behaviour of Al, by means of simulating the \dAl distribution in the ocean with
a reversible scavenging and general circulation model, and comparing it to
observations.  In turn, these results can be used for further developing this
model to simulate the Al distribution more precisely.  A more realistic
simulation could then be used for constraining dust deposition fields to more
precisely derive nutrient input rates.  It can also be used to study further the
influence of sediments on the water column [\dAl], e.g.\ by means of coupling
the ocean model to a sediment model.

The biogeochemical model PISCES is run off-line, forced by a climatological
velocity field with a temporal resolution of five days and a monthly dust
deposition field.  In one of the sensitivity experiments a margin sediment
source is included, which has not been done before for Al.
We are able to simulate the main features of the global \dAl distribution in
accordance with available observations.  Specifically we are able to simulate
reasonably well the distribution in the West Atlantic compared to observations
from the West Atlantic Geotraces cruises.
%
%The comparison with previous model studies \citep{gehlen2003,han2008} shows an
%improved representation of DAl distributions in our model.
% TODO?: examples of improvements
%
Since our results are close to observations, we can assume that the Al
distribution is indeed mainly controlled by advection and reversible scavenging,
with \bsi as the main scavenger.

It is possible to improve specific features of the distribution by changing
certain parameters of the scavenging process, or adjusting the Al sources.
Increasing dust Al dissolution results in an overall aluminium increase
proportional to the increased dust factor everywhere in the ocean.  Adding
dissolution in the water column gives higher concentrations, especially near
dust deposition sites around one or two km depth.  Decreasing the partition
coefficient $k_d$ results in a higher concentration of \dAl everywhere in the
ocean.  Especially the relative increase in the \dAl concentration in the
Southern Ocean is large.  This parameter highlights the importance of the
influence of spatial variability in biogenic silica on [\dAl], since $k_d$
signifies the amount of \dAl which can adsorb onto \bsi.  When including
sediments as an input to \dAl, elevated concentrations are seen, especially near
the margins, as expected.  [\dAl] is strongly elevated in the Arctic Ocean,
which is not reconcilable with the observations, and is possibly due to the
lack of oceanic-margin scavengers in the model.  The model should be modified
such that a realistically constrained ocean margin source of \dAl is present.

By means of a timescale analysis based on our model equations, we have shown
that the importance of scavenging versus advection is highly location
dependent.  The local residence time is strongly space dependent as
well.  Especially latitudinal gradients are large.  This residence time varies
between less than a week at high [\bsi] and strong advection, to many years in
the oligotrophic gyres.

%interplay between scavenging and advection can explain the observed meridional
%distribution of \dAl in intermediate waters in the Atlantic Ocean.
%Even though Al and Si are partly coupled in the internal processes in the ocean,
%this meridional distribution is more or less opposite to that of Si.  This is
%partly because of strong scavenging of Al at specific locations and the
%recycling of Si.
%
%
%The cause of the opposite ocean distributions of [Si$_\text{diss}$] versus
%[\dAl] is a combination of the sources of Si$_\text{diss}$ (remineralised
%diatoms in the Southern Ocean, taken with the AABW) and \dAl (dust in the centre
%Atlantic Ocean), and the Meridional Overturning Circulation.
% FIXME: The above paragraph breaks up the flow of the text.  Maybe I should
% remove this; is it trivial?; is it correct?
% AT: my concern is that you don't spend a lot of time discussing Al vs Si, so it seems 
% strange to bring this up in the concl?

Even though the most relevant features of the \dAl distribution are captured,
some fine-tuning of the key parameters is necessary.  There is clearly a
localised source missing in the West Atlantic between 45--50\degree N near the
sediment.  As already hypothesised by \citet{moran1991}, resuspension of nepheloid
layers along the western boundary of the North Atlantic Ocean is a source of
\dAl.  A more detailed analysis of the locally elevated [\dAl] in the West
Atlantic near 45\degree N is in progress \citep{middag_prep}.

Another location of interest is the Arctic Ocean, where we could not simulate
the observations.  This is possibly because we miss a sediment source, or an
important process is missing in the model, namely the biological incorporation
of Al into the diatom's frustules \citep{caschetto1979,gehlen2002}.  If this
process would be implemented in the model, dissolved Al could be taken up by diatoms
in the surface ocean and released in the deep ocean where the biogenic silica is
remineralised together with the incorporated Al, yielding a strong Al\,:\,Si
correlation as present in the observations \citep{middag2009}.

Good progress is made in simulating the distribution of dissolved aluminium in
the world ocean, and our approach confirms that dust deposition is the main
source of aluminium and reversible scavenging is the main process in removing
it.  However, significant improvement might be possible by (1) developing a
more sophisticated model for ocean sediment source of Al, (2) adding diatom Al
incorporation, and (3) fine tuning the free parameters and constraining them by
means of field studies and laboratory experiments to better understand the role
for these processes.

%\end{linenumbers}
% Acknowledgements, version 1.0
% History:
%   23 aug  0.0 initial version adapted from Liege colloquium abstract
%   28 sept 0.1 better version, and included NWO acknowledgement
%   30 aug  1.0 submission
%   21 dec  1.1 added acknowledgement of use of Ferret
%   3 febr  2.0 submission
%   5 juni  2.1 some reasonable changes
% 
\section*{Acknowledgements}
The comments of two anonymous reviewers significantly improved this manuscript.
The authors are grateful to those who have been proved useful in discussion,
among which Micha Rijkenberg, Laurent Bopp, Jack Middelburg and Wilco Hazeleger.
Furthermore, we want to thank our colleagues from the IMAU institute for their
critical and useful feedback to the model configuration and output.  We also
want to thank the people who kindly provided their datasets that are mentioned
in this paper.  Christoph Heinze kindly dug up the model data from
\cite{gehlen2003} and made it available to us.
Jasmijn van Huis improved the language by proofreading several article sections.

The authors wish to acknowledge use of the Ferret free software program for
analysis and graphics in this paper. Ferret is a product of NOAA's Pacific
Marine Environmental Laboratory (\url{http://ferret.pmel.noaa.gov/Ferret/}).
Other noteworthy free software used is the \emph{GNU Operating System} and
\emph{Climate Data Operators} (cdo).
This research is funded by the Netherlands Organisation for Scientific Research
(NWO), grant Nr.\ 839.08.414.

This work is licensed under \emph{Creative Commons Attribution-ShareAlike 3.0 Unported License} (CC BY-SA). This note overrides other licenses, giving you true freedom for this preprint.

\appendix
\section{Supplementary data}
Supplementary data to this article can be found online at \url{http://dx.doi.org/10.1016/j.jmarsys.2012.05.005}.

%\section*{References}
\bibliographystyle{elsart-harv}
\bibliography{articles,books}

\begin{thebibliography}{92}
\expandafter\ifx\csname natexlab\endcsname\relax\def\natexlab#1{#1}\fi
\expandafter\ifx\csname url\endcsname\relax
  \def\url#1{\texttt{#1}}\fi
\expandafter\ifx\csname urlprefix\endcsname\relax\def\urlprefix{URL }\fi

\bibitem[{Anderson(2006)}]{anderson2006}
Anderson, R., 2006. Chemical tracers of particle transport. In: Elderfield, H.
  (Ed.), The Oceans and Marine Geochemistry. Vol.~6. Elsevier, Ch. 6.09, pp.
  247--274.
\newline\urlprefix\url{http://dx.doi.org/10.1016/B0-08-043751-6/06111-9}

\bibitem[{Arsouze et~al.(2009)Arsouze, Dutay, Lacan, and Jeandel}]{arsouze2009}
Arsouze, T., Dutay, J.-C., Lacan, F., Jeandel, C., 2009. {Reconstructing the Nd
  oceanic cycle using a coupled dynamical--biogeochemical model}.
  Biogeosciences 6~(12), 2829--2846.
\newblock \doi{10.5194/bg-6-2829-2009}.

\bibitem[{Aumont and Bopp(2006)}]{aumont2006}
Aumont, O., Bopp, L., 2006. {Globalizing results from ocean in situ iron
  fertilization studies}. Global Biogeochemical Cycles 20~(2).
\newblock \doi{10.1029/2005GB002591}.

\bibitem[{Aumont et~al.(2008)Aumont, Bopp, and Schulz}]{aumont2008}
Aumont, O., Bopp, L., Schulz, M., apr 2008. What does temporal variability in
  aeolian dust deposition contribute to sea-surface iron and chlorophyll
  distributions? Geophys. Res. Lett. 35~(7), L07607--.
\newblock \doi{10.1029/2007GL031131}.

\bibitem[{Bacon and Anderson(1982)}]{bacon1982}
Bacon, M., Anderson, R., 1982. Distribution of thorium isotopes between
  dissolved and particulate forms in the deep sea. J. Geophys. Res. 87~(C3),
  2045--2056.
\newblock \doi{10.1029/JC087iC03p02045}.

\bibitem[{Bacon et~al.(1988)Bacon, Roether, and Elderfield}]{bacon1988}
Bacon, M., Roether, W., Elderfield, H., 1988. Tracers of chemical scavenging in
  the ocean: Boundary effects and large-scale chemical fractionation [and
  discussion]. Philosophical Transactions of the Royal Society of London.
  Series A, Mathematical and Physical Sciences 325~(1583), 147.
\newblock \doi{10.1098/rsta.1988.0048}.

\bibitem[{Baker et~al.(2006)Baker, Jickells, Witt, and Linge}]{baker2006}
Baker, A., Jickells, T., Witt, M., Linge, K., 2006. Trends in the solubility of
  iron, aluminium, manganese and phosphorus in aerosol collected over the
  {A}tlantic {O}cean. Marine Chemistry 98~(1), 43 -- 58.
\newblock \doi{10.1016/j.marchem.2005.06.004}.

\bibitem[{Boyd et~al.(2007)Boyd, Jickells, Law, Blain, Boyle, Buesseler, Coale,
  Cullen, De~Baar, Follows, et~al.}]{boyd2007}
Boyd, P., Jickells, T., Law, C., Blain, S., Boyle, E., Buesseler, K., Coale,
  K., Cullen, J., De~Baar, H., Follows, M., et~al., 2007. Mesoscale iron
  enrichment experiments 1993-2005: Synthesis and future directions. science
  315~(5812), 612--617.
\newblock \doi{10.1126/science.1131669}.

\bibitem[{Broecker and Peng(1982)}]{broecker1982}
Broecker, W., Peng, T.-H., 1982. Tracers in the Sea. Eldigio Press.

\bibitem[{Brown et~al.(2010)Brown, Lippiatt, and Bruland}]{brown2010aluminium}
Brown, M., Lippiatt, S., Bruland, K., 2010. Dissolved aluminum, particulate
  aluminum, and silicic acid in northern {Gulf of Alaska} coastal waters:
  Glacial/riverine inputs and extreme reactivity. Marine Chemistry 122~(1),
  160--175.
\newblock \doi{10.1016/j.marchem.2010.04.002}.

\bibitem[{Bruland and Lohan(2006)}]{bruland2006}
Bruland, K., Lohan, M., 2006. Controls of trace metals in seawater. In:
  Elderfield, H. (Ed.), The Oceans and Marine Geochemistry. Vol.~6. Elsevier,
  Ch. 6.02, pp. 23--47.

\bibitem[{Burd and Jackson(2009)}]{burd2009}
Burd, A., Jackson, G., 2009. Particle aggregation. Annual review of marine
  science 1, 65--90.
\newblock \doi{10.1146/annurev.marine.010908.163904}.

\bibitem[{Caschetto and Wollast(1979)}]{caschetto1979}
Caschetto, S., Wollast, R., 1979. {Vertical distribution of dissolved aluminium
  in the Mediterranean Sea}. Marine Chemistry 7~(2), 141--155.
\newblock \doi{10.1016/0304-4203(79)90006-9}.

\bibitem[{Chou and Wollast(1997)}]{chou1997}
Chou, L., Wollast, R., 1997. Biogeochemical behavior and mass balance of
  dissolved aluminum in the western {Mediterranean Sea}. Deep Sea Research Part
  II: Topical Studies in Oceanography 44~(3-4), 741 -- 768.
\newblock \doi{10.1016/S0967-0645(96)00092-6}.

\bibitem[{{d}e Baar et~al.(2005){d}e Baar, Boyd, Coale, Landry, Tsuda, Assmy,
  Bakker, Bozec, Barber, Brzezinski, et~al.}]{debaar2005}
{d}e Baar, H., Boyd, P., Coale, K., Landry, M., Tsuda, A., Assmy, P., Bakker,
  D., Bozec, Y., Barber, R., Brzezinski, M., et~al., 2005. Synthesis of iron
  fertilization experiments: from the iron age in the age of enlightenment. J.
  Geophys. Res 110, C09S16.
\newblock \doi{10.1029/2004JC002601}.

\bibitem[{Dittert et~al.(2005)Dittert, Corrin, Bakker, Bendtsen, Gehlen,
  Heinze, Maier-Reimer, Michalopoulos, Soetaert, and Tol}]{dittert2005}
Dittert, N., Corrin, L., Bakker, D., Bendtsen, J., Gehlen, M., Heinze, C.,
  Maier-Reimer, E., Michalopoulos, P., Soetaert, K., Tol, R., 2005. Integrated
  Data Sets of the EU FP5 Research Project: ORFOIS: Origin and Fate of Biogenic
  Particle Fluxes in the Ocean and Their Interactions with Atmospheric CO2
  Concentrations as Well as the Marine Sediment. Vol.~1. WDC-MARE Reports 0002,
  Alfred Wegener Institute for Polar and Marine Research.

\bibitem[{Dixit and Cappellen(2002)}]{dixit2002}
Dixit, S., Cappellen, P.~V., 2002. Surface chemistry and reactivity of biogenic
  silica. Geochimica et Cosmochimica Acta 66~(14), 2559 -- 2568.
\newblock \doi{10.1016/S0016-7037(02)00854-2}.

\bibitem[{Dixit et~al.(2001)Dixit, Van~Cappellen, and van Bennekom}]{dixit2001}
Dixit, S., Van~Cappellen, P., van Bennekom, A., 2001. {Processes controlling
  solubility of biogenic silica and pore water build-up of silicic acid in
  marine sediments}. Marine Chemistry 73~(3-4), 333--352.
\newblock \doi{10.1016/S0304-4203(00)00118-3}.

\bibitem[{Dutay et~al.(2002)Dutay, Bullister, Doney, Orr, Najjar, Caldeira,
  Campin, Drange, Follows, Gao, et~al.}]{dutay2002}
Dutay, J., Bullister, J., Doney, S., Orr, J., Najjar, R., Caldeira, K., Campin,
  J., Drange, H., Follows, M., Gao, Y., et~al., 2002. Evaluation of ocean model
  ventilation with cfc-11: comparison of 13 global ocean models. Ocean
  Modelling 4~(2), 89--120.
\newblock \doi{10.1016/S1463-5003(01)00013-0}.

\bibitem[{Dutay et~al.(2009)Dutay, Lacan, Roy-Barman, and Bopp}]{dutay2009}
Dutay, J., Lacan, F., Roy-Barman, M., Bopp, L., 2009. {Influence of particle
  size and type on 231Pa and 230Th simulation with a global coupled
  biogeochemical-ocean general circulation model: A first approach}.
  Geochemistry Geophysics Geosystems 10~(1), Q01011.
\newblock \doi{10.1029/2008GC002291}.

\bibitem[{Emerson and Hedges(2006)}]{emerson2006}
Emerson, S., Hedges, J., 2006. Sediment diagenesis and benthic flux. In:
  Elderfield, H. (Ed.), The Oceans and Marine Geochemistry. Vol.~6. Elsevier,
  Ch. 6.11, pp. 293--319.

\bibitem[{Eth\'e et~al.(2006)Eth\'e, Aumont, Foujols, and L\'evy}]{ethe2006}
Eth\'e, C., Aumont, O., Foujols, M.-A., L\'evy, M., 2006. {NEMO reference
  manual, tracer component : NEMO-TOP. Preliminary version}. Note du Pole de
  Mod{\'e}lisation, Institut Pierre-Simon Laplace.

\bibitem[{Filella(2007)}]{colloids:filella2007}
Filella, M., 2007. Colloidal properties of submicron particles in natural
  waters. In: Wilkinson, K., Leas, J. (Eds.), Environmental Colloids and
  Particles: Behaviour, Separation and Characterisation. Vol.~10. John Wiley \&
  Sons, Ltd, Ch.~2, pp. 17--94.

\bibitem[{Gary et~al.(2011)Gary, Lozier, B\"oning, and Biastoch}]{gary2011}
Gary, S.~F., Lozier, M.~S., B\"oning, C.~W., Biastoch, A., 2011. Deciphering
  the pathways for the deep limb of the {Meridional Overturning Circulation}.
  Deep Sea Research Part II: Topical Studies in Oceanography 58~(17-18), 1781
  -- 1797.
\newblock \doi{10.1016/j.dsr2.2010.10.059}.

\bibitem[{Gehlen et~al.(2002)Gehlen, Beck, Calas, Flank, Bennekom, and
  Beusekom}]{gehlen2002}
Gehlen, M., Beck, L., Calas, G., Flank, A., Bennekom, A.~v., Beusekom, J.~v.,
  2002. {Unraveling the atomic structure of biogenic silica: evidence of the
  structural association of Al and Si in diatom frustules}. Geochimica et
  Cosmochimica Acta 66~(9), 1601--1609.
\newblock \doi{10.1016/S0016-7037(01)00877-8}.

\bibitem[{Gehlen et~al.(2006)Gehlen, Bopp, Emprin, Aumont, Heinze, and
  Ragueneau}]{gehlen2006}
Gehlen, M., Bopp, L., Emprin, N., Aumont, O., Heinze, C., Ragueneau, O., 2006.
  Reconciling surface ocean productivity, export fluxes and sediment
  composition in a global biogeochemical ocean model. Biogeosciences 3~(4),
  521--537.
\newline\urlprefix\url{http://www.biogeosciences.net/3/521/2006/}

\bibitem[{Gehlen et~al.(2007)Gehlen, Gangst{\o}, Schneider, Bopp, Aumont, and
  Ethe}]{gehlen2007}
Gehlen, M., Gangst{\o}, R., Schneider, B., Bopp, L., Aumont, O., Ethe, C.,
  2007. The fate of pelagic {CaCO3} production in a high {CO2} ocean: a model
  study. Biogeosciences 4~(4), 505--519.

\bibitem[{Gehlen et~al.(2003)Gehlen, Heinze, Maier-Reimer, et~al.}]{gehlen2003}
Gehlen, M., Heinze, C., Maier-Reimer, E., et~al., 2003. {Coupled Al-Si
  geochemistry in an ocean general circulation model: A tool for the validation
  of oceanic dust deposition fields?} Global Biogeochemical Cycles 17~(1),
  1028.
\newblock \doi{10.1029/2001GB001549}.

\bibitem[{Goldberg(1954)}]{goldberg1954}
Goldberg, E., 1954. {Marine geochemistry 1. Chemical scavengers of the sea}.
  The Journal of Geology 62~(3), 249--265.

\bibitem[{Guerzoni et~al.(1997)Guerzoni, Molinaroli, and
  Chester}]{guerzoni1997}
Guerzoni, S., Molinaroli, E., Chester, R., 1997. Saharan dust inputs to the
  western {Mediterranean Sea}: depositional patterns, geochemistry and
  sedimentological implications. Deep Sea Research Part II: Topical Studies in
  Oceanography 44~(3-4), 631--654.
\newblock \doi{10.1016/S0967-0645(96)00096-3}.

\bibitem[{Han(2010)}]{han2010}
Han, Q., 2010. Crustal tracers in the atmosphere and ocean: Relating their
  concentrations, fluxes, and ages. Ph.D. thesis, University of California,
  Irvine.
\newline\urlprefix\url{http://adsabs.harvard.edu/abs/2010PhDT.......174H}

\bibitem[{Han et~al.(2008)Han, Moore, Zender, and Hydes}]{han2008}
Han, Q., Moore, J., Zender, C., Hydes, D., 2008. {Constraining oceanic dust
  deposition using surface ocean dissolved Al}. Global Biogeochemical Cycles
  22~(2).
\newblock \doi{10.1029/2007GB002975}.

\bibitem[{Hydes et~al.(1988)Hydes, de~Lange, and {d}e Baar}]{hydes1988}
Hydes, D., de~Lange, G., {d}e Baar, H., 1988. Dissolved aluminium in the
  {M}editerranean. Geochimica et Cosmochimica Acta 52~(8), 2107 -- 2114.
\newblock \doi{10.1016/0016-7037(88)90190-1}.

\bibitem[{Hydes et~al.(1986)Hydes, Statham, and Burton}]{hydes1986}
Hydes, D., Statham, P., Burton, J., 1986. {A vertical profile of dissolved
  trace metals (Al, Cd, Cu, Mn, Ni) over the median valley of the mid Atlantic
  ridge, 43°N: Implications for Hydrothermal activity}. Science of The Total
  Environment 49~(0), 133 -- 145.
\newblock \doi{10.1016/0048-9697(86)90236-6}.

\bibitem[{Jickells(1995)}]{jickells1995}
Jickells, T., 1995. Atmospheric inputs of metals and nutrients to the oceans:
  their magnitude and effects. Marine Chemistry 48~(3-4), 199--214.
\newblock \doi{10.1016/0304-4203(95)92784-P}.

\bibitem[{Jickells et~al.(2005)Jickells, An, Andersen, Baker, Bergametti,
  Brooks, Cao, Boyd, Duce, Hunter, et~al.}]{jickells2005}
Jickells, T., An, Z., Andersen, K., Baker, A., Bergametti, G., Brooks, N., Cao,
  J., Boyd, P., Duce, R., Hunter, K., et~al., 2005. {Global iron connections
  between desert dust, ocean biogeochemistry, and climate}. Science 308~(5718),
  67.
\newblock \doi{10.1126/science.1105959}.

\bibitem[{Jickells et~al.(1994)Jickells, Church, Veron, and
  Arimoto}]{jickells1994}
Jickells, T., Church, T., Veron, A., Arimoto, R., 1994. Atmospheric inputs of
  manganese and aluminium to the {S}argasso {S}ea and their relation to surface
  water concentrations. Marine chemistry 46~(3), 283--292.
\newblock \doi{10.1016/0304-4203(94)90083-3}.

\bibitem[{Kramer et~al.(2004)Kramer, Laan, Sarthou, Timmermans, and
  De~Baar}]{kramer2004}
Kramer, J., Laan, P., Sarthou, G., Timmermans, K., De~Baar, H., 2004.
  {Distribution of dissolved aluminium in the high atmospheric input region of
  the subtropical waters of the North Atlantic Ocean}. Marine Chemistry
  88~(3-4), 85--101.
\newblock \doi{10.1016/j.marchem.2004.03.009}.

\bibitem[{Krauss and Beckmann(1996)}]{krauss1996}
Krauss, W., Beckmann, A., 1996. {The warmwatersphere of the North Atlantic
  Ocean}. Borntraeger.
\newline\urlprefix\url{http://books.google.nl/books?id=qL4PAQAAIAAJ}

\bibitem[{Kremling(1985)}]{kremling1985}
Kremling, K., 1985. {The distribution of cadmium, copper, nickel, manganese,
  and aluminium in surface waters of the open Atlantic and European shelf
  area}. Deep Sea Research Part A. Oceanographic Research Papers 32~(5),
  531--555.
\newblock \doi{10.1016/0198-0149(85)90043-3}.

\bibitem[{Kriest and Evans(1999)}]{kriest1999}
Kriest, I., Evans, G., 1999. Representing phytoplankton aggregates in
  biogeochemical models. Deep Sea Research Part I: Oceanographic Research
  Papers 46~(11), 1841--1859.
\newblock \doi{10.1016/S0967-0637(99)00032-1}.

\bibitem[{Lam(2011)}]{lam2011}
Lam, P., 2011. {The dynamic ocean biological pump: Insights from a global
  compilation of particulate organic carbon, CaCO3, and opal concentration
  profiles from the mesopelagic}. Global Biogeochem. Cycles 25~(3), GB3009.
\newblock \doi{10.1029/2010GB003868}.

\bibitem[{Lead and Wilkinson(2007)}]{colloids:lead2007}
Lead, J., Wilkinson, K., 2007. Environmental colloids and particles: Current
  knowledge and future developments. In: Wilkinson, K., Leas, J. (Eds.),
  Environmental Colloids and Particles: Behaviour, Separation and
  Characterisation. Vol.~10. John Wiley \& Sons, Ltd, Ch.~1, pp. 1--16.

\bibitem[{Lewin(1961)}]{lewin1961}
Lewin, J., 1961. The dissolution of silica from diatom walls. Geochimica et
  Cosmochimica Acta 21~(3-4), 182 -- 198.
\newblock \doi{10.1016/S0016-7037(61)80054-9}.

\bibitem[{Loucaides et~al.(2010)Loucaides, Behrends, and van
  Cappellen}]{loucaides2010}
Loucaides, S., Behrends, T., van Cappellen, P., 2010. Reactivity of biogenic
  silica: Surface versus bulk charge density. Geochimica et Cosmochimica Acta
  74~(2), 517 -- 530.
\newblock \doi{10.1016/j.gca.2009.10.038}.

\bibitem[{Lozier(2010)}]{lozier2010:deconstructing}
Lozier, M., 2010. {Deconstructing the Conveyor Belt}. Science 328~(5985), 1507.
\newblock \doi{10.1126/science.1189250}.

\bibitem[{MacKenzie et~al.(1978)MacKenzie, Stoffyn, and
  Wollast}]{mackenzie1978}
MacKenzie, F., Stoffyn, M., Wollast, R., 1978. Aluminum in seawater: control by
  biological activity. Science 199~(4329), 680.
\newblock \doi{10.1126/science.199.4329.680}.

\bibitem[{Mackin(1986)}]{mackin1986control}
Mackin, J., 1986. Control of dissolved al distributions in marine sediments by
  clay reconstitution reactions: experimental evidence leading to a unified
  theory. Geochimica et Cosmochimica Acta 50~(2), 207--214.
\newblock \doi{10.1016/0016-7037(86)90170-5}.

\bibitem[{Mackin and Aller(1986)}]{mackin1986}
Mackin, J., Aller, R., 1986. The effects of clay mineral reactions on dissolved
  al distributions in sediments and waters of the amazon continental shelf.
  Continental Shelf Research 6~(1-2), 245--262.
\newblock \doi{10.1016/0278-4343(86)90063-4}.

\bibitem[{Madec(2008)}]{madec2008}
Madec, G., 2008. {NEMO ocean engine}. Note du Pole de Mod{\'e}lisation,
  Institut Pierre-Simon Laplace.

\bibitem[{Madec et~al.(1998)Madec, Delecluse, Imbard, and L{\'e}vy}]{madec1998}
Madec, G., Delecluse, P., Imbard, M., L{\'e}vy, C., 1998. {OPA 8.1 Ocean
  General Circulation Model reference manual}. Note du Pole de
  Mod{\'e}lisation, Institut Pierre-Simon Laplace 11, 91p.
\newline\urlprefix\url{http://hal.archives-ouvertes.fr/hal-00154217/en/}

\bibitem[{Mahowald et~al.(1999)Mahowald, Kohfeld, Hansson, Balkanski, Harrison,
  Prentice, Schulz, and Rodhe}]{mahowald1999}
Mahowald, N., Kohfeld, K., Hansson, M., Balkanski, Y., Harrison, S., Prentice,
  I., Schulz, M., Rodhe, H., 1999. {Dust sources and deposition during the last
  glacial maximum and current climate: A comparison of model results with
  paleodata from ice cores and marine sediments}. Journal of Geophysical
  Research 104~(D13), 15895--15.
\newblock \doi{10.1029/1999JD900084}.

\bibitem[{Maier-Reimer et~al.(1993)Maier-Reimer, Mikolajewicz, and
  Hasselmann}]{maier1993}
Maier-Reimer, E., Mikolajewicz, U., Hasselmann, K., 1993. {Mean circulation of
  the Hamburg LSG OGCM and its sensitivity to the thermohaline surface
  forcing}. Journal of Physical Oceanography 23~(4), 731--757.

\bibitem[{Maring and Duce(1987)}]{maring1987}
Maring, H., Duce, R., 1987. The impact of atmospheric aerosols on trace metal
  chemistry in open ocean surface seawater, 1. aluminum. Earth and Planetary
  Science Letters 84~(4), 381 -- 392.
\newblock \doi{10.1016/0012-821X(87)90003-3}.

\bibitem[{Martin(1990)}]{martin1990}
Martin, J., 1990. Glacial-interglacial {CO}2 change: The iron hypothesis.
  Paleoceanography 5~(1), 1--13.
\newblock \doi{10.1029/PA005i001p00001}.

\bibitem[{McAlister and Orians(2011)}]{mcalister2011}
McAlister, J., Orians, K., 2011. Controls on hydroxide-speciated trace metals
  in the ocean.
\newline\urlprefix\url{http://modb.oce.ulg.ac.be/colloquium/2011/abstracts2011%
.pdf}

\bibitem[{McDonnell and Buesseler(2010)}]{mcdonnell2010}
McDonnell, A., Buesseler, K., 2010. Variability in the average sinking velocity
  of marine particles. Limnol. Oceanogr 55~(5), 2085--2096.
\newblock \doi{10.4319/lo.2010.55.5.2085}.

\bibitem[{McDonnell et~al.(2011)}]{mcdonnell2011}
McDonnell, A., et~al., 2011. Marine particle dynamics: sinking velocities, size
  distributions, fluxes, and microbial degradation rates. Ph.D. thesis,
  Massachusetts Institute of Technology.
\newline\urlprefix\url{http://hdl.handle.net/1721.1/65326}

\bibitem[{Measures et~al.(2005)Measures, Brown, and Vink}]{measures2005}
Measures, C., Brown, M., Vink, S., sep 2005. Dust deposition to the surface
  waters of the western and central north pacific inferred from surface water
  dissolved aluminum concentrations. Geochem. Geophys. Geosyst. 6~(9),
  Q09M03--.
\newblock \doi{10.1029/2005GC000922}.

\bibitem[{Measures et~al.(2010)Measures, Sato, Vink, Howell, and
  Li}]{measures2010}
Measures, C., Sato, T., Vink, S., Howell, S., Li, Y., 2010. The fractional
  solubility of aluminium from mineral aerosols collected in hawaii and
  implications for atmospheric deposition of biogeochemically important trace
  elements. Marine Chemistry 120~(1-4), 144--153.
\newblock \doi{10.1016/j.marchem.2009.01.014}.

\bibitem[{Measures and Vink(2000)}]{measures2000}
Measures, C., Vink, 2000. On the use of dissolved aluminum in surface waters to
  estimate dust deposition to the ocean. Global Biogeochem. Cycles 14~(1),
  317--327.
\newblock \doi{10.1029/1999GB001188}.

\bibitem[{Middag(2010)}]{middag2010}
Middag, R., 2010. Dissolved aluminium and manganese in the polar oceans. Ph.D.
  thesis, University of Groningen.

\bibitem[{Middag et~al.(2009)Middag, {de Baar}, Laan, and Bakker}]{middag2009}
Middag, R., {de Baar}, H., Laan, P., Bakker, K., 2009. {Dissolved aluminium and
  the silicon cycle in the Arctic Ocean}. Marine Chemistry 115~(3-4), 176--195.
\newblock \doi{10.1016/j.marchem.2009.08.002}.

\bibitem[{Middag et~al.(2012)Middag, de~Baar, Laan, and Huhn}]{middag2012}
Middag, R., de~Baar, H., Laan, P., Huhn, O., 2012. The effects of continental
  margins and water mass circulation on the distribution of dissolved aluminum
  and manganese in drake passage. J. Geophys. Res. 117~(C1), C01019.
\newblock \doi{10.1029/2011JC007434}.

\bibitem[{Middag et~al.(in preparation)Middag, van Aken, van Hulten, and
  de~Baar}]{middag_prep}
Middag, R., van Aken, H., van Hulten, M., de~Baar, H., in preparation.
  Aluminium in the ocean: unique mirror image of the biological cycle.

\bibitem[{Middag et~al.(2011)Middag, van Slooten, {de Baar}, and
  Laan}]{middag2011}
Middag, R., van Slooten, C., {de Baar}, H., Laan, P., 2011. Dissolved aluminium
  in the southern ocean. Deep Sea Research Part II: Topical Studies in
  Oceanography\doi{10.1016/j.dsr2.2011.03.001}.

\bibitem[{Moore et~al.(2008)Moore, Braucher, et~al.}]{moore2008}
Moore, J., Braucher, O., et~al., 2008. Sedimentary and mineral dust sources of
  dissolved iron to the world ocean. Biogeosciences 5~(3), 631--656.
\newline\urlprefix\url{http://hal.archives-ouvertes.fr/hal-00297688/}

\bibitem[{Moran and Moore(1988)}]{moran1988}
Moran, S., Moore, R., 1988. Temporal variations in dissolved and particulate
  aluminum during a spring bloom. Estuarine, Coastal and Shelf Science 27~(2),
  205 -- 215.
\newblock \doi{10.1016/0272-7714(88)90090-X}.

\bibitem[{Moran and Moore(1989)}]{moran1989}
Moran, S., Moore, R., 1989. {The distribution of colloidal aluminum and organic
  carbon in coastal and open ocean waters off Nova Scotia}. Geochimica et
  Cosmochimica Acta 53~(10), 2519--2527.
\newblock \doi{10.1016/0016-7037(89)90125-7}.

\bibitem[{Moran and Moore(1991)}]{moran1991}
Moran, S., Moore, R., 1991. The potential source of dissolved aluminum from
  resuspended sediments to the north atlantic deep water. Geochimica et
  Cosmochimica Acta 55~(10), 2745 -- 2751.
\newblock \doi{10.1016/0016-7037(91)90441-7}.

\bibitem[{Moran et~al.(1992)Moran, Moore, and Westerlund}]{moran1992}
Moran, S., Moore, R., Westerlund, S., 1992. Dissolved aluminum in the weddell
  sea. Deep Sea Research Part A. Oceanographic Research Papers 39~(3-4),
  537--547.
\newblock \doi{10.1016/0198-0149(92)90087-A}.

\bibitem[{Obata et~al.(2007)Obata, Alibo, and Nozaki}]{obata2007}
Obata, H., Alibo, D., Nozaki, Y., dec 2007. {Dissolved aluminum, indium, and
  cerium in the Sea of Japan and the Sea of Okhotsk: Comparison to the marginal
  seas of the western North Pacific}. J. Geophys. Res. 112~(C12), C12003--.
\newblock \doi{10.1029/2006JC003944}.

\bibitem[{Orians and Bruland(1985)}]{orians1985}
Orians, K., Bruland, K., 1985. {Dissolved aluminium in the central North
  Pacific}. Nature 316~(6027), 427--429.
\newblock \doi{10.1038/316427a0}.

\bibitem[{Orians and Bruland(1986)}]{orians1986}
Orians, K., Bruland, K., 1986. {The biogeochemistry of aluminum in the Pacific
  Ocean}. Earth and planetary science letters 78~(4), 397--410.
\newblock \doi{10.1016/0012-821X(86)90006-3}.

\bibitem[{Ragueneau et~al.(2000)Ragueneau, Tr{\'e}guer, Leynaert, Anderson,
  Brzezinski, DeMaster, Dugdale, Dymond, Fischer, Francois,
  et~al.}]{ragueneau2000}
Ragueneau, O., Tr{\'e}guer, P., Leynaert, A., Anderson, R., Brzezinski, M.,
  DeMaster, D., Dugdale, R., Dymond, J., Fischer, G., Francois, R., et~al.,
  2000. {A review of the Si cycle in the modern ocean: recent progress and
  missing gaps in the application of biogenic opal as a paleoproductivity
  proxy}. Global and Planetary Change 26~(4), 317--365.
\newblock \doi{10.1016/S0921-8181(00)00052-7}.

\bibitem[{Rijkenberg et~al.(2008)Rijkenberg, Powell, Dall'Osto, Nielsdottir,
  Patey, Hill, Baker, Jickells, Harrison, and Achterberg}]{rijkenberg2008}
Rijkenberg, M., Powell, C., Dall'Osto, M., Nielsdottir, M., Patey, M., Hill,
  P., Baker, A., Jickells, T., Harrison, R., Achterberg, E., 2008. {Changes in
  iron speciation following a Saharan dust event in the tropical North Atlantic
  Ocean}. Marine Chemistry 110~(1-2), 56 -- 67.
\newblock \doi{10.1016/j.marchem.2008.02.006}.

\bibitem[{Rijkenberg et~al.(2012)Rijkenberg, Steigenberger, Powell, van Haren,
  Patey, Baker, and Achterberg}]{rijkenberg2012}
Rijkenberg, M., Steigenberger, S., Powell, C., van Haren, H., Patey, M.~D.,
  Baker, A.~R., Achterberg, E., 2012. Fluxes and distribution of dissolved iron
  in the eastern (sub-)tropical north atlantic ocean. Global Biogeochem.
  Cycles\doi{10.1029/2011GB004264}.

\bibitem[{Sarmiento and Gruber(2006)}]{sarmiento2006}
Sarmiento, J., Gruber, N., 2006. {Ocean biogeochemical dynamics}. Princeton
  University Press.

\bibitem[{Schulz et~al.(2009)Schulz, Cozic, and Szopa}]{schulz2009}
Schulz, M., Cozic, A., Szopa, S., 2009. {LMDzT-INCA dust forecast model
  developments and associated validation efforts}. In: IOP Conference Series:
  Earth and Environmental Science. Vol.~7. IOP Publishing, p. 012014.

\bibitem[{Slemons et~al.(2010)Slemons, Murray, Resing, Paul, and
  Dutrieux}]{slemons2010}
Slemons, L., Murray, J., Resing, J., Paul, B., Dutrieux, P., 2010. {Western
  Pacific coastal sources of iron, manganese, and aluminum to the Equatorial
  Undercurrent}. Global Biogeochemical Cycles 24~(3).
\newblock \doi{10.1029/2009GB003693}.

\bibitem[{Stoffyn(1979)}]{stoffyn1979}
Stoffyn, M., 1979. Biological control of dissolved aluminum in seawater:
  experimental evidence. Science 203~(4381), 651.
\newblock \doi{10.1126/science.203.4381.651}.

\bibitem[{Stoffyn and Mackenzie(1982)}]{stoffyn1982}
Stoffyn, M., Mackenzie, F., 1982. Fate of dissolved aluminum in the oceans.
  Marine Chemistry 11~(2), 105 -- 127.
\newblock \doi{10.1016/0304-4203(82)90036-6}.

\bibitem[{Tagliabue et~al.(2010)Tagliabue, Bopp, Dutay, Bowie, Chever,
  Jean-Baptiste, Bucciarelli, Lannuzel, Remenyi, Sarthou, Aumont, Gehlen, and
  Jeandel}]{tagliabue2010}
Tagliabue, A., Bopp, L., Dutay, J.-C., Bowie, A., Chever, F., Jean-Baptiste,
  P., Bucciarelli, E., Lannuzel, D., Remenyi, T., Sarthou, G., Aumont, O.,
  Gehlen, M., Jeandel, C., 2010. Hydrothermal contribution to the oceanic
  dissolved iron inventory. Nature Geoscience 3~(4), 252--256.
\newblock \doi{10.1038/NGEO818}.
\newpage
\bibitem[{Tagliabue et~al.(2011)Tagliabue, Bopp, and Gehlen}]{tagliabue2011}
Tagliabue, A., Bopp, L., Gehlen, M., 2011. The response of marine carbon and
  nutrient cycles to ocean acidification: Large uncertainties related to
  phytoplankton physiological assumptions. Global Biogeochemical Cycles 25~(3),
  GB3017.
\newblock \doi{10.1029/2010GB003929}.

\bibitem[{Textor et~al.(2006)Textor, Schulz, Guibert, Kinne, Balkanski, Bauer,
  Berntsen, Berglen, Boucher, Chin, et~al.}]{textor2006}
Textor, C., Schulz, M., Guibert, S., Kinne, S., Balkanski, Y., Bauer, S.,
  Berntsen, T., Berglen, T., Boucher, O., Chin, M., et~al., 2006. {Analysis and
  quantification of the diversities of aerosol life cycles within AeroCom}.
  Atmos. Chem. Phys 6~(7), 1777.
\newblock \doi{10.5194/acp-6-1777-2006}.

\bibitem[{van Aken(2011)}]{vanaken2011}
van Aken, H., 2011. {GEOTRACES, the hydrography of the Western Atlantic Ocean}.

\bibitem[{van Bennekom et~al.(1991)van Bennekom, Buma, and
  Nolting}]{vanbennekom1991}
van Bennekom, A., Buma, A., Nolting, R., 1991. {Dissolved aluminium in the
  Weddell-Scotia Confluence and effect of Al on the dissolution kinetics of
  biogenic silica}. Marine Chemistry 35~(1-4), 423 -- 434.
\newblock \doi{10.1016/S0304-4203(09)90034-2}.

\bibitem[{van Beusekom et~al.(1997)van Beusekom, Van~Bennekom, Tr{\'e}guer, and
  Morvan}]{vanbeusekom1997}
van Beusekom, J., Van~Bennekom, A., Tr{\'e}guer, P., Morvan, J., 1997.
  Aluminium and silicic acid in water and sediments of the enderby and crozet
  basins. Deep Sea Research Part II: Topical Studies in Oceanography 44~(5),
  987--1003.
\newblock \doi{10.1016/S0967-0645(96)00105-1}.

\bibitem[{Vink and Measures(2001)}]{vink2001}
Vink, S., Measures, C., 2001. {The role of dust deposition in determining
  surface water distributions of Al and Fe in the South West Atlantic}. Deep
  Sea Research Part II: Topical Studies in Oceanography 48~(13), 2787 -- 2809.
\newblock \doi{10.1016/S0967-0645(01)00018-2}.

\bibitem[{Walker et~al.(1988)Walker, Cronan, and Patterson}]{walker1988}
Walker, W., Cronan, C., Patterson, H., 1988. A kinetic study of aluminum
  adsorption by aluminosilicate clay minerals. Geochimica et Cosmochimica Acta
  52~(1), 55--62.
\newblock \doi{10.1016/0016-7037(88)90056-7}.

\bibitem[{Wedepohl(1995)}]{wedepohl1995}
Wedepohl, K., 1995. The composition of the continental crust. Geochimica et
  Cosmochimica Acta 59~(7), 1217--1232.
\newblock \doi{10.1016/0016-7037(95)00038-2}.

\bibitem[{Whitehead et~al.(1998)Whitehead, Wilson, and
  Butterfield}]{whitehead1998}
Whitehead, P., Wilson, E., Butterfield, D., 1998. {A semi-distributed
  Integrated Nitrogen model for multiple source assessment in Catchments
  (INCA): Part I — model structure and process equations}. Science of The
  Total Environment 210-211~(0), 547 -- 558.
\newblock \doi{10.1016/S0048-9697(98)00037-0}.

\end{thebibliography}
\end{document}